\documentclass{aa}  

\usepackage{graphicx}
\usepackage{txfonts}
\usepackage[breaklinks]{hyperref}
\usepackage{xurl}
\usepackage{siunitx}
\usepackage[version=3]{mhchem} 
\usepackage{xcolor}
\usepackage{caption}
\usepackage{subcaption}
\usepackage{booktabs}
\usepackage{multirow}

\DeclareSIUnit\year{yr}
\DeclareSIUnit\bar{bar}
\DeclareSIUnit{\wtpercent}{wt\%}
\DeclareSIUnit\molar{M}

\newcommand{\PB}[1]{\textcolor{black}{#1}}

\begin{document} 

\title{The effect of lightning on the atmospheric chemistry of exoplanets and potential biosignatures}
\titlerunning{Lightning in exoplanet atmospheres}

    \author{Patrick Barth\inst{1,2,3,4,5}
        \and
            Eva E. Stüeken\inst{2,3}
        \and
            Christiane Helling\inst{1,6} 
        \and
            Edward W. Schwieterman\inst{7,8}
        \and 
            Jon Telling\inst{9}
            }

    \institute{Space Research Institute, Austrian Academy of Sciences, Schmiedlstrasse 6, Graz, A-8042, Austria 
            \and
                Centre for Exoplanet Science, University of St Andrews, North Haugh, St Andrews, KY16 9SS, UK
            \and
                School of Earth \& Environmental Sciences, University of St Andrews, Bute Building, Queen's Terrace, St Andrews, KY16 9TS, UK
            \and
                SUPA, School of Physics \& Astronomy, University of St Andrews, North Haugh, St Andrews, KY16 9SS, UK
            \and
                Stuttgart Center for Simulation Science, University of Stuttgart, Pfaffenwaldring 5a, 70569 Stuttgart, Germany\\
                \email{patrick.barth@simtech.uni-stuttgart.de}
            \and
                Fakultät für Mathematik, Physik und Geodäsie, TU Graz, Petersgasse 16, Graz, A-8010, Austria
            \and
                Department of Earth and Planetary Sciences, University of California, Riverside, CA, USA
            \and
                Blue Marble Space Institute of Science, Seattle, WA, USA
            \and
                School of Natural and Environmental Sciences, Newcastle University, Newcastle upon Tyne, NE1 7RU, UK
                }


   \date{Received 26.06.23; accepted 19.02.2024}

 
  \abstract
   {Lightning has been suggested to play a role in triggering the occurrence of bio-ready chemical species.
   Future missions like PLATO, ARIEL, HWO, and LIFE but also the ground-based ELTs will investigate the atmospheres of potentially habitable exoplanets.}
   {We therefore aim to study the effect of lightning on the atmospheric chemistry, how it affects false-positive and false-negative biosignatures, and if its effect would be observable on an exo-Earth and on TRAPPIST-1 planets.}
   {A combination of laboratory experiments, photochemical and radiative transfer modelling is utilised.
    Spark discharge experiments are conducted in \ce{N2-CO2-H2} gas mixtures, representing a range of possible rocky-planet atmospheres. 
    The production of potential lightning signatures (CO, NO), possible biosignature gases (\ce{N2O}, \ce{NH3}, \ce{CH4}), and important prebiotic precursors (HCN, Urea) is investigated.
    Using the measured CO and NO production rates, photochemical simulations are conducted for oxygen-rich and anoxic atmospheres for rocky planets orbiting in the habitable zones of the Sun and TRAPPIST-1 fosr a range of lightning flash rates.
    Synthetic spectra are calculated using SMART to study the atmosphere's reflectance, emission and transmission spectra.}
   {Lightning enhances, through direct production, the spectral features of NO, \ce{NO2}, and, in some cases, CO; \ce{CH4} and \ce{C2H6} may be enhanced indirectly.
    Lightning at a flash rate slightly higher than on modern Earth is able to mask the ozone features of an oxygen-rich, biotic atmosphere, making it harder to detect the biosphere of such a planet.
    Similarly, lightning at a flash rate at least ten times higher than on modern Earth is also able to mask the presence of ozone in the anoxic, abiotic atmosphere of a planet orbiting a late M dwarf, reducing the potential for a false-positive life-detection.}
   {The threshold lightning flash rates to eliminate oxygen ($>0.1\%$) and ozone false positive biosignatures on planets orbiting ultra-cool dwarfs is up to ten times higher than the modern flash rate, suggesting that lightning cannot always prevent these false-positive scenarios.}


   \maketitle
%
\section{Introduction}

In the past decades, more than 5500 extrasolar planets have been confirmed, the majority of which are located in the solar neighbourhood.
Of these, 69 planets are potentially habitable\footnote{\url{https://phl.upr.edu/hwc}, accessed 23.01.2024}. 
These rocky planets orbit their host star at a distance where water may exist in liquid form, the circumstellar habitable zone. 
Including the stellar and galactic environments, the number of potentially habitable known extrasolar planets decreases to just five \citep{spinelli_ultraviolet_2023}. 
The limiting factor is the high-energy radiation which enables or disables the necessary chemical pathways to form complex molecules.
{JWST}, launched in December 2021, is the first telescope that enables detailed observations of exoplanet atmosphere composition through its infrared (IR) instruments. 
The first step towards detecting habitable planets, however, is the verification that a rocky planet has an atmosphere \citep[e.g.,][]{turbet_trappist-1_2022,ih_constraining_2023}.
Assuming an atmosphere is indeed present, extensive studies have been conducted to assess the detectability of the biosignature pairs \ce{CO2-CH4} or \ce{CH4-O3} \citep{lin_high-resolution_2022,rotman_general_2023} with {JWST} and the future {Extremely Large Telescopes (ELTs)} in the atmosphere of, for example, the TRAPPIST-1 planets.

To interpret observations from present ({JWST}) and future missions and telescopes ({PLAnetary Transits and Oscillations of stars, PLATO}, {Atmospheric Remote-sensing Infrared Exoplanet Large-survey, ARIEL}, {ELTs}), so-called `biosignatures' have been postulated.
Biosignatures are gases or other planetary features that --- singly or in combination --- are potentially indicative of life \citep[e.g.,][]{seager_astrophysical_2012,grenfell_review_2017,schwieterman_exoplanet_2018}.
To avoid misinterpreting such signatures, other processes that can lead to an observable abundance of these gases need to be quantified.
One such group of processes is electrostatic discharges in atmospheres, the largest of which is lightning. 
Lightning is expected to be present in various kinds of environments, including planetary atmospheres and planet-forming disks \citep[e.g.,][]{helling_atmospheric_2016}. 
Lightning in the cloudy atmosphere of a potentially habitable exoplanet is presenting a strong energy source for disequilibrium chemistry to take place.
Lightning is a significant though small source of fixed nitrogen on modern Earth \citep[][and references therin]{Schumann2007}.
Previous studies have shown that lightning can also produce fixed nitrogen in an \ce{N2-CO2} atmosphere, similar to the early Archean \citep{NnaMvondo2001,Navarro-Gonzalez2001,Summers2007,barth_isotopic_2023}, albeit \citet{hu_stability_2019} suggest that lightning-fixed nitrogen will be quickly returned into to the atmosphere as \ce{N2}. 
Further, lightning has also been postulated to have played an important role in the origin of life itself \citep{Miller1953}.

The most abundant biological-produced gas on modern Earth and therefore prime candidate for a biosignature is \ce{O2} and its byproduct ozone \ce{O3} that is more easily detectable with a prominent absorption feature at $\SI{9.6}{\micro\metre}$ (\ce{O2} only has weak absorption features in the mid-infrared within the $\SI{6.3}{\micro\metre}$ water band) \citep{segura_ozone_2003,meadows_exoplanet_2018}.
On modern Earth, \ce{O2} is produced by photosynthesis, but in other circumstances, large amounts of \ce{O2} can be produced abiotically.
For example when a rocky planet around an M dwarf loses a large part of its water inventory during the early, active phase of its host star.
Selective escape of lighter H and retention of heavier O could overwhelm reductant sinks and cause \ce{O2} accumulation in the atmosphere \citep{ramirez_habitable_2014,wordsworth_abiotic_2014,Luger2015,Meadows2017b,Wordsworth2018,johnstone_hydrodynamic_2020,Barth2021a}.
Particularly on planets orbiting M dwarfs with a large atmospheric \ce{CO2} concentration, the increased intensity of the X-ray \& UV (XUV) radiation can robustly dissociate the \ce{CO2} to produce CO and O, as we will discuss below.
The recombination of CO and O is restricted by deficient near UV (NUV) radiation from the M dwarf host star, which is key to generating photochemical catalysts that facilitate this reaction (the direct CO + O reaction is spin-forbidden), potentially producing a false positive \ce{O2} biosignature \citep{segura_abiotic_2007,harman_abiotic_2015}. 
\citet{harman_abiotic_2018} suggest that lightning-produced NO might act as a catalyst to prevent the buildup of \ce{O2} in such an atmosphere. 
We will discuss this possibility later.

Nitrous oxide (\ce{N2O}) has been stipulated as another potential biosignature \citep{rauer_potential_2011,grenfell_review_2017,schwieterman_evaluating_2022}.
In the Earth's spectral energy distribution, \ce{N2O} produces detectable peaks in the near- and mid-IR \citep{sagan_search_1993,gordon_hitran2020_2022}.
Further, \ce{N2O} in the Earth's atmosphere is mainly from biological origin and there are only a few abiotic sources.
Mainly, stellar radiation or lightning can photochemically produce NO, which in an anoxic and weakly reducing atmosphere can undergo further reactions to produce \ce{N2O}.
To distinguish biotically from abiotically produced \ce{N2O}, spectral discriminants can be used, such as HCN and \ce{NO2}, which are abiotically produced together with the \ce{N2O} \citep{Airapetian2016,airapetian_impact_2020,schwieterman_evaluating_2022}.

In many circumstances, the detection of a single biosignature gas such as \ce{O2} would be insufficient evidence to claim the detection of life. 
Extensive planetary context to rule out false positives --- see \citet{Sousa-Silva2019} for \ce{PH3} or \citet{thompson_case_2022} for \ce{CH4} --- and/or additional biosignature gases would be required for a confident biosignature claim, for example \ce{O2} in combination with \ce{CH4} \citep{lovelock_thermodynamics_1975,sagan_search_1993}, which in equilibrium would react to \ce{CO2} and \ce{H2O} \citep{segura_biosignatures_2005}.
Or the combination of \ce{N2} and \ce{O2}, i.e. modern Earth's atmosphere, as this gas mixture would likely not be stable over geological timescales without the constant production of both \ce{O2} and \ce{N2} by life \citep{stueken_modeling_2016,Krissansen-Totton2018,lammer_role_2019,spros_life_2021}.
However, it still needs to be assessed whether photolysis of aqueous nitrite and nitrate could provide enough \ce{N2} to the atmosphere to stabilise the \ce{N2} concentration abiotically \citep{zafiriou_nitrate_1979,zafiriou_nitrite_1979,carpenter_chemistry_2015,Ranjan2019}.
\citet{Wogan2020} discuss the potential of chemical disequilibria as biosignatures: only an `inedible' disequilibrium, where a high activation energy is needed to move the system to equilibrium, can be considered a biosignature.
\citeauthor{Krissansen-Totton2018} (\citeyear{Krissansen-Totton2018,Krissansen-Totton2019}) suggest the combination of \ce{CO2} and \ce{CH4}, which was present in the Archean atmosphere, as such a disequilibrium biosignature.
This biosignature would be strengthened by the absence of CO which has been suggested as an antibiosignature for exoplanets \citep{wang_detection_2016}.

In contrast to a false-positive biosignature, where life can still be present on a planet, an antibiosignature suggests that the planet is not inhabited and is usually defined as the evidence for free energy not being exploited by life \citep{schwieterman_rethinking_2019}.
CO provides a source of chemical free energy and reduced carbon to life in metabolisms such as the Wood-Ljungdahl pathway \citep{ragsdale_life_2004}.
Previous simulations of the atmospheric chemistry during the early Archean represented several metabolisms that govern the concentration of CO in early Earth's atmosphere \citep{kharecha_coupled_2005}:
Methanogens provide a source of CO, as the \ce{CH4} they produce will be photochemically oxidised to CO if irradiated by far UV (FUV) radiation.
Acetogens, on the other hand, provide a biological sink of CO that is limited by the rate at which CO is deposited in the ocean (assuming immediate consumption of CO by acetogens).
A major abiotic CO source is the photolysis of atmospheric \ce{CO2}, and an abiotic sink is the oxidation of CO by hydroxyl radicals (OH) which are mainly produced by photochemical reactions \citep{schwieterman_rethinking_2019}.

Past studies demonstrated that many exoplanets will be covered in clouds for an extended period during their evolution such that it is reasonable to expect lightning to occur also in extrasolar planets \citep{Woitke2003,helling_dust_2008,Helling2013a,Helling2013b,Hodosan2021}. 
Moreover, lightning will contribute to the formation of a global electric circuit \citep{helling_lightning_2019} and produce chemical tracers of a convectively active atmosphere of any planet \citep{hodosan_is_2016}.
However, the only planets where in-situ measurements can be conducted are those within the Solar System, and the only planet for which lightning can be studied to a reasonable degree of completeness concerning flash density and energy range is modern Earth \citep{Hodosan2016}.
The global lightning flash rate on modern Earth is estimated to be $44\pm \SI{5}{\per\second}$ \citep{Christian2003} with an energy of $\SI{6.7}{\giga\joule}$ per flash \citep{Price1997}, but much uncertainty remains for the lightning flash rate on the early Earth, terrestrial planets, and exoplanets in general \citep{Hodosan2021}.

\citet{Wong2017} used climate simulations and the convective available potential energy to estimate the lightning flash rate in potential Archean atmospheres with varying \ce{CO2} partial pressure.
They found the lightning flash rate to peak at 3.4 times the modern Earth's flash rate at $\SI{1}{\bar}$ of \ce{CO2}, with values lower than on modern Earth for \ce{CO2} partial pressures of 0.1 and $\SI{10}{\bar}$ (with $1-\SI{2}{\bar}$ of \ce{N2}).
\citet{braam_lightning-induced_2022} suggest that the lightning flash rate on tidally locked exoplanets such as Proxima Centauri~b is less than 10\% of modern Earth's.
An additional factor that can influence the occurrence of lightning is cosmic rays.
Planets orbiting M dwarfs will experience more frequent and intense stellar flares that are associated with flares of charged particles, known to enhance the ionisation in the planet's atmosphere \citep{stozhkov_role_2003, Rimmer2013a, Griessmeier2015, Fraschetti2019, Scheucher2020, Barth2021}.
Comparison between lightning flash rates and cosmic ray ionisation rates in Earth's atmosphere has shown a strong correlation between these two quantities \citep{stozhkov_role_2003}.
However, the increased influx of charged particles into the atmospheres of planets orbiting M dwarfs is likely only more efficiently enabling lightning discharges in already existing electric fields in the clouds of these atmospheres.
We might therefore only find a slightly enhanced lightning activity on these planets, but more detailed studies on the connection between cosmic ray ionisation and lightning are necessary to fully understand these processes.

This paper adopts an approach of combining laboratory experiments and modelling to investigate the impact of lightning on the atmospheric chemistry of exoplanets.
We present results from spark-discharge experiments with different gas mixtures initially containing \ce{N2}, \ce{CO2}, and \ce{H2}.
The setup of the experiments and the photochemical model are described in Section~\ref{Sec_Methods}.
In Section~\ref{Sec_Results}, we present the results of our experiments and their implications.
We apply our experimental results to two different hypothetical exoplanets and use photochemical simulations and calculated spectra to determine potentially observable signatures from lightning and the prospect for false-positive or false-negative biosignatures (Section~\ref{Sec_Results_Photochem}).
Due to the wide range of work presented in this paper and the large amount of results, we decided to discuss the implications of individual results in the same sections.
We discuss our assumptions on lightning flash rates and atmospheric composition in Section~\ref{Sec_Discussion} and present our conclusions in Section~\ref{Sec_Conclusion}.

\section{Methods}
\label{Sec_Methods}

\subsection{Spark Experiments}

\begin{figure}[ht]
    \centering
    \includegraphics[width=\columnwidth]{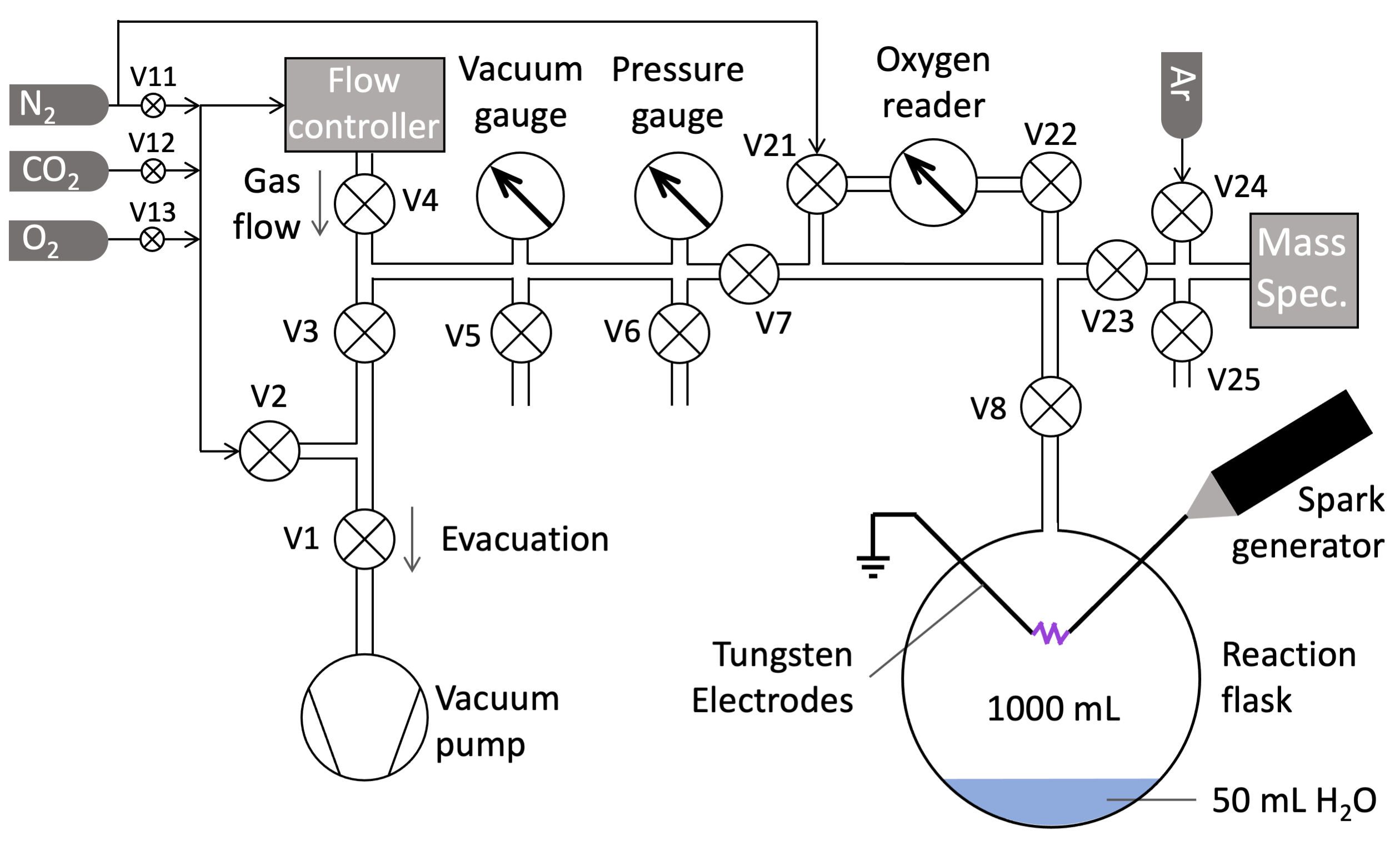}
    \caption{Schematic of the experimental setup of the discharge experiment. First published in \citet{barth_isotopic_2023} by Springer Nature.}
    \label{Fig_ExpSetup}%
\end{figure}

All experiments were carried out at the University of St Andrews in the St Andrews Isotope Geochemistry lab (StAIG). 
We used the experimental setup described in \citet{barth_isotopic_2023}, similar to the one described by \citet{Parker2014} (Fig.~\ref{Fig_ExpSetup}). 
The spark discharge (generated by a BD-50E heavy-duty spark generator with a maximum voltage of $\SI{49}{\kilo\volt}$) was contained in the 1-litre reaction flask (Pyrex glass), which contained $\SI{50}{\milli\litre}$ of water at the bottom and the spark electrodes (tungsten metal) secured in the headspace.
The water was agitated with a magnetic stir bar. 
The system was evacuated and purged with \ce{N2} three times before adding the desired gas mixture and starting the experiment.
To investigate the effect of water vapour in the gas phase on the final results, a set of dry experiments was run.
For these, we added the water with a syringe through the septum port on the flask only after the spark had been turned off.
The water was previously flushed with pure \ce{N2} (10~min at $\sim \SI{50}{\milli\litre\per\minute}$) to remove dissolved oxygen.
We then let the experiment with the water sit for 3~hours (with the spark still switched off) to allow for the gaseous and liquid phases to equilibrate.
The water was continuously stirred with the magnetic stir bar to facilitate gas exchange between the headspace and the liquid phase.

Before and after each experiment, gas from the flask was analysed by a quadrupole mass spectrometer gas analyser.
After the experiment, a gas sample was extracted from the flask with a gas-tight, lockable syringe to determine the concentration of \ce{CH4} and \ce{N2O} with a gas chromatograph.
From a limited sample of experiments, multiple gas samples were extracted for analyses of CO.
The fluid phase was transferred into a 50ml Falcon centrifuge tube for subsequent analyses of dissolved nitrite, nitrate, ammonium, urea, and cyanide (see below).
The analytical methods used to determine the concentrations of these species are detailed in Appendix~\ref{Sec_Appendix_Methods}.
From these concentrations, the energy yield ($\si{molecules\per\joule}$) can be calculated, using the energy of the spark $E = 1/2 U I t$ with the applied voltage $U = \SI{49}{\kilo\volt}$, the current $I = \SI{1}{\milli\ampere}$, and the duration $t$ of the spark.
To extrapolate this yield to the annual, global production, we used an estimate for the global lightning flash rate on modern Earth of $44\pm \SI{5}{\per\second}$ \citep{Christian2003} with an energy of $\SI{6.7}{\giga\joule}$ per flash \citep{Price1997}.

\subsection{Photochemical Simulations}

\begin{figure}[ht]
    \centering
    \includegraphics[width=\columnwidth]{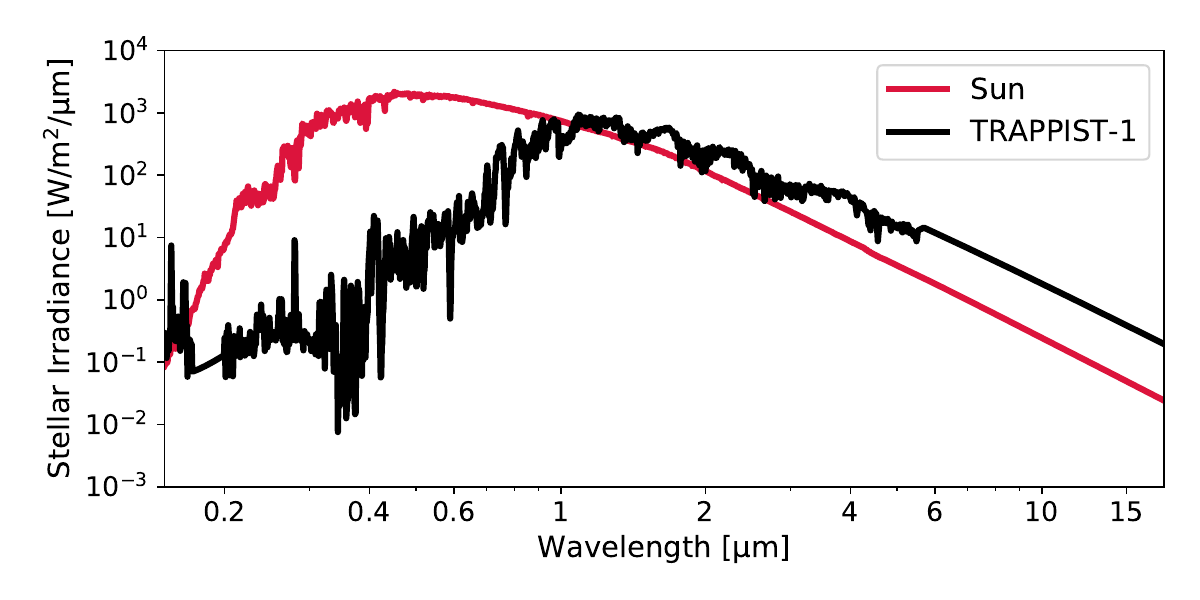}
    \caption{Spectral Energy Distributions at the top of the planets' atmospheres of the Sun (\textit{red}) and TRAPPIST-1 (\textit{black}) as used in the photochemical model and for the planetary spectra.}
    \label{Fig_Spectra_Stars}%
\end{figure}

We conducted photochemical simulations to calculate the mixing ratios of CO, NO, and \ce{NO2} in the atmosphere of different test planets for a range of NO and CO production rates, corresponding to a range of lightning flash rates. 
To conduct these tests, we used the photochemical model component of the Atmos coupled climate-photochemistry code \citep{arney_pale_2016, Lincowski2018}\footnote{\url{https://github.com/VirtualPlanetaryLaboratory/atmos}}. 
We conducted all simulations in uncoupled mode (no climate adjustment) to isolate the specific chemical impact of varying CO and NO fluxes from lightning. 
When simulating anoxic atmospheres, we adopt an Archean-Earth planet template with 74 chemical species and 392 photochemical reactions. 
For \ce{O2}-rich atmospheres, we adopt a modern Earth-like template with 50 species and 239 reactions. 
We have incorporated the latest \ce{H2O} cross-sections and corrected sulfur reaction rate as recommended by \citealt{ranjan_photochemistry_2020}. 
The model normally includes the impact of Earth-like lightning by injecting NO into the troposphere \citep{harman_abiotic_2018}; 
however, we have removed this feature and replaced it with a variable NO injection rate to assess the impact of varying NO production from lightning. 
The model includes diffusion-limited hydrogen escape \citep{harman_abiotic_2015}.

We ran the simulations for an Earth-sized planet, orbiting the Sun at $\SI{1}{au}$ and the M dwarf TRAPPIST-1 at an instellation identical to that of TRAPPIST-1~e \citep{Agol2020}.
The initial atmospheric composition was set to be similar to our high-\ce{CO2} experiments (4.6\% \ce{CO2} with \ce{N2} as filler gas).
We also calculated the CO and \ce{NO2} mixing ratios in a corresponding oxic atmosphere (21\% \ce{O2}, 4.6\% \ce{CO2} with \ce{N2} as filler gas).
In both cases, the \ce{CO2} concentration is fixed to the initial value of 4.6\%.
The tropospheric water vapour concentration is governed by the surface temperature of the planet: 0.5\% \ce{H2O} for $T = \SI{275}{\kelvin}$, 1.2\% \ce{H2O} for $T = \SI{288}{\kelvin}$, and 2.5\% \ce{H2O} for $T = \SI{300}{\kelvin}$.
The pressure-temperature profiles for the anoxic simulations assume surface temperatures of $T_\mathrm{surf} = \SI{275}{\kelvin}$ and $T_\mathrm{surf} = \SI{300}{\kelvin}$, adiabatic cooling throughout the troposphere, and then an isothermal temperature of $T_\mathrm{gas} = \SI{180}{\kelvin}$.
For the oxic simulations, the pressure-temperature profile of modern Earth is used, with a surface temperature of $T_\mathrm{surf} = \SI{288}{\kelvin}$.
The spectra of the Sun and TRAPPIST-1 (Fig.~\ref{Fig_Spectra_Stars}) are used for the photochemistry simulations and to simulate the planetary spectra. 
The spectrum of the Sun was sourced from \citet{thuillier_solar_2004} while the TRAPPIST-1 spectrum represents an average of the three activity models presented in \citet{peacock_predicting_2019}. 
Each spectrum is re-interpolated onto the standard Atmos base grid (Lgrid=0 setting). 

We ran the model for different scenarios:
\textit{Abiotic scenario:} Volcanic fluxes of \ce{CH4} ($10^8 \, \si{molecules\per\centi\metre\squared\per\second}$), \ce{H2} ($10^{10} \, \si{molecules\per\centi\metre\squared\per\second}$, deposition velocity into the ocean: $v_\mathrm{dep} = 2.4 \times 10^{-4}\, \si{\centi\metre\per\second}$), \ce{H2S} ($3.5 \times 10^{8}\, \si{molecules\per\centi\metre\squared\per\second}$, $v_\mathrm{dep} = 2 \times 10^{-2} \, \si{\centi\metre\per\second}$), and \ce{SO2} ($3.5 \times 10^{9} \, \si{molecules\per\centi\metre\squared\per\second}$, $v_\mathrm{dep} = 1 \si{\centi\metre\per\second}$) are included and distributed over the bottom $\SI{10}{\kilo\metre}$ of the atmosphere profile (as is the variable CO flux).
The CO deposition velocity is $v_\mathrm{dep} = 10^{-8} \, \si{\centi\metre\per\second}$ which is the limit for the abiotic formation of formate \citep{kharecha_coupled_2005}.
\textit{Biotic scenario:} In addition to the volcanic sources of the abiotic scenario, a biological methane production of $10^{11} \, \si{molecules\per\centi\metre\squared\per\second}$ is included, which corresponds to the Earth's current biogenic methane flux \citep{jackson_increasing_2020}.
The CO deposition velocity in the biotic case is $v_\mathrm{dep} = 1.2 \times 10^{-4} \, \si{\centi\metre\per\second}$ which is the maximum deposition velocity for an ocean with a CO concentration of 0, i.e. where all CO is immediately consumed by acetogens \citep{kharecha_coupled_2005}.
A detailed table containing the parameters for all the different scenarios can be found in the Appendix.

\subsection{Spectral Simulations}

To simulate reflectance, emission, and transmission spectra we used the Spectral Mapping Atmospheric Radiative Transfer code \citep[SMART,][]{meadows_ground-based_1996,crisp_absorption_1997} with transit updates as described in \citet{robinson_theory_2017}. 
SMART is a versatile and well-validated line-by-line, multi-stream, multiple scattering, and absorption model that can produce planetary spectra from the far-UV to far-IR. 
SMART relies on the DISORT Fortran code \citep{stamnes_numerically_1988} to solve the radiative transfer equation via the discrete ordinate method. 
SMART includes opacity data from HITRAN \citep{gordon_hitran2016_2017} that are preprocessed by their Line-By-Line Absorption Coefficients (LBLABC) companion model. 
SMART has been used to simulate spectra of planets inside and outside of the Solar System including the TRAPPIST-1 planets \citep{tinetti_disk-averaged_2005,robinson_earth_2011,arney_spatially_2014,lustig-yaeger_detectability_2019}. 
For the spectra presented in Section~\ref{Sec_Results_Spectra}, we show reflectance spectra degraded to a spectral resolving power of R=400, transmission spectra degraded to R=200, and emission spectra with a spectra resolution of 1 cm$^{-1}$. 


\section{Results and Implications of Experimental Measurements}
\label{Sec_Results}

\begin{table*}[ht]
    \begin{center}
    \caption{Initial gas compositions in our experiments in bar${}^{a}$.}
    \renewcommand{\arraystretch}{1.3}
    \begin{tabular}{w{c}{1.5cm}w{c}{3.5cm}w{c}{3.5cm}w{c}{3.5cm}w{c}{3.5cm}}
    	\hline
    	\multirow{2}{*}{Species} & \multicolumn{4}{c}{Gas mixture} \\ \cline{2-5}
                 & pure \ce{N2} & \ce{N2-H2} & + low \ce{CO2} & + high \ce{CO2}\\ 
    	\hline
    	\hline
    	\multicolumn{5}{l}{$\SI{120}{\minute}$ experiments (wet)} \\
    	\hline
    	\ce{N2}  & $0.99 \pm 0.00$            & $0.98 \pm 0.00$              & $0.98 \pm 0.00$              & $0.93 \pm 0.01$            \\
    	\ce{H2}  & $0.00 \pm 0.00$            & $(9.98\pm0.01)\times10^{-3}$ & $(9.98\pm0.01)\times10^{-3}$ & $(9.3\pm0.1)\times10^{-3}$ \\
    	\ce{CO2} & $(2.6\pm0.9)\times10^{-5}$ & $(1.1\pm0.1)\times10^{-4}$   & $(2.5\pm0.1)\times10^{-3}$   & $(4.6\pm0.6)\times10^{-2}$ \\
    	\ce{O2}  & $(1.3\pm0.4)\times10^{-4}$ & $(8\pm1)\times10^{-5}$       & $(1.2\pm0.2)\times10^{-4}$   & $(1.3\pm0.3)\times10^{-4}$ \\
    	\ce{H2O} & $(1.1\pm0.1)\times10^{-2}$ & $(1.4\pm0.1)\times10^{-2}$   & $(1.0\pm0.1)\times10^{-2}$   & $(1.21\pm0.03)\times10^{-2}$ \\
    	\hline
    	\multicolumn{5}{l}{$\SI{120}{\minute}$ experiments (dry)} \\ 
    	\hline
    	\ce{N2}  & $0.99 \pm 0.00$        & $0.99 \pm 0.00$            & $0.98 \pm 0.01$             & $0.94 \pm 0.01$            \\
    	\ce{H2}  & $0.00 \pm 0.00$        & $(9.85\pm0.02)\times10^{-3}$ & $(9.77\pm0.03)\times10^{-3}$  & $(9.37\pm0.04)\times10^{-3}$ \\
    	\ce{CO2} & $(2\pm1)\times10^{-6}$ & $(2\pm2)\times10^{-5}$     & $(6\pm2)\times10^{-3}$      & $(4.7\pm0.5)\times10^{-2}$ \\
    	\ce{O2}  & $(6\pm2)\times10^{-5}$ & $(1.4\pm0.3)\times10^{-4}$ & $(5\pm3)\times10^{-4}$      & $(2.6\pm0.1)\times10^{-4}$ \\
    	\ce{H2O} & $(6\pm3)\times10^{-3}$ & $(4\pm2)\times10^{-3}$     & $(6\pm2)\times10^{-3}$      & $(6\pm2)\times10^{-3}$     \\
    	\hline
    	\multicolumn{5}{l}{Overnight experiments (wet)} \\ %
    	\hline
    	\ce{N2}  & $0.99 \pm 0.00$            & $0.97 \pm 0.00$              & $0.97 \pm 0.00$              & $0.94 \pm 0.01$            \\
    	\ce{H2}  & $0.00 \pm 0.00$            & $(9.74\pm0.01)\times10^{-3}$ & $(9.71\pm0.02)\times10^{-3}$ & $(9.4\pm0.1)\times10^{-3}$ \\
    	\ce{CO2} & $(1.1\pm0.1)\times10^{-5}$ & $(4\pm3)\times10^{-5}$       & $(4\pm1)\times10^{-3}$       & $(3.8\pm0.5)\times10^{-2}$ \\
    	\ce{O2}  & $(9\pm3)\times10^{-5}$     & $(5\pm1)\times10^{-5}$       & $(2.5\pm1.2)\times10^{-4}$   & $(1.4\pm0.4)\times10^{-4}$ \\
    	\ce{H2O} & $(1.0\pm0.5)\times10^{-2}$ & $(1.6\pm0.1)\times10^{-2}$   & $(1.5\pm0.2)\times10^{-2}$   & $(1.8\pm0.1)\times10^{-2}$ \\
    	\hline
    \end{tabular}
    \\
    ${}^{a}$ values are averaged from multiple sets of experiments, errors are standard errors of the mean ($\sigma / \sqrt{N}$)
    \label{Tab_GasComp}
    \end{center}
\end{table*}

We performed spark discharge experiments in gas mixtures with different combinations of \ce{N2}, \ce{CO2}, and \ce{H2} to investigate the effect of lightning on the chemistry of lightly reducing gas mixtures with varying fractions of \ce{CO2}. 
Our goal is to give a complete picture of the most abundant gaseous and aqueous compounds produced in spark discharge experiments in such gas mixtures.
These compounds can be grouped into three categories:
(1) CO and NO are, as we will see later, the most abundant products and are directly produced in the spark channel. Because of that, they might present signatures for lightning activity in exoplanetary atmospheres as we will discuss in Section~\ref{Sec_Results_Photochem}.
(2) \ce{N2O}, \ce{NH3}, and \ce{CH4} are potential biosignatures \citep[e.g.,][]{seager_biosignature_2013,phillips_detecting_2021,huang_assessment_2022,schwieterman_evaluating_2022,thompson_case_2022} and we want to investigate whether lightning might present a significant source of these gases.
(3) \ce{NH4^+}, \ce{NO2^-}, \ce{NO3^-}, HCN, and Urea are important prebiotic compounds, either as precursors for the formation of biological macromolecules or as nutrients for early life forms \citep[e.g.,][]{miller_formation_1957,Miller1959,abelson_chemical_1966,sanchez_studies_1967,miller_atmosphere_1983,schopf_evidence_2007,ducluzeau_was_2009,Wong2017,das_insights_2019}.
We investigated how the gas composition changes the yields of these products and what effect the presence of water vapor in the gas mixture has.
We conducted a range of short, 120-minute experiments as well as longer, overnight experiments to study the production of molecules with low yields that could not be detected in our short experiments.
First, we present the results from our short experiments in individual sections for each compound and discuss their implications on the importance of lightning as a source of the specific molecule in the same section (Sections~\ref{Subsec_ammonium}--\ref{Subsec_nitrate}).
We then present and discuss the results from our overnight experiments (Section~\ref{Sec_Results_Overnight}).
The different gas compositions for all experiments are compiled in Table~\ref{Tab_GasComp}.

\begin{figure*}
    \centering
    \includegraphics[width=\textwidth]{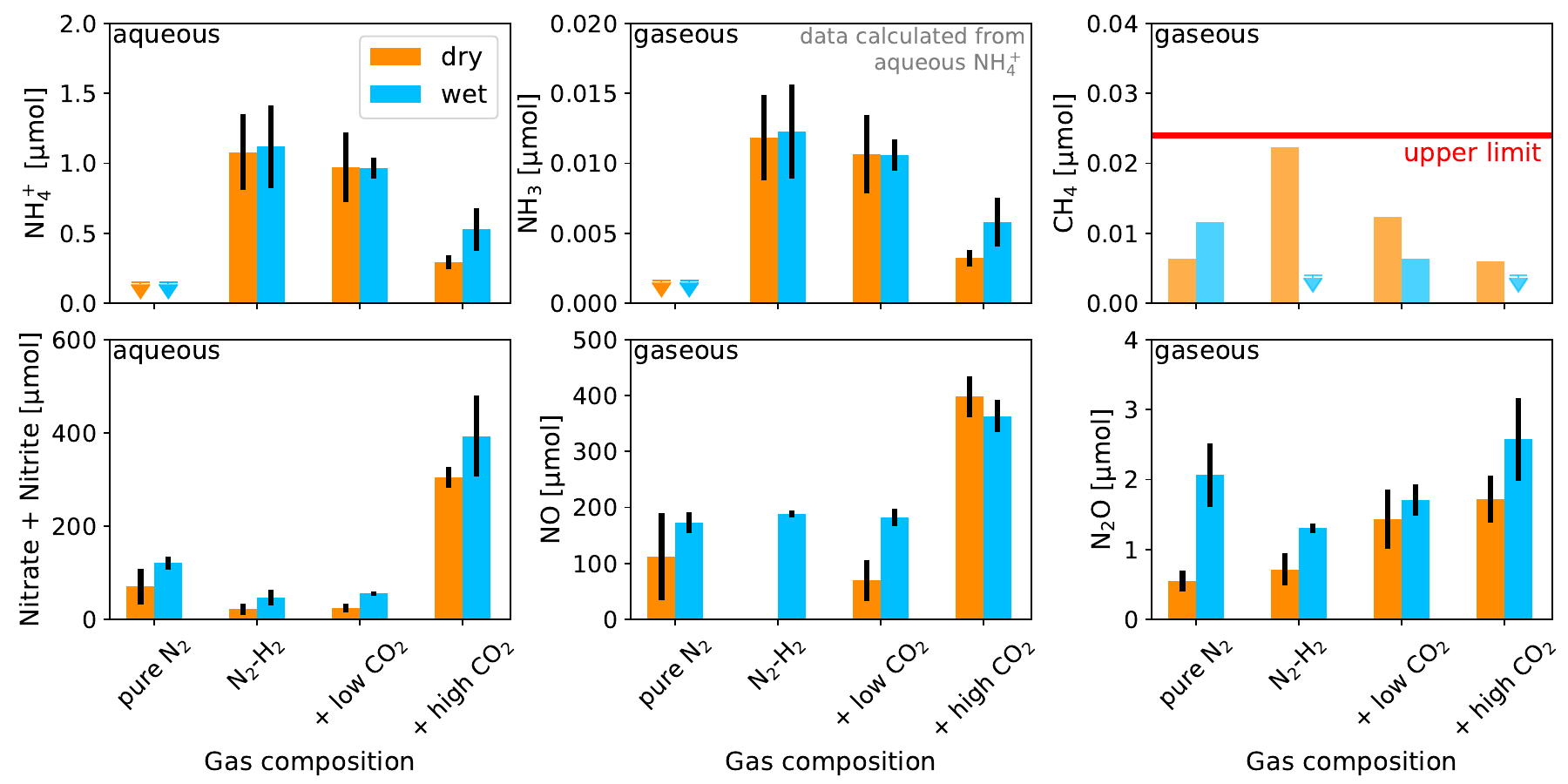}
    \caption{Final abundance of ammonium (\ce{NH4^+}), ammonia (\ce{NH3}), methane (\ce{CH4}), nitrate (\ce{NO3^-}) and nitrite (\ce{NO2^-}), nitrous oxide (\ce{N2O}), and nitric oxide (\ce{NO}) in $\SI{120}{\minute}$ spark experiments with different gas compositions (see Tab.~\ref{Tab_GasComp} for individual gas compositions). 
    Results from experiments with (\textit{blue}) and without (\textit{orange}) water are shown separately. 
    If no concentrations were measurable, the detection limit is indicated with an arrow. 
    The data presented here are based on averages of multiple measurements with the error bars representing the individual standard error of the mean ($\sigma / \sqrt{N}$).
    Data for \ce{NH3} is calculated from aqueous \ce{NH4^+} and not measured directly. 
    Measurements of methane abundance are within $1\sigma$ of 0 (\textit{red line}) and are therefore shown without error bars and semi-transparent.}
    \label{Fig_Bar_120min}
\end{figure*}

For our short, 120-minute experiments, we performed experiments both with (wet) and without water (dry) in the flask during the spark to investigate the effect of water vapor on the yields of our products. 
As described in Section~\ref{Sec_Methods}, we added $\SI{50}{\milli\litre}$ of water to the dry experiments after turning off the spark.
We then analysed the gas and water for the concentrations of NO, CO, \ce{N2O}, \ce{CH4}, \ce{NH4^+}, \ce{NO2^-}, and \ce{NO3^-}.
Figure~\ref{Fig_Bar_120min} shows the combined results from our short experiments.
The individual measurements of each gas composition are averaged over 3 to 7 replicates (a detailed table with individual results is available online\footnote{\url{https://doi.org/10.17630/8b72510f-62a8-43dc-94f1-af9b7766f817}}), the results for wet and dry experiments are shown in different colors next to each other.
As expected, one of the main results is that reduced nitrogen and carbon species (ammonium, ammonia, and methane) are more abundant in the pure \ce{N2-H2} and low-\ce{CO2} experiments.
Oxidised forms of nitrogen (nitrate, nitrite, nitric oxide, and nitrous oxide) are more abundant in the high~\ce{CO2} experiments where the dissociation of \ce{CO2} provides the necessary oxygen source for NO and subsequent nitrogen oxides.

\PB{This follows the results from different studies in the past:
Experiments with electric discharges in reducing \ce{CH4-NH3} and \ce{CH4-N2} gas mixtures produced a variety of reduced nitrogen and carbon products, including \ce{H2}, \ce{NH3}, hydrocarbons, and nitriles such as HCN \citep{toupance_formation_1975}.
Similar experiments simulating corona discharges in Titan's reducing atmosphere (10\% \ce{CH4} and 2\% Ar in \ce{N2}) have shown show the production of reduced nitrogen and carbon in the form of various hydrocarbons and nitriles \citep{reid_thompson_plasma_1991,navarro-gonzalez_corona_1997,navarro-gonzalez_production_2001}.
Spark experiments with different \ce{N2-CO2} gas mixtures have shown that predominantly oxidised forms of nitrogen and carbon are produced, such as NO, \ce{N2O}, \ce{NO2}, \ce{HNO3}, or CO \citep[e.g.,][]{levine_production_1982,nna_mvondo_nitrogen_2005,heays_nitrogen_2022}.
\citet{Navarro-Gonzalez2001} and \citet{NnaMvondo2001} studied the production of NO by lightning in gas mixtures with varying \ce{CO2} concentration and found a clear trend of decreasing nitrogen fixation efficiency with decreasing availability of oxygen from \ce{CO2}.}

We found increased efficiency of nitrogen oxides production in the experiments that contained water in the flask during the spark.
The presence of liquid water resulted in approximately 1\% of water vapour in the gas phase at room temperature (Table~\ref{Tab_GasComp}), which was also dissociated in the spark, providing additional oxygen for the NO production.
The individual products shown in Fig.~\ref{Fig_Bar_120min} as well as CO are discussed in the following sections.

\subsection{Ammonium (\ce{NH4^+})}
\label{Subsec_ammonium}

Ammonium is an important nutrient for microbial life on Earth and was the product of the first developed pathways of biological nitrogen fixation \citep{schopf_evidence_2007,dodd_evidence_2017}.
Lightning-produced ammonium would therefore present a potential nutrient source for life before the onset of biological nitrogen fixation.
We found that maximum ammonium production happens in experiments with no or $< 1\%$ of \ce{CO2}.
A higher \ce{CO2} concentration limits the efficiency of ammonium production.
However, except for a few individual experiments, we found the final ammonium concentration to be lower than the concentration of nitrite and nitrate, in particular for the wet experiments.
Our highest ammonium production rate (the wet, \ce{CO2}-free case) is $(3.8 \pm 1.7) \times 10^{12}\, \si{molecules\per\joule}$ which, using the modern Earth flash rate, corresponds to a yearly production of $(1.1 \pm 0.5) \times 10^{-3}\, \si{\tera\gram\per\year}$.
With a higher \ce{CO2} concentration, more likely resembling early Earth's atmosphere, this reduces to $(1.8 \pm 1.2) \times 10^{12} \, \si{molecules\per\joule}$ and $(0.5 \pm 0.3) \times 10^{-3}\, \si{\tera\gram\per\year}$.
This is much lower than the production of nitrite and nitrate (see below), \PB{though nitrite can subsequently be reduced to ammonium by \ce{Fe^{+2}} and FeS in the ocean, while FeS can also reduce nitrate to ammonium albeit with a lower yield \citep{Summers1993,summers_ammonia_2005}}.
Direct production of ammonium by lightning is therefore not significant on planets with a substantial concentration of \ce{CO2} and/or water vapour in the atmosphere.

\subsection{Ammonia (\ce{NH3})}

Ammonium is not directly produced by the spark.
Instead, ammonia is the gaseous product of the spark discharge that then equilibrates with the water where it reacts to ammonium.
Ammonia can be produced biologically and has been previously suggested as a biosignature for planets with a significant \ce{H2} fraction in the atmosphere \citep{seager_biomass-based_2013,seager_biosignature_2013,huang_assessment_2022}.
\citet{ranjan_photochemical_2022} describe a scenario where \ce{NH3} is produced in such high quantities that it saturates its photochemical sinks, such as \ce{O2} in modern Earth's atmosphere, and goes into a runaway mode.
The surface flux above which the \ce{NH3} concentration enters into this runaway state, depends on the level of UV-radiation.
In their simulations of atmospheres with 10\% \ce{H2} and 90\% \ce{N2} (with 1\% water vapour) on an Earth-sized planet orbiting an M dwarf, Ranjan et al. find this flux to be approximately $10^{11} \, \si{\per\centi\metre\squared\per\second}$, and as low as $10^{8} \, \si{\per\centi\metre\squared\per\second}$ for an elevated stratospheric temperature (from $T_\mathrm{strat} = \SI{170}{\kelvin}$ to $\SI{210}{\kelvin}$).
The modern Earth \ce{NH3} flux is  $(1.1-1.8) \times 10^{10}\, \si{\per\centi\metre\squared\per\second}$ \citep{bouwman_global_1997} and the pre-industrial flux  $(2-9) \times10^9 \, \si{\per\centi\metre\squared\per\second}$ \citep{zhu_sources_2015}.
The maximum lightning-induced \ce{NH3} flux we can extrapolate from our experiments with an atmosphere of 1\% \ce{H2} and 1\% \ce{H2O} (in \ce{N2}), including the aqueous \ce{NH4^+}, is approximately  $(2.2 \pm 1.0) \times 10^5 \, \si{\per\centi\metre\squared\per\second}$ (wet, pure \ce{N2}-\ce{H2}).
This is many orders of magnitude below the flux necessary to enter into a runaway state, suggesting that lightning cannot be responsible for a false-positive biosignature detection with \ce{NH3}.

\subsection{Methane (\ce{CH4})}

Methane is frequently discussed as a potential biosignature, in particular for Archean Earth-like worlds \citep[e.g.,][]{thompson_case_2022}.
If lightning could produce significant amounts of methane in an early Earth-like atmosphere, this would present an important restriction on methane's role as a biosignature.
However, in our experiments, only small amounts of methane are produced.
Most of the measurements are below the error of an individual measurement ($\SI{0.6}{ppm}$) which means that they are within $1\sigma$ of 0.
We therefore assume $\SI{0.6}{ppm}$ (or a total production of $\SI{0.024}{\micro mol}$) to be the upper limit for methane production in our experiments, as indicated by the red line in Fig.~\ref{Fig_Bar_120min}.
This limit corresponds to an annual methane production with a modern lightning flash rate of less than $2 \times 10^{-5} \, \si{\tera\gram\per\year}$ or $1.3 \times 10^{-6} \, \si{\tera mol\per\year}$.
The annual methane production on modern Earth is $\SI{37}{\tera mol\per\year} = \SI{596}{\tera\gram\per\year}$, of which $\sim 40 \%$ is natural (not anthropogenic) \citep{jackson_increasing_2020} while the possible Archean biological methane production was between 9 and $\SI{20}{\tera mol\per\year}$ \citep{sauterey_co-evolution_2020}.
The upper limit for abiotic methane production by serpentinising systems (hydrothermal alteration of crustal mafic rocks) is between 0.02 and $\SI{10}{\tera mol\per\year}$ \citep{Krissansen-Totton2018}.
This suggests that in the atmospheric conditions explored in this study, methane production by lightning is negligible compared to other abiotic and biotic sources.
The lightning flash rate in such an atmosphere would have to be 5 -- 6 orders of magnitude higher than on modern Earth to produce a comparable amount of methane. 

\subsection{Carbon Monoxide (CO)}

\begin{figure}
    \centering
    \includegraphics[width=\columnwidth]{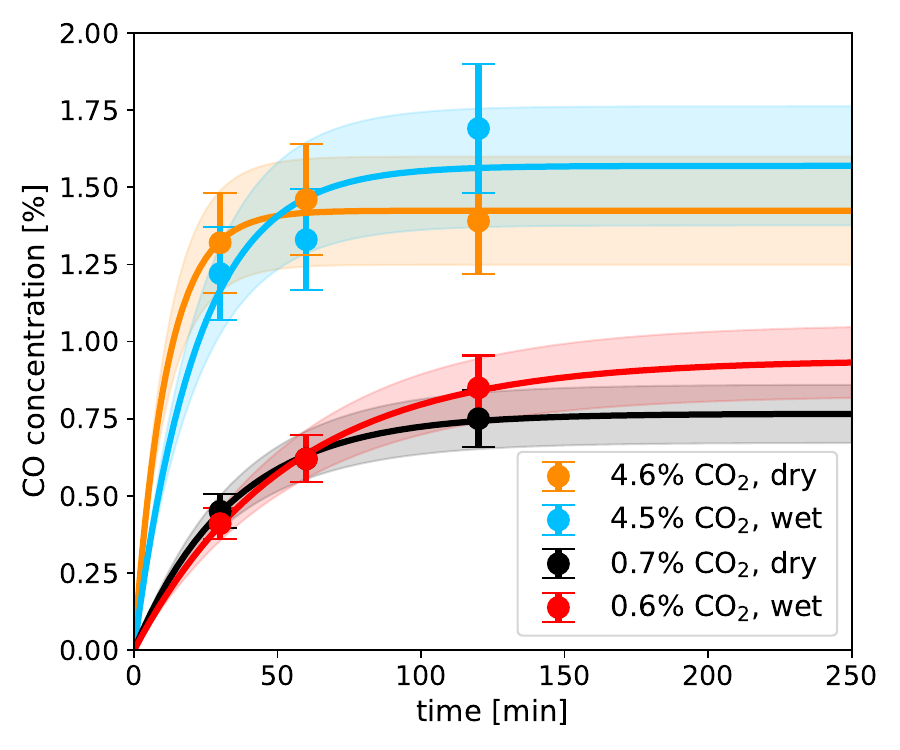}
    \caption{CO concentration in 4 different spark experiments (high and low-\ce{CO2}, wet and dry), measured at 3 different points in time during the experiment. 
    Fitted lines give the CO concentration assuming CO is the product of an equilibrium reaction, following $c_{\ce{CO2}} = (1 - ( k_1 + k_2 \cdot \exp{(-(k_1+k_2)t)}) /(k_1 + k_2)) \cdot c_1$.
    Shaded areas represent errors in CO measurements.
    }
    \label{Fig_CO_time}
\end{figure}

\begin{figure}
    \centering
    \includegraphics[width=\columnwidth]{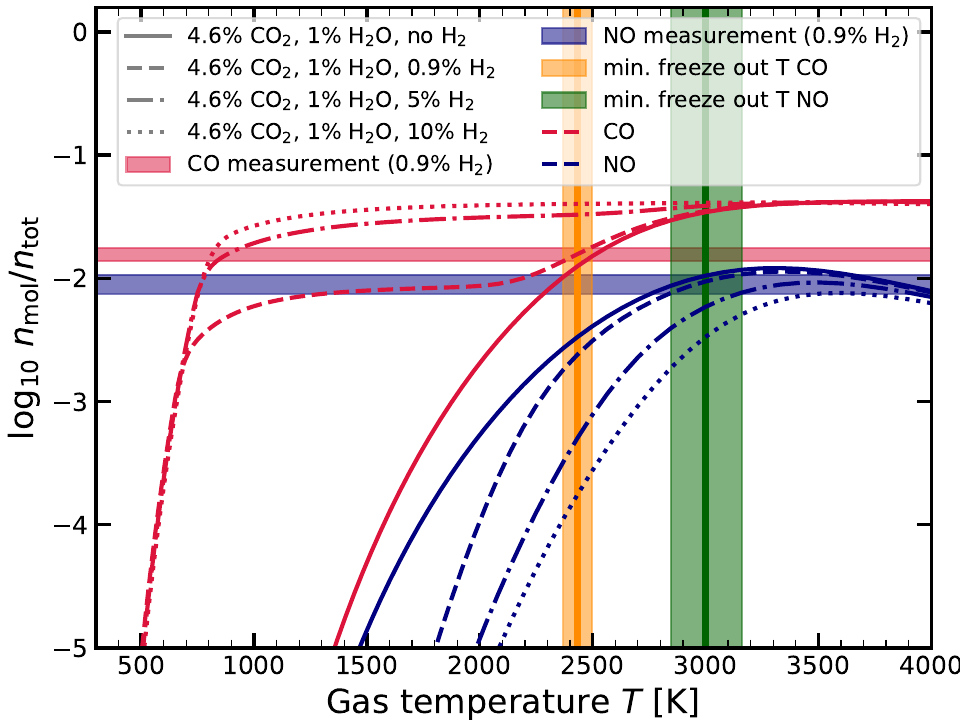}
    \caption{CO and NO concentrations in chemical equilibrium as a function of gas temperature \citep[with GGChem,][]{woitke_equilibrium_2018} for high-\ce{CO2} gas mixtures. The different cases considered are: 0\% (\textit{solid lines}), 0.9\% (\textit{dashed}), 5\% (\textit{dash dotted}), and 10\% (\textit{dotted}) of \ce{H2}. 
    As \ce{H2} is binding free oxygen into \ce{H2O}, increasing the \ce{H2} concentration prevents the recombination of CO to \ce{CO2} as well as the production of NO.
    CO (\textit{red}) and NO (\textit{blue shaded area}) measurements with uncertainties from wet $\SI{120}{\minute}$ experiments. 
    Minimum freeze-out temperature estimates for CO (\textit{orange}) and NO (\textit{green}) with uncertainties.
    }
    \label{Fig_CO_equi}
\end{figure}

In contrast to methane, carbon monoxide has been discussed as a potential antibiosignature \citep{wang_detection_2016,schwieterman_rethinking_2019}. 
If on an inhabited planet, lightning were to produce detectable amounts of CO by overwhelming biological sinks, this would provide a limitation on the use of CO as an antibiosignature.
We analysed the CO concentrations of four experiments, each at three different points in time during the experiment (after 30, 60, and $\SI{120}{\minute}$). 
These results are shown in Fig.~\ref{Fig_CO_time}.
The CO production rate decreased with time, likely because CO concentration reaches equilibrium.
We extrapolated the CO concentration to estimate the equilibrium concentration (Fig.~\ref{Fig_CO_time}).
Following the equilibrium reaction \ce{A1 <=> A2} (\ce{A2} is CO in this case) with the reaction coefficients $k_1$ and $k_2$, the change in $c_2$ (the concentration of \ce{A2}) is given by
\begin{equation}
    \frac{dc_2}{dt} = - k_2 c_2 + k_1 c_1
\end{equation}
Starting with all molecules in \ce{A1} ($c_{1,0} = c_\mathrm{tot}, c_{2,0} = 0$), $c_2$ increases with time:
\begin{equation}
    \label{Reac_Equi_Fit}
    c_2 = \left(1 - \frac{k_2 + k_1 \cdot \exp{(-(k_1+k_2)t)}}{k_1 + k_2} \right) \cdot c_\mathrm{tot}.
\end{equation}
The resulting equilibrium concentration of $c_2$ (for $t \rightarrow \infty$) is then given by
\begin{equation}
    c_{2,\mathrm{eq}} = \left( \frac{k_1}{k_1 + k_2} \right) \cdot c_\mathrm{tot}.
\end{equation}
For our experiments, the estimates for CO equilibrium concentrations are 1.5\% for the high-\ce{CO2} experiments (4.5-4.6\% \ce{CO2}) and 0.85\% for the low-\ce{CO2} experiments (0.6-0.7\% \ce{CO2}).

\subsection{Extrapolating CO-Production to Different Atmospheres}
\label{Sec_Results_extra}

In past studies, the lightning production rates of different gases like NO have been determined by assuming the equilibrium concentration of said gases at the so-called `freeze-out temperature' \citep[e.g., ][]{Chameides1977,Hill1980,borucki_lightning_1984}.
The freeze-out temperature is the temperature at which the equilibrium timescale is larger than the cooling timescale of the gas mixture, freezing in the gas composition at that temperature. 
Every time a gas parcel has been heated up by a lightning strike and is cooling down again, the concentration of NO, CO, and other lightning products in the air parcel is increased to its equilibrium concentration at the respective freeze-out temperature.
Comparison to experimental results of NO production rates by lightning show that this method is a valid approach to determine the production of NO etc. by lightning \citep[see reviews by ][]{Schumann2007,heays_nitrogen_2022}.

In this study, we used our experimental results and chemical equilibrium calculations to determine the freeze-out temperatures of NO and CO, allowing us in the next step to extrapolate our results to different gas compositions.
To evaluate the effect of \ce{H2} on the CO and NO production, we carried out equilibrium calculations with GGChem \citep[Fig.~\ref{Fig_CO_equi};][]{woitke_equilibrium_2018}.
We used a subset of the species within GGChem with 87 charged and neutral species containing the elements H, C, N, and O, as well as electrons.
We calculated the thermochemical equilibrium composition of gas mixtures with varying \ce{H2} fractions resembling our wet, high-\ce{CO2} experiments (4.6\% \ce{CO2}, 1\% \ce{H2O}, $0-10\%$ \ce{H2}, with the rest to 100\% \ce{N2}) for temperatures between 300 and $\SI{4000}{\kelvin}$.
By comparing our CO and NO measurements from the experiments with 0.9\% \ce{H2} (red and blue shaded areas, respectively) to the corresponding calculations (dashed lines), we can estimate a minimum freeze-out temperature for the respective gas.
Here, the CO measurement is from only one experiment, the NO concentration is the average of multiple experiments.

While the CO measurements with time allow us to estimate the equilibrium composition of CO in the flask (see Fig.~\ref{Fig_CO_time}), we only have one measurement of NO after the full $\SI{120}{\minute}$ run time of the experiment, which might increase slightly further. 
We therefore can only give a minimum estimate on the freeze-out temperature for NO.
Our estimates are $T_\mathrm{f} (\mathrm{CO}) = (2430\pm 65)\, \si{\kelvin}$ and $T_\mathrm{f} (\mathrm{NO}) \geq (3000\pm 160)\, \si{\kelvin}$.
The latter is similar to previous estimates by of $T_\mathrm{f} (\mathrm{NO}) = \SI{2300}{\kelvin}$ \citep{gilmore_production_1975,Chameides1977}, $\SI{2660}{\kelvin}$ \citep{Picone1981,borucki_lightning_1984}, and $\SI{3500}{\kelvin}$ \citep{Kasting1981}.
These estimates are based on different methods and assumptions on the cooling timescale of the gas heated in the lightning channel, showing that extrapolating from lab experiments and theoretical calculations to real-world lightning strikes comes with large uncertainties.
The errors we report with our freeze-out temperatures represent the range of temperatures necessary to explain the range of CO and NO concentrations in our experiments but do not include further uncertainties that arise when extrapolating our experimental conditions to real-world lightning strikes.
Our estimate of the NO freeze-out temperature also allows us to determine a lower limit of the maximum temperature reached within the spark channel of approximately $\SI{3000}{\kelvin}$.

There are much fewer estimates for the CO freeze-out temperature (compared to NO), with $\SI{3500}{\kelvin}$ when the CO is produced in the expanding shock front around the lightning channel \citep{Levine1979,chameides_implications_1979} and $\SI{2000}{\kelvin}$ if the CO is produced in the inner core of the cooling channel \citep{chameides_implications_1979}. 
\citet{Hill1980} point out that the cooling channel is more important for the production of NO, CO, and other gases because of its larger volume compared to the expanding shock front. According to \citet{Stark1996}, this is particularly true for spark experiments. 
Our estimate for the freeze-out temperature is thus slightly higher than the estimate of $\SI{2000}{\kelvin}$ for the cooling channel, potentially explained by an additional contribution of the higher freeze-out temperature of the shock front.

The estimates for the freeze-out temperatures allow us to extrapolate the CO and NO production to atmospheres with different \ce{H2} fractions.
As the freeze-out temperature depends on the cooling timescale of the spark and the equilibrium timescale of the gas mixture, changing the gas composition can lead to a change in the freeze-out temperature.
In our case, we only slightly vary the atmospheric composition, which allows us to assume a constant freeze-out temperature.
Figure~\ref{Fig_CO_equi} shows additional equilibrium calculations for gas mixtures without \ce{H2} (solid lines), with 5\% (dash-dotted) and with 10\% \ce{H2} (dotted).
The addition of \ce{H2} is increasing the abundance of CO in the lower temperature regime where (in the absence of \ce{H2}) \ce{CO2} would be more stable.
The \ce{H2} reacts with the atomic oxygen produced by the dissociation of the \ce{CO2} to form \ce{H2O}, preventing the CO from recombining with the O to \ce{CO2}.
Eventually, the abundance of CO is limited by the availability of C from the initial \ce{CO2}.
At the freeze-out temperature of $\sim\SI{2400}{\kelvin}$ this means an increase of the CO abundance by a factor of 2.5 when increasing the \ce{H2} fraction from 0.9\% to 10\% (in a background of 4.6\% \ce{CO2} and in the context of the limited chemistry applied in these calculations).
At the same time, the presence of \ce{H2} (and subsequent production of \ce{H2O}) reduces the availability of oxygen for the production of NO.
At $\SI{3000}{\kelvin}$, we find that the NO concentration for 10\% \ce{H2} is only $1/3$ of the concentration expected in the 0.9\%-\ce{H2} case.

From the initial slope of our fit to the CO concentration with time (Fig.~\ref{Fig_CO_time}) a global, annual production of $(6-18) \, \si{\tera\gram\per\year}$ and a surface flux of $(7.6 - 24) \times 10^8 \, \si{\per\centi\metre\squared\per\second}$ can be estimated for the different \ce{H2} concentrations (assuming modern Earth's lightning flash rate).
Using those estimates for surface fluxes, a grid of photochemical simulations were performed, similar to those presented by \citet{schwieterman_rethinking_2019}.
This allowed us to test under which conditions a lightning contribution to CO production could be observable (Section~\ref{Sec_Results_Photochem}).

Previous experimental results and calculations for the CO production by lightning in modern Earth's atmosphere are $\SI{0.01}{\tera\gram\per\year}$ \citep{green_production_1973}, $\SI{0.04}{\tera\gram\per\year}$ \citep{Levine1979}, and $(0.004-0.2)\, \si{\tera\gram\per\year}$ \citep{chameides_implications_1979}.
If we use the approach outlined above, using the equilibrium concentration of CO at $\SI{2430}{\kelvin}$ in a gas mixture resembling modern Earth's atmosphere, our estimate for the global CO production by lightning is $(0.01 \pm 0.003)\,\si{\tera\gram\per\year}$ which is in agreement with the values presented in the literature.

\subsection{Nitrogen Oxide (NO) Production and Extrapolation to Different Atmospheres}

\citet{barth_isotopic_2023} have shown that lightning can produce large amounts of nitrogen oxide in both \ce{N2}-\ce{O2} and \ce{N2}-\ce{CO2} atmospheres.
This NO provides the precursor of other nitrogen oxides in the gas phase as well as nitrite and nitrate in the aqueous phase (see below).
From our equilibrium calculations, we find that the maximum possible NO concentration (at $\sim \SI{3300}{\kelvin}$) is only slightly higher than our measurement for the wet, high-\ce{CO2} experiments (Fig.~\ref{Fig_CO_equi}).
This suggests, that at the time we took the NO measurement, the NO concentration in the flask had (nearly) reached equilibrium.
We cannot fit the production law we used for CO (Reaction~\ref{Reac_Equi_Fit}) to this single data point, but, assuming our data point represents the equilibrium NO concentration, we can find the slowest production that will reach equilibrium after $\SI{120}{\minute}$ (within 1\%).
The slope of this production curve at the origin provides a lower limit for the NO production rate.
For the wet experiments with a high \ce{CO2} concentration, this returns a lower limit of $(5.6 \pm 1.0) \times 10^{15} \, \si{molecules\per\joule}$ or a yearly production of $(2.6 \pm 0.4) \, \si{\tera\gram\per\year}$ with modern Earth's lightning flash rate.

For the other three sets of wet experiments (with \ce{N2}, \ce{N2-H2}, and low \ce{CO2} gas mixtures), the minimum NO production is independent from the \ce{CO2} and \ce{H2} fraction in the gas $(3 \pm 1) \times 10^{15} \, \si{molecules\per\joule}$ or $(1.2 \pm 0.3) \times 10^{-2} \, \si{\tera\gram\per\year}$.
In these experiments, water vapour in the gas mixture ($1 - 1.4 \%$) provides the necessary oxygen to oxidise nitrogen to NO.
In the corresponding dry experiments, the NO production is lower than in the wet experiments, but not 0 due to traces of \ce{CO2} (in particular in the low-\ce{CO2} experiments), water vapour, and \ce{O2} being present in the gas mixture.
\citet{barth_isotopic_2023} found in their wet experiments with only trace amounts of \ce{O2} (0.06\%) that $\sim 3 \times 10^{15} \, \si{molecules\per\joule}$ of NO are produced.
This production rate is similar to our wet experiments without any or with only small amounts of \ce{CO2}, suggesting that dissociation of water vapour is the main production pathway for NO in all of these experiments where the concentration of \ce{CO2} and \ce{O2} is low.

From the high-\ce{CO2} experiments and subsequent equilibrium calculations (Fig.~\ref{Fig_CO_equi}) we find that the presence of \ce{H2} in the gas mixture decreases the NO production.
Instead, more \ce{H2O} is produced.
We used these results for the NO production rate as input for photochemical simulations of the \ce{NO2} concentration in the atmospheres of different potential exoplanets (Section~\ref{Sec_Results_Photochem}).
\ce{NO} and \ce{NO2} are spectrally active at $\SI{5.3}{\micro\metre}$ and $\SI{6}{\micro\metre}$, respectively, and therefore potential signatures for lightning activity in exoplanet atmospheres \citep{gordon_hitran2020_2022}.

\subsection{Nitrous Oxide (\ce{N2O})}

\ce{N2O} is another potential biosignature, making an investigation of the possibility of a false-positive signature from lightning important.
The maximum \ce{N2O} production found in our experiments is $\SI{2.5}{\micro\mol}$ which corresponds to an energy yield of $8.5 \times 10^{12} \, \si{molecules\per\joule}$ or $1.3 \times 10^{-4} \, \si{\tera\mol\per\year}$ with modern Earth's lightning flash rate (see Section~\ref{Sec_Methods}). 
This value is similar to experimental results for Earth's atmosphere \citep{Levine1979,Hill1984,Chameides1986}.
The total biological \ce{N2O} emission on Earth is much larger with $\SI{0.45}{\tera\mol\per\year}$ \citep{bouwman_emissions_2002,tian_comprehensive_2020}.
This \ce{N2O} is produced by incomplete denitrification of Nitrate to \ce{N2} \citep{schwieterman_evaluating_2022}.
In the Proterozoic, the \ce{N2O} flux might have been significantly higher due to the limited availability of copper catalysts, preventing the last step of denitrification from \ce{N2O} to \ce{N2} \citep{buick_did_2007}.

In addition to the \ce{N2O} directly produced by lightning, other forms of fixed nitrogen (e.g. NO, \ce{NO2}, or \ce{HNO3}) can be deposited into the ocean, converted to \ce{N2O} by \ce{Fe^{2+}} \citep{Ranjan2019}, and again outgassed into the atmosphere.
If all lightning-produced NO (based on our experimental results) were to be converted to \ce{N2O} eventually, that would correspond to an annual production of $\SI{0.09}{\tera\mol\per\year}$.
This would only be a factor of 5 lower than the modern Earth \ce{N2O} flux and potentially detectable in the emission spectrum of an Earth-like planet orbiting a K dwarf \citep{schwieterman_evaluating_2022}.
This would be particularly true if the lightning flash rate in the atmosphere of such a planet is larger than that on modern Earth.

However, in the anoxic atmosphere of an Archean Earth-like planet, the \ce{N2O} abundance will be reduced by the missing \ce{O2}-shielding, decreasing the probability of a detectable signal and strengthening the case for the \ce{O2}/\ce{O3} + \ce{N2O} biosignature.
Moreover, an \ce{O2}-rich planet probably does not have \ce{Fe^{2+}}-rich oceans.
In addition, UV photolysis of nitrate and nitrite, releasing nitrogen back into the atmosphere as \ce{N2} \citep{zafiriou_nitrate_1979,zafiriou_nitrite_1979,carpenter_chemistry_2015,Ranjan2019} would reduce the feed-stock to produce abiotic \ce{N2O}.
A detectable \ce{NO2} signature from lightning (see Section~\ref{Sec_Results_Spectra}) might help to distinguish a lightning-produced \ce{N2O} signature from a biogenic source \citep{schwieterman_evaluating_2022}.
We use our results from the photochemical simulations to estimate the maximum potential \ce{N2O} production and compare it to the \ce{NO2} concentration (Section~\ref{Subsec_Ndep}).

\subsection{Nitrate (\ce{NO3^-}) \& Nitrite (\ce{NO2^-})}
\label{Subsec_nitrate}

Lightning-produced nitrite and nitrate have been hypothesised as potential nutrients for life on early Earth before the onset of biological nitrogen fixation \citep[e.g.,][]{ducluzeau_was_2009,Canfield2010,nitschke_beating_2013,shibuya_free_2016,Wong2017,Ranjan2019}.
Even though \citet{barth_isotopic_2023} have shown that life likely became independent from lightning as a nutrient source very early, it might have still contributed to support Earth's earliest biosphere.
Further, this nitrate and nitrite could present a nutrient source for life on exoplanets.

We combined the concentrations of nitrate and nitrite (Fig.~\ref{Fig_Bar_120min}) because part of the nitrite will oxidise to nitrate in the solution during the experiment.
The extent of this oxidation varies depending on the individual experiment setup.
As was the case for NO, we find that in wet experiments, the production of nitrite and nitrate increases compared to the dry experiments. 
Also the experiments without \ce{CO2} show some nitrite and nitrate production: 
in the wet experiments, this is due to the presence of water, but also in the dry experiments, some nitrite and nitrate are present.
This small residue may reflect the presence of trace amounts of gaseous \ce{H2O}, \ce{CO2}, or \ce{O2} in the initial gas mixture due to the limits of our vacuum.
Overall, in the high-\ce{CO2} experiments the nitrite and nitrate yield is very similar to the NO yield, with a maximum production of $\sim \SI{400}{\micro\mol}$. 
With modern Earth's lightning flash rate, that corresponds to $\sim \SI{0.02}{\tera\mol\per\year}$ or $\sim \SI{0.3}{\tera\gram N\per\year}$ of fixed nitrogen.
This yield is very similar to the value presented by \citet{barth_isotopic_2023} for a potential Archean atmosphere (0.16\% \ce{CO2} in \ce{N2}).

\PB{Our dry experiments suggest that the oxidation from lightning produced NO to \ce{NO_x^-} takes place very fast, while the NO- and HNO-rich air (HNO is also a lightning product) and the droplets of the cloud deck are still in contact to each other, even at some time after the flash.
This is in contradiction to \citet{hu_stability_2019} who argue that the separation of HNO and NO before the equilibration with the ocean and the subsequent aqueous chemistry quickly return lightning-fixed nitrogen into to the atmosphere as \ce{N2}.}

\subsection{Overnight Experiments}
\label{Sec_Results_Overnight}

\begin{figure*}
    \centering
    \includegraphics[width=\textwidth]{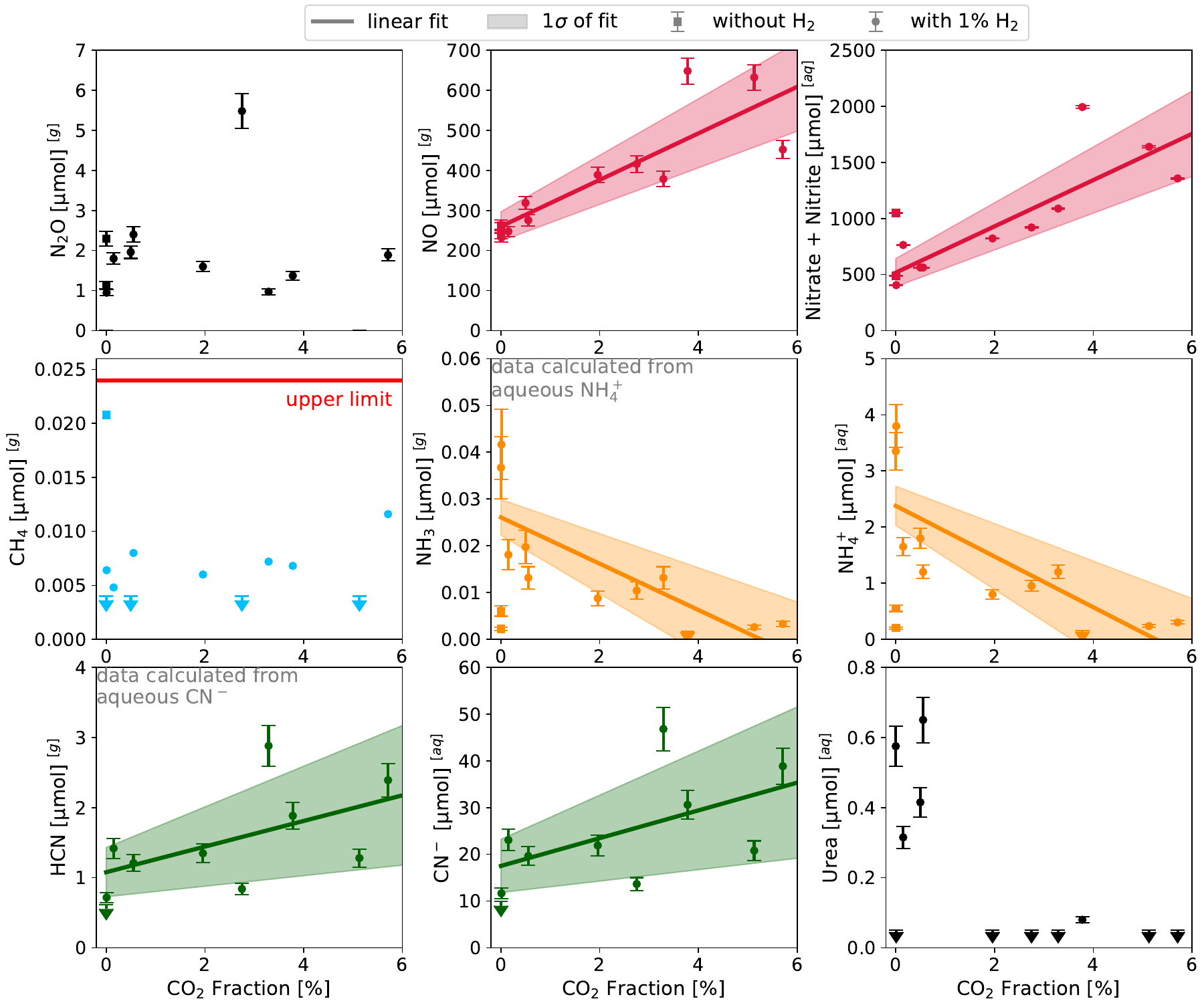}
    \caption{Final abundances of gaseous ($^{[g]}$; nitrous oxide (\ce{N2O}), methane (\ce{CH4}), hydrogen cyanide (HCN), nitric oxide (NO), and ammonia (\ce{NH3})) and aqueous ($^{[aq]}$) products (\ce{CN^-}, nitrite + nitrate, \ce{NH4^+}, and urea) in overnight experiments. 
    Data points represent individual experiments with varying \ce{CO2} fractions. 
    Experiments without any \ce{H2} and \ce{CO2} are indicated by squares and not included in fits.
    Lines are best linear fits for dependency between \ce{CO2} concentration and the final abundance of products. 
    Measurements below the detection limit (arrows) are included as (Detection Limit $/ \sqrt{2}$) in the fitting process.
    Shaded areas give $1\sigma$-range.
    No fit is shown for methane since all measured values are within $1\sigma$ of 0 (\textit{red line}), for nitrous oxide since there is no trend visible in the data, and for urea as most measurements for \ce{CO2} $>1\%$ below the detection limit.
    Abundances of HCN and \ce{NH3} are calculated from measured aqueous abundances of \ce{CN^-} and \ce{NH4^+}, respectively.}
    \label{Fig_Overnight}
\end{figure*}

In addition to the short 120-min experiments, we also conducted overnight experiments with an average total spark time of $(925 \pm 35)\, \si{\minute}$ to investigate the production of compounds that were not detectable in the short experiments or had only very small yields.
Figure~\ref{Fig_Overnight} shows the abundances of gaseous and aqueous products after the overnight experiments and linear fits to show the relation between \ce{CO2} concentration and the final abundance of the product.
Again, the abundances of HCN and \ce{NH3} are calculated from the measured aqueous abundances of \ce{CN^-} and \ce{NH4^+}, respectively, with their respective Henry's law constants as described in Appendix~\ref{Sec_Appendix_Methods}.
Similar to the short experiments, we see an increasing production of oxidised nitrogen (NO, nitrate and nitrite) and a decreasing production of reduced nitrogen (ammonium) with increasing \ce{CO2} concentration.
The maximum ammonium concentration (at 0\% \ce{CO2}) is only about $2-4$ times the corresponding ammonium concentration in the short experiments, even though the spark was running for $\sim 8$ times as long, suggesting that the ammonium concentration reached equilibrium during the experiment.
For a \ce{CO2} concentration of $\gtrsim 5 \%$ basically no ammonium was produced.
Instead, the dissociated nitrogen was likely oxidised to NO and eventually nitrite and nitrate.
A higher \ce{H2} concentration might allow the production of ammonium at higher \ce{CO2} concentrations, but eventually, if there is significantly more \ce{CO2} (or a different oxygen source) than \ce{H2} in the gas mixture, the ammonium production would probably still be suppressed.
As discussed above, significant amounts of ammonium can still be produced by subsequent reduction of nitrite by \ce{Fe^{+2}} and FeS in the ocean, \PB{and to a lesser extent by reduction of nitrate by FeS} \citep{Summers1993,summers_ammonia_2005}.

The concentration of nitrite and nitrate as well as NO in our discharge experiments shows a very clear trend with \ce{CO2} concentration in the initial gas mixture.
For \ce{CO2} concentrations around 0, water vapour in the gas mixture is a significant source of oxygen for the production of nitrogen oxides. The slightly increased scatter of measurements at that point can be explained by the additional uncertainty of the concentration of water vapour and other trace gases like \ce{O2}.
We also find the nitrite and nitrate concentration in the overnight experiments to be $5-10$ times the concentration in the corresponding short experiments, suggesting that even though NO equilibrium in the gas phase is reached rather quickly (the final NO concentration in the overnight experiments is similar to the short experiments), the subsequent oxidation to \ce{NO2} and equilibration with the aqueous phase takes more time, in particular in the experiments where relatively small amounts of nitrite and nitrate are produced.
All of these experiments were run with $\SI{50}{\milli\litre}$ of water in the flask, so even when the \ce{CO2} content is 0, there was sufficient oxygen available from the water to provide oxygen for the production of some NO and subsequently nitrite and nitrate.
The presence of water as an oxygen source is likely also the reason that we again see only very small concentrations of methane, below the measurement error.
Unlike the other forms of oxidised nitrogen, we do not see a clear trend in the production of \ce{N2O} with \ce{CO2} concentration.
This follows the trend in our short, wet experiments where also no clear trend was visible, suggesting that if enough oxygen is available, nitrogen oxides with a higher oxidation state are preferred. 
\ce{N2O} has an oxidation state of $+1$ while NO, \ce{NO2}, nitrite and nitrate have oxidation states of $\ge 2$.

Urea (\ce{CO(NH2)2}) is an important precursor for cyanamide (\ce{CH2N2}) which itself is a precursor for RNA \citep{das_insights_2019}.
In our experiments, urea follows a similar trend as ammonium, with abundances of roughly one order of magnitude lower.
However, because most measurements for \ce{CO2} concentrations above 1\% were below the detection limit, we did not try to fit a line to the urea data.
It thus seems that the production of (detectable levels of) urea is only possible under reducing conditions.

\begin{figure}
    \centering
    \includegraphics[width=\columnwidth,page=1]{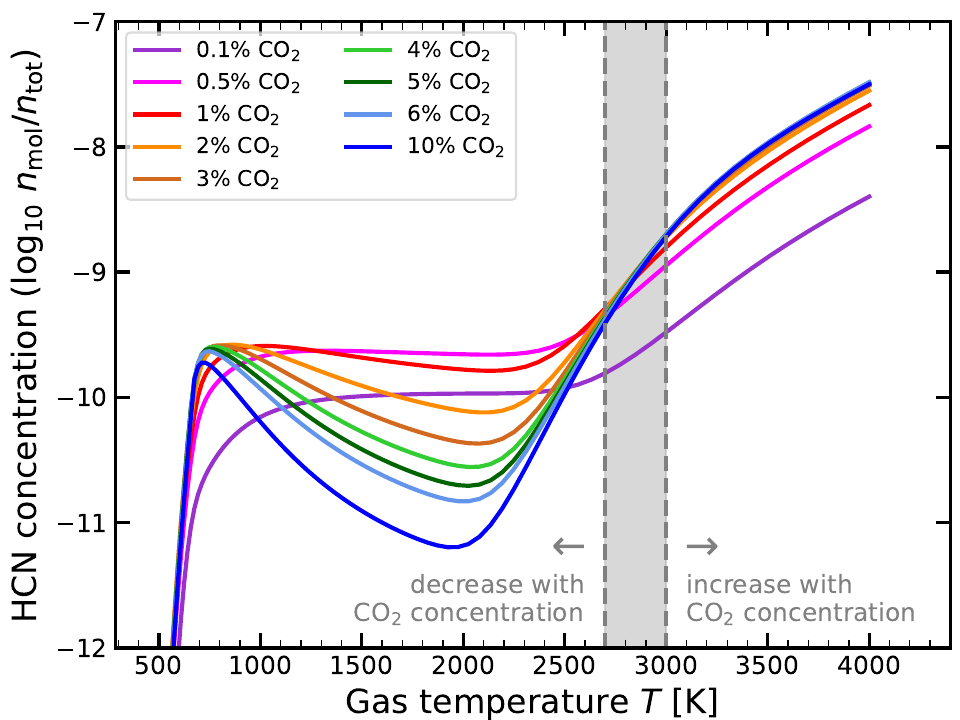}
    \caption{HCN concentration in chemical equilibrium \citep[with GGChem,][]{Woitke2003} as function of gas temperature for different \ce{CO2} fractions in the gas mixture (other gases: 1\% \ce{H2}, 1\% \ce{H2O}, rest \ce{N2}). For $T \gtrsim \SI{3000}{\kelvin}$ (\textit{right dashed grey line}), HCN concentration increases with increasing \ce{CO2} concentration (until $\sim 3-5\%$ \ce{CO2}). For $T \lesssim \SI{2700}{\kelvin}$ HCN concentrations decreases with increasing \ce{CO2} concentration (except for 0.1\% \ce{CO2}).}
    \label{Fig_HCN_GGChem}
\end{figure}

HCN is an important precursor molecule for the formation of RNA and has been hypothesised to be produced by lightning in reduced atmospheres \citep{miller_formation_1957,miller_atmosphere_1983,Pearce2017,pearce_experimental_2022}.
In our experiments, we can monitor the HCN production by its dissolved form, cyanide \ce{CN^-}.
The cyanide abundance increases with increasing \ce{CO2} abundance, though much more slowly than the abundance of nitrite and nitrate.
This follows the calculations performed by \citet{chameides_rates_1981}, who predicted an increase in HCN production when decreasing the C/O-ratio (at constant \ce{H2} concentration), which is happening when increasing the \ce{CO2} concentration (without adding other forms of carbon this limits the C/O ratio to $1/2$).
\citet{chameides_rates_1981} predict an HCN production rate of $\sim 3 \times 10^{9}$ to $4 \times 10^{11} \, \si{molecules\per\joule}$ for the range of C/O-ratios equivalent to our experiments (for a gas mixture with $\SI{0.9}{\bar}$ \ce{N2} and $\SI{0.05}{\bar}$ \ce{H2}, C + O = $\SI{0.1}{\bar}$).
We find our results for HCN production to be $\sim 2$ to 3 orders of magnitude larger than the values calculated by \citet{chameides_rates_1981}. 
The reason for this might be that the water in our experiments acts as a buffer: produced HCN dissolves in the water, lowering the concentration in the gas phase and allowing for more HCN to be produced. 
This suggests that we over-predict the production of HCN even though lightning occurs in the water-saturated atmosphere (clouds).
In real lightning conditions, HCN is produced with the chemical equilibrium composition at the specific freeze-out temperature.
After the HCN is produced, equilibration with water droplets happens. 
The next lightning flash produces new HCN in a different air parcel. 
In contrast, in our experiments, we keep adding HCN to the same gas mixture which is constantly equilibrating with the water phase.
For other gases, this is less of a problem as we are measuring most of them in the gas phase and their Henry’s law constants are orders of magnitude lower, meaning less of the gas is absorbed into the water.
\PB{In conclusion, these experimental limitations suggest that our results do not apply to the real atmosphere for HCN.}
Chemical equilibrium calculations (Fig.~\ref{Fig_HCN_GGChem}) show that the concentration of HCN increases with \ce{CO2} concentration for gas temperatures above $\sim \SI{3000}{\kelvin}$, while temperatures below $\sim \SI{2700}{\kelvin}$ the HCN concentration decreases.
This suggests that the freeze-out temperature for HCN in our experiment is $\gtrsim \SI{3000}{\kelvin}$.
Other experiments studied the effect of \ce{CH4} concentration on HCN production: an increase in \ce{CH4} increases the C/O-ratio, leading to a strong increase in HCN production \citep{chameides_rates_1981,Tian2011,pearce_experimental_2022}.

\section{Results and Implications of Photochemical Simulations}
\label{Sec_Results_Photochem}

Now that we know the production rates for CO and NO, by far the most important direct products of lightning in \ce{N2-CO2} gas mixtures, we want to know how this influx of CO and NO changes the composition of different planetary atmospheres.
We are particularly interested in whether lightning can produce observable signatures in transmission, emission, and reflected light spectroscopy.
To answer these questions, we used the photochemical model of the Atmos coupled climate-photochemistry code \citep{arney_pale_2016,Lincowski2018} (see Section~\ref{Sec_Methods}) to calculate the atmospheric mixing ratio of CO, NO, \ce{NO2}, and other gases for a large range of lightning flash rates.
We modelled both oxygen-poor (anoxic) and oxygen-rich (oxic) atmospheres on Earth-sized planets orbiting the Sun (G-type star) and TRAPPIST-1 (M-type star) at the inner edge of their respective habitable zones.

\begin{figure*}
    \centering
    \includegraphics[width=0.7\textwidth]{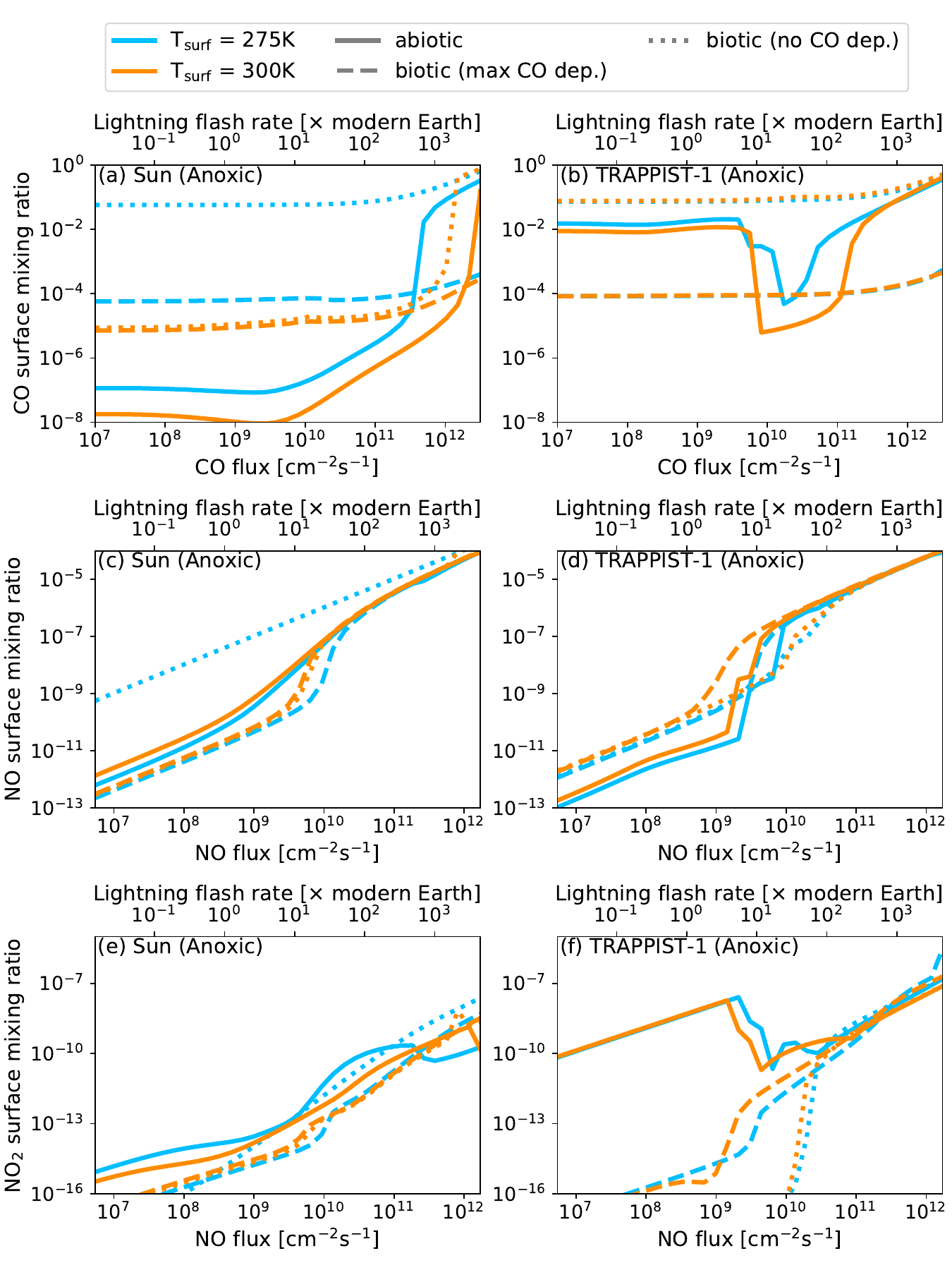}
    \caption{Photochemically simulated CO (a \& b), NO (c \& d), and \ce{NO2} (e \& f) mixing ratios in the anoxic atmosphere (4.6\% \ce{CO2}) of an Earth-sized planet orbiting the Sun (a, c, e) and TRAPPIST-1 (b, d, f) for varying CO and NO production rates and different scenarios: 
    abiotic (\textit{solid}), biotic with CO deposition (\textit{dashed}), and biotic without CO deposition (\textit{dotted}). 
    The surface temperature (\textit{blue}: $T_\mathrm{surf} = \SI{275}{\kelvin}$, \textit{orange}: $T_\mathrm{surf} = \SI{300}{\kelvin}$) controls the water vapour concentration. 
    The second $x$-axis gives the corresponding lightning flash rate (in units of modern Earth's flash rate), as estimated for the respective atmospheric composition.}
    \label{Fig_CO_NO_simulations_anoxic}
\end{figure*}

\begin{figure}
    \centering
    \includegraphics[width=\columnwidth]{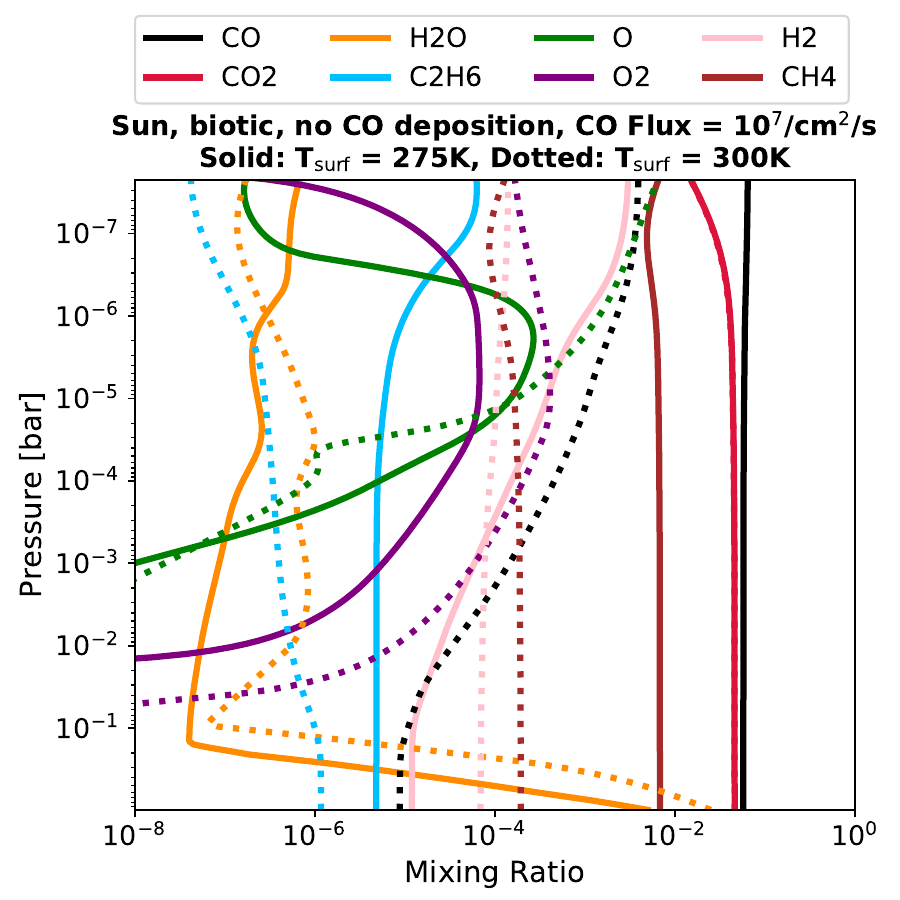}
    \caption{Mixing ratios for biotic scenario (Sun) without CO deposition (CO flux $\sim 10^{7} \, \si{molecules\per\centi\metre\squared\per\second}$) for two different surface temperatures (\textit{solid lines}: $T_\mathrm{surf} = \SI{275}{\kelvin}$, \textit{dotted lines}: $T_\mathrm{surf} = \SI{300}{\kelvin}$) as result of a kinetic gas-phase calculation with photochemistry.}
    \label{Fig_Sun_Anoxic_275_biotic_nodep_PMR_vs300}
\end{figure}

\begin{figure*}
    \centering
    \includegraphics[width=0.7\textwidth]{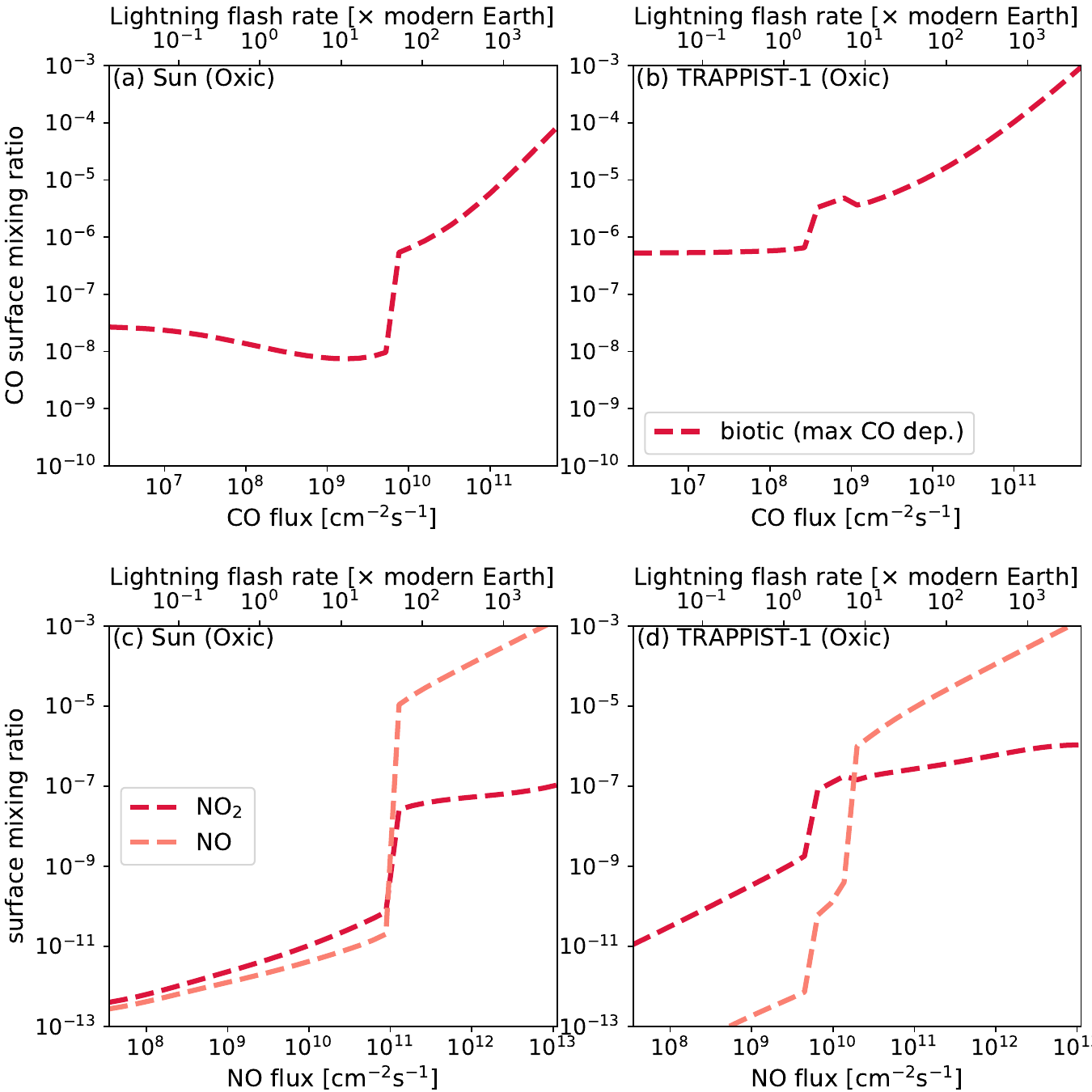}
    \caption{Photochemically simulated CO (a \& b), NO and \ce{NO2} (c \& d) mixing ratios in the oxic atmosphere (4.6\% \ce{CO2}, 21\% \ce{O2}) of an Earth-sized planet orbiting the Sun (a \& c) and TRAPPIST-1 (b \& d) for varying CO and NO production rates and the biotic scenario with CO deposition ($T_\mathrm{surf} = \SI{288}{\kelvin}$).
    The second $x$-axis gives the corresponding lightning flash rate (in units of modern Earth's flash rate), as estimated for the respective atmospheric composition.}
    \label{Fig_CO_NO_simulations_oxic}
\end{figure*}

\subsection{Photochemistry in Anoxic Atmospheres}

Figure~\ref{Fig_CO_NO_simulations_anoxic} (a, b) shows the resulting CO mixing ratios for the anoxic atmosphere (4.6\% \ce{CO2} in \ce{N2} filler gas), equivalent to our experiments and necessary to keep the surface of the investigated planets clement \citep{meadows_habitability_2018}, for abiotic and biotic scenarios with different \ce{CH4} fluxes and CO deposition rates and a range of lightning flash rates (Table~\ref{Tab_Photochem_Parameters}).
Using equilibrium chemistry calculations as described in Section~\ref{Sec_Results_extra}, we can estimate the production of CO and NO with modern Earth's lightning flash rate to be $7.0 \times 10^8$ and $3.8 \times 10^8 \si{molecules\per\centi\metre\squared\per\second}$, respectively, in this anoxic gas mixture.
The results look very differently for the two different host stars with different spectra:
On the planet orbiting the Sun, the CO mixing ratio in the biotic scenario is up to three orders of magnitude larger than in the abiotic scenario.
On the planet orbiting TRAPPIST-1, this trend is reversed.
The CO ratio in the biotic scenario is 100 times smaller than in the abiotic scenario for most of the CO flux range.
the high CO concentrations in the TRAPPIST-1 simulations are due to efficient \ce{CO2} photolysis:
The XUV flux is concentrated near the Ly-$\alpha$ line and the FUV continuum and therefore absorbed by the abundant \ce{CO2} molecules that are dissociated into CO and O.
Thus, the \ce{CO2} is shielding water molecules from the XUV radiation and the photolysis of water in the atmosphere of a planet orbiting an M dwarf is less efficient and the concentration of the OH radical is lower \citep{segura_biosignatures_2005,schwieterman_rethinking_2019}.
Since the deposition velocity with which CO is deposited in the ocean is lower in the abiotic than in the biotic scenario where acetogens in the ocean consume the CO, the abiotic CO concentration remains higher than in the biotic case.

In the simulations for the planet irradiated by the Sun, the OH concentration in the atmosphere is higher and the CO lifetime is shorter.
In the biotic scenario, the increased, biogenic flux of \ce{CH4} which is photo-oxidised to CO leads to a higher CO concentration than in the abiotic scenario.
In a test case for the biotic scenario, where we assume no CO to be deposited into the ocean (dotted lines in Fig.~\ref{Fig_CO_NO_simulations_anoxic}), the final CO concentration is increased to well above 1\% in all but one case despite OH being present.
An increase in surface temperature increases the concentration of water vapour in the atmosphere and subsequently the concentration of OH, decreasing the lifetime of CO and its final concentration.
This is especially apparent for the biotic planet around the Sun where we assume no CO deposition:
while for a surface temperature of $T_\mathrm{surf} = \SI{275}{\kelvin}$ (blue) the CO concentration is enhanced to above 1\%, for a slightly higher surface temperature of $T_\mathrm{surf} = \SI{300}{\kelvin}$ the subsequently higher water and OH concentrations allow for more efficient removal of CO from the atmosphere (see Fig.~\ref{Fig_Sun_Anoxic_275_biotic_nodep_PMR_vs300} for comparison of mixing ratios).
This suggests that for the biotic scenario for the Sun-orbiting planet with $T_\mathrm{surf} = \SI{300}{\kelvin}$, recombination of CO with OH is the major CO sink.
Overall, the major sink for CO in both biotic scenarios for TRAPPIST-1 and the biotic scenario with $T_\mathrm{surf} = \SI{275}{\kelvin}$ for the Sun is deposition in the ocean.

In nearly all anoxic simulations, the CO concentration is relatively high due to \ce{CO2} dissociation and \ce{CH4} photo-oxidation and additional contributions of CO by lightning are not able to further increase the CO concentration.
Only for lightning flash rates around 1000 times the modern Earth lightning flash rate we can see the CO concentration increase slightly beyond the background level.
The only scenario, where the background CO level is low enough to see an increase in CO concentration at lower flash rates, is the abiotic scenario for the planet orbiting the Sun.
For CO production rates higher than $\sim 10^{11}$ or $\sim 10^{12} \, \si{molecules\per\centi\metre\squared\per\second}$ for $T_\mathrm{surf} = \SI{275}{\kelvin}$ and $T_\mathrm{surf} = \SI{300}{\kelvin}$ respectively, the atmosphere enters into a CO runaway: the CO flux overwhelms the CO deposition and the CO concentration jumps up to several per cent.
\citet{kasting_bolide_1990} showed that also an increased NO concentration (in our simulations, the NO flux and thus concentration increases with increasing flash rate) can lead to a longer lifetime of CO in the atmosphere and thus to a higher CO concentration.
The CO flux where the atmosphere enters into a CO runaway is similar to the total CO emission on the modern Earth of $\sim 2 \times 10^{11} \, \si{molecules\per\centi\metre\squared\per\second}$ which is mainly from anthropogenic sources (68\%) and wildfire and deforestation \citep[32\%;][]{zhong_global_2017}.
However, on the uninhabited Archean Earth, the volcanic CO flux is estimated to be much lower at $\sim (1-2) \times 10^{8} \, \si{molecules\per\centi\metre\squared\per\second}$ \citep{Kasting1981}.
We also should note that for CO concentrations larger than the \ce{CO2} concentration (4.6\%) which occur in our simulations only for extremely high lightning flash rates, our assumption of constant \ce{CO2} concentration might not be applicable anymore, as it would require very high volcanic fluxes of \ce{CO2} to replenish the \ce{CO2} dissociated by lightning and would greatly enhance the carbon budget in the atmosphere.

For the planet orbiting the Sun, we find a very similar trend for the \ce{NO2} concentration for all scenarios (Fig.~\ref{Fig_CO_NO_simulations_anoxic}, e \& f):
Very low abundances ($< 10^{-8}$ for nearly all simulations), but a steady increase with increasing NO flux.
For the planet orbiting TRAPPIST-1, this looks a bit different:
in the biotic scenario with CO deposition, the \ce{NO2} concentration follows a very similar trend to the same case for the Sun-planet, but with slightly higher mixing ratios of up to 1~ppm for the highest NO fluxes.
In the abiotic case, though, the \ce{NO2} abundance is more than six orders of magnitude higher for low NO fluxes compared to the biotic case, before it rapidly decreases, parallel to the CO abundance, at an NO input flux of $\sim 3 \times 10^{9} \, \si{molecules\per\centi\metre\squared\per\second}$.
At higher fluxes, the \ce{NO2} abundance increases again, very similar to the biotic scenario.

We can also observe this behaviour in the CO concentration for the abiotic scenario on the TRAPPIST-1 planet (Fig.~\ref{Fig_CO_NO_simulations_anoxic}b, solid lines): 
here the background CO concentration (for low CO lightning fluxes) is very high at approximately 1\%.
Once the CO and the corresponding NO flux cross a certain threshold ($\sim 10^{10} \, \si{molecules\per\centi\metre\squared\per\second}$), the CO concentration drops by more than two orders of magnitude before rapidly climbing again in a CO runaway.
This drop in the CO and \ce{NO2} concentration is accompanied by a drop in \ce{O2} concentration and a strongly increasing NO concentration (Fig.~\ref{Fig_CO_NO_simulations_anoxic}, c \& d).
In the anoxic gas mixture in our experiments as well as in lightning strikes in the anoxic Archean atmosphere, where NO and CO are produced from dissociated \ce{CO2} and \ce{N2}, free oxygen is produced if not the same amount of NO and CO is produced to match the stoichiometric ratio of the split \ce{CO2} \citep{kasting_bolide_1990}. 
When using the CO and NO production fluxes in our photochemical simulations, we included an additional O flux, which is variable alongside the NO and CO fluxes, to balance the ratio between the elements N, C, and O.
Together, these three fluxes replace the lightning module originally included in Atmos \citep{harman_abiotic_2018} which we turned off for these simulations.
The background concentrations of \ce{O2} and ozone as byproducts from \ce{CO2} photolysis are very high in this scenario and even slightly enhanced further by the addition of free oxygen from lightning, up to an \ce{O2} surface mixing ratio of $1.4 \times 10^{-3}$ and a peak ozone concentration of $\sim \SI{5}{ppm}$ at a height of $\SI{32}{\kilo\metre}$.
The high concentration of oxygen allows for efficient oxidation of NO to \ce{NO2} (\ce{NO + O -> NO2}), resulting in a higher concentration of \ce{NO2} than NO. 
Other efficient destruction channels for NO and \ce{NO2} are \ce{NO + HCO -> HNO + CO} and \ce{NO2 + O -> NO + O2}, respectively. 
At high NO fluxes, the channels \ce{H + NO -> HNO},  \ce{NO + OH -> HNO2}, \ce{NO + O -> NO2}, and \ce{NO + HO2 -> NO2 + OH} become the limiting factors controlling the NO concentration, while \ce{NO2} photolysis and the reaction \ce{NO2 + H -> NO + OH} control the destruction of \ce{NO2}. 
At high NO fluxes, the overall balance of the HO$_x$ reservoir shifts by decreasing steady-state \ce{HO2} while increasing steady-state OH. 
This shift allows for increased recombination of \ce{CO + OH -> CO2 + H}, leading to a sharp decrease in the concentration of CO and oxygen species, entering a different photochemical regime \citep{ranjan_photochemical_2022}. 
We note that the HO$_x$ reservoir ultimately depends on \ce{H2O} photolysis \citep{harman_abiotic_2015}, and all NO-mediated catalytic cycles that net recombine CO and O to \ce{CO2} require the presence of HO$_x$ species \citep[see][their Table 1]{harman_abiotic_2018}. 
For anoxic atmospheres with a Sun host, the dominant photochemical channels are more consistent throughout the range of NO fluxes, which is reflected in the lessened discontinuous behaviour in the flux-abundance relationships shown in the left panels of Fig~\ref{Fig_CO_NO_simulations_anoxic}. 
For these atmospheres, NO is primarily destroyed by \ce{NO + O -> NO2} and \ce{NO2} is primarily destroyed by \ce{NO2} photolysis.

This process was described by \citet{harman_abiotic_2018} who suggested that lightning-produced NO can provide a catalyst for the recombination of CO and O to \ce{CO2}, preventing a false-positive \ce{O2}-biosignature.
They assumed an NO production of $\sim 6 \times 10^{8} \, \si{molecules\per\centi\metre\squared\per\second}$ (for modern Earth's lightning flash rate) in an atmosphere with 5\% \ce{CO2} and found this to strongly reduce the CO and \ce{O2} concentrations (for an M dwarf host star).
This NO production rate is about 50\% higher than what we assume for modern Earth's lightning flash rate, but even if we correct for this difference, the additional production by lightning of CO and O in our model moves this behaviour to flash rates at least three times that of modern Earth.
Using our assumption for the NO production per flash, a flash rate of more than six times modern Earth's is necessary to prevent the build-up of significant amounts of oxygen in the atmosphere.
This difference is likely due to differences in the UV spectrum used for the simulations. 
Larger stellar FUV/NUV flux ratios tend to drive higher abiotic \ce{O2} production rates \citep{harman_abiotic_2015} and TRAPPIST-1 (M8V) has a larger FUV/NUV ratio than the latest host star considered by \citet{harman_abiotic_2018}, an M4V dwarf.
Importantly, this suggests that lightning may not eliminate all \ce{O2}-false positive scenarios for \ce{CO2}-rich terrestrial planets orbiting ultra-cool dwarf hosts, at least at the $\sim 0.1 \%$ level. 
Figure~\ref{Fig_A_Anoxic_abiotic_ncol} shows the column density of the atmospheric constituents for the range of simulated CO and NO fluxes, Fig.~\ref{Fig_A_TR1_Anoxic_275_abiotic_PMR_spec} the atmospheric mixing ratio profiles for simulations before and after the drop in CO and \ce{O2} concentration for the scenario with $T_\mathrm{surf} = \SI{275}{\kelvin}$.
The spectra for these two cases are discussed below (Fig.~\ref{Fig_Spectra_TR1_Anoxic_275_abiotic}).

\subsection{Photochemistry in Oxygen-rich Atmospheres}

Earth's atmosphere was anoxic for the first $\sim \SI{2}{\giga\year}$ of its evolution. 
After that, the oxygen concentration increased drastically, but it was only $\sim \SI{0.5}{\giga\year}$ ago that the oxygen concentration reached today's level \citep[e.g.,][]{Catling2020}.
Assuming that the evolution of oxygenic photosynthesis takes a similar amount of time on other worlds if it happens at all, it is therefore most likely that if we find an inhabited planet, it will have an anoxic or low-oxygen atmosphere \citep{Krissansen-Totton2018}.
However, as we have discussed in the introduction, photochemistry and hydrogen escape can lead to the abiotic build-up of \ce{O2} in an otherwise anoxic atmosphere.
To fully investigate the impact of lightning on observable oxygen and ozone features, we therefore also conducted simulations for oxygen-rich atmospheres with a biosphere (4.6\% \ce{CO2}, 21\% \ce{O2} in \ce{N2} filler gas; biotic scenario with maximum CO deposition; Fig.~\ref{Fig_CO_NO_simulations_oxic}).
In addition, the \ce{O2} and \ce{O3} features of an \ce{N2-O2} atmosphere are likely not detectable with \textit{JWST} \citep{krissansen-totton_detectability_2018}.
Thus, the atmosphere of a modern-Earth-like, inhabited planet might resemble the Archean Earth instead, with \ce{CH4} and \ce{CO2} features.
In this case, it is important to understand the role of lightning in potentially producing a CO signature.

In such an oxic gas mixture, our estimate for the CO and NO production rates at modern Earth's lightning flash rate are $1.5 \times 10^8$ and $2.5 \times 10^9 \, \si{molecules\per\centi\metre\squared\per\second}$, respectively, using the method described in Section~\ref{Sec_Results_extra}.
Our simulations find, for both host stars, a significantly lower CO mixing ratio than in the anoxic atmospheres, independent from the CO deposition velocity.
The high abundance of \ce{O2} in these atmospheres offers an additional sink for atmospheric CO: excited atomic oxygen from ozone photolysis increases the production of the OH radical: \ce{O3 + $h\nu (\lambda < \SI{320}{\nano\metre})$ -> O(^1D) + O2} and \ce{H2O + O(^1D) -> 2OH}.
Since the NUV radiation necessary for this pathway is lower for M dwarfs, the CO mixing ratio on the planet around TRAPPIST-1 is higher than on the planet around the Sun \citep{segura_biosignatures_2005,schwieterman_rethinking_2019}.
However, for CO fluxes $> 3\times 10^{9} \, \si{molecules\per\centi\metre\squared\per\second}$
for the planet orbiting the Sun and $> 10^{8} \, \si{molecules\per\centi\metre\squared\per\second}$
for the planet orbiting TRAPPIST-1 we find an increase of the CO mixing ratio.
This corresponds to lightning flash rates of $\sim 10$ and $\sim 0.7$ times modern Earth's flash rate, respectively.
In the Sun case, the CO mixing ratio increases to $\sim 10^{-7}$ for a CO flux of $3\times 10^{9} \, \si{molecules\per\centi\metre\squared\per\second}$ and then steadily to $\sim 10^{-4}$ for a CO flux of $5 \times 10^{11} \, \si{molecules\per\centi\metre\squared\per\second}$.
In the TRAPPIST-1 case, the CO mixing ratio increases to $\sim 10^{-6}$ for a CO flux of $10^{8} \, \si{molecules\per\centi\metre\squared\per\second}$ and then steadily to $\sim 10^{-4}$ for a CO flux of $10^{11} \, \si{molecules\per\centi\metre\squared\per\second}$.
To estimate the corresponding lightning flash rate, we used again equilibrium calculations to determine the CO concentration at the freeze-out temperature of $\SI{2430}{\kelvin}$ for the simulated, oxic gas mixture. 
We find that at modern Earth's lightning flash rate, the CO concentration is approximately 6 and 2 orders of magnitude smaller than in the corresponding anoxic atmosphere for the Sun and TRAPPIST-1 planet, respectively (same \ce{CO2} concentration of 4.6\%), as the abundant \ce{O2} increases the recombination of CO to \ce{CO2}.
This means lightning flash rates of 3000 and 500 times the modern Earth's flash rate are needed to achieve CO concentrations similar to the corresponding anoxic scenarios (biotic with maximum CO deposition).

The \ce{NO2} concentration follows a very similar trend to the CO concentrations in the photochemical calculations with a sharp increase at CO fluxes of $\sim 3\times 10^{9} \si{molecules\per\centi\metre\squared\per\second}$ and $\sim 10^{8} \si{molecules\per\centi\metre\squared\per\second}$, and corresponding NO fluxes of $\sim 6\times 10^{10} \si{molecules\per\centi\metre\squared\per\second}$ and $\sim 2 \times 10^{9} \si{molecules\per\centi\metre\squared\per\second}$, for the Sun and TRAPPIST-1 case, respectively.
The changes in the slope of the NO and \ce{NO2} abundances vs. NO flux are caused by shifts in their main photochemical destruction channels as these species overwhelm the photochemical sinks with the fastest kinetic rates. 
For NO, the \ce{NO + O3 -> NO2 + O2} channel saturates near this threshold flux, and the slower \ce{NO + O -> NO2} and \ce{NO + OH -> HNO2} channels become comparable above it. 
This tracks with the depletion in \ce{O3} we also see in our photochemical simulations at increasing NO flux. 
For \ce{NO2}, the \ce{NO2 + O -> NO + O2} channel saturates, and \ce{NO2} photolysis becomes the dominant destruction channel. 
The flux-abundance relationships vary between host stars due to the different distributions of radical species generated by differences in stellar spectra. 
We also emphasise that these shifts in photochemical channels with increasing NO flux differ in the oxic vs. anoxic cases.

Similar to the anoxic case, in the oxic atmosphere, lightning will also produce free oxygen which we expect to mostly recombine to \ce{O2}.
However, it is hard to determine how much atomic oxygen is produced.
Stoichiometrically, all NO and CO can be produced from dissociated \ce{N2}, \ce{O2}, and \ce{CO2} without any remaining O.
To test whether the addition of free O might change the chemistry significantly, we ran simulations with an O flux similar to the lightning NO flux.
We only found small enhancements in the final \ce{NO2} concentrations for the highest NO fluxes ($> 10^{12} \, \si{molecules\per\centi\metre\squared\per\second}$).

\subsection{Nitrogen Deposition into an Ocean}
\label{Subsec_Ndep}

\begin{figure}
    \centering
    \includegraphics[width=\columnwidth]{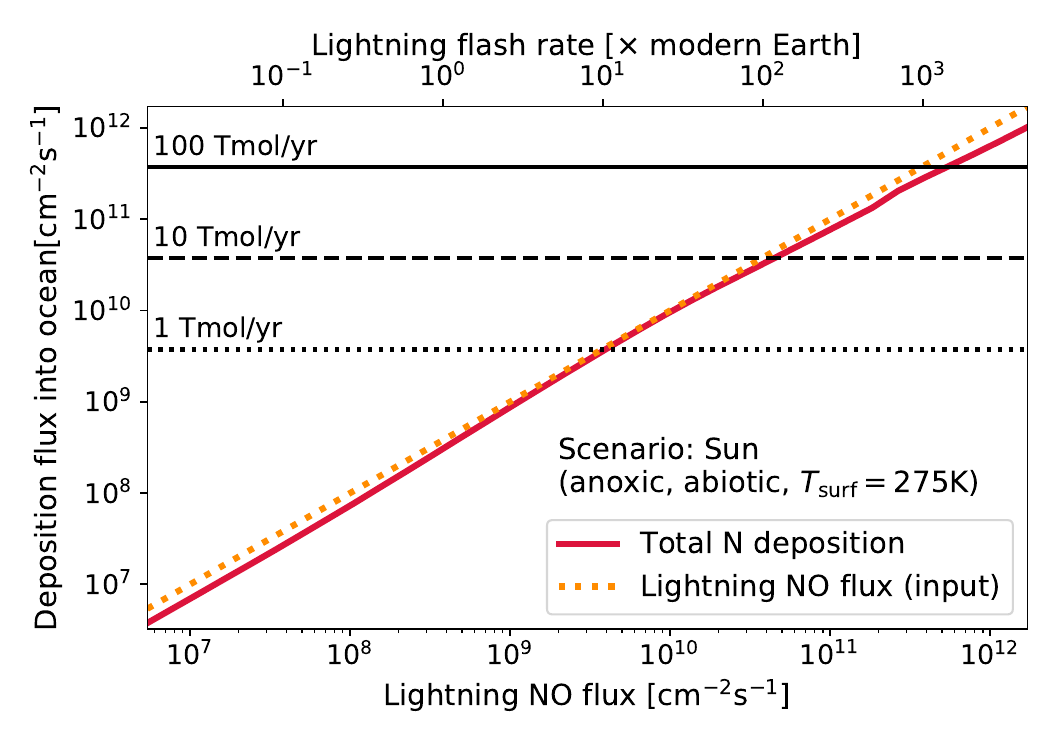}
    \caption{Deposition flux of fixed nitrogen into an ocean (\textit{red}) for the anoxic, abiotic planet around the Sun ($T_\mathrm{surf} = \SI{275}{\kelvin}$) for a range of lightning NO fluxes and corresponding lightning flash rates. 
    For comparison, the total lightning-produced NO flux is shown (\textit{orange}).
    The corresponding deposition rates in $\si{\tera\mol\per\year}$ are shown as black lines (converted into $\si{molecules\per\centi\metre\squared\per\second}$).}
    \label{Fig_N_deposition}
\end{figure}

To investigate the possibility of lightning-produced NO to be converted to \ce{N2O} in the ocean, which could in return be outgassed into the atmosphere where it might produce a false-positive biosignature, we calculate the nitrogen deposition flux into the surface of the planet.
In addition to NO, also \ce{NO2}, HNO, and \ce{HNO3} are deposited from the atmosphere. 
The combined nitrogen deposition flux is shown in Fig.~\ref{Fig_N_deposition} in comparison to the NO production from lightning for the anoxic and abiotic scenario for the planet orbiting the Sun ($T_\mathrm{surf} = \SI{275}{\kelvin}$).
A large part of the introduced NO is deposited into the surface, mainly in the form of HNO: up to $\sim95\%$ at NO fluxes of $\sim 10^{10} \, \si{molecules\per\centi\metre\squared\per\second}$, decreasing to $\sim60\%$ at NO fluxes of $\sim 10^{12} \, \si{molecules\per\centi\metre\squared\per\second}$.
This behaviour is similar to the other anoxic-abiotic scenarios.
Even if all of this deposited nitrogen were to be converted to \ce{N2O} and outgassed into the atmosphere, a lightning flash rate of more than 100 times modern Earth's would be necessary to produce an \ce{N2O} flux of $\SI{10}{\tera\mol\per\year}$ which would still likely be not enough to be detectable with {JWST} \citep{schwieterman_evaluating_2022}.
On an Earth-like planet orbiting a K dwarf, an \ce{N2O} flux of $\SI{1}{\tera\mol\per\year}$ or even less might be detectable, though, potentially producing a false-positive biosignature.
These calculations present an upper limit, assuming the whole surface is covered by an ocean, as is suggested for the exoplanet Kepler-138~d \citep{piaulet_evidence_2023}, but on a planet with a lower ocean-to-land surface ratio, deposition of nitrogen oxides into the ocean will be less efficient.
This and other uncertainties that will likely decrease the \ce{N2O} flux significantly, such as the efficiency of the conversion of deposited nitrogen into \ce{N2O}, are the reasons why we are not further investigating this potential signature.

\subsection{Observational Signatures from Lightning in Transmission, Emission, and Reflected Light}
\label{Sec_Results_Spectra}

\begin{figure*}
     \centering
     \begin{subfigure}[b]{\columnwidth}
         \centering
         \includegraphics[width=\textwidth]{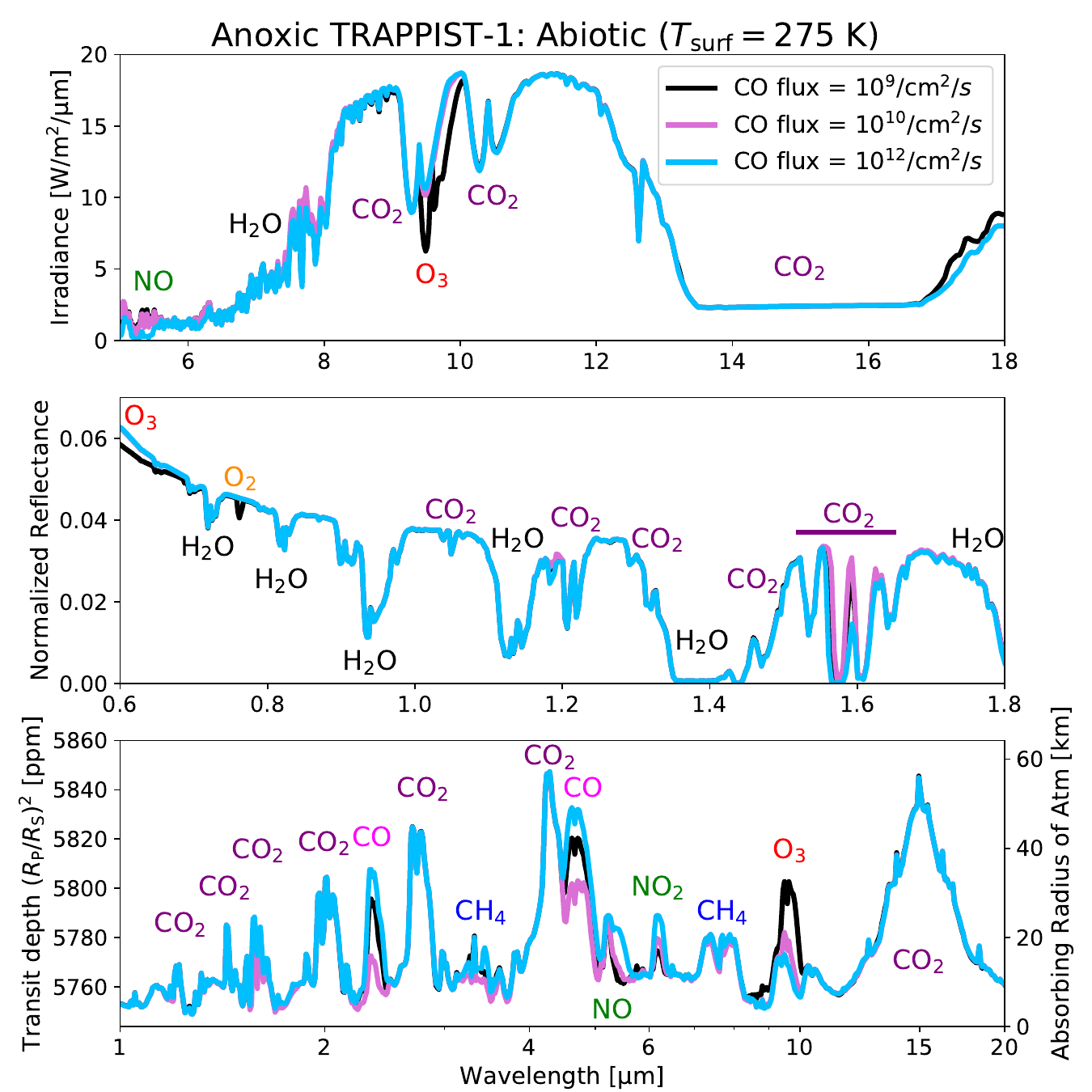}
         \caption{Anoxic, abiotic scenario ($T_\mathrm{surf} = \SI{275}{\kelvin}$): CO flux of $10^9$ (\textit{black}), $10^{10}$ (\textit{magenta}), and $10^{12} \, \si{molecules\per\centi\metre\squared\per\second}$ (\textit{blue}), corresponding to lightning flash rates of $\sim 2$, $\sim 35$ and $\sim 1500$ times modern Earth's, respectively.}
         \label{Fig_Spectra_TR1_Anoxic_275_abiotic}
     \end{subfigure}
     \hfill
     \begin{subfigure}[b]{\columnwidth}
         \centering
         \includegraphics[width=\textwidth]{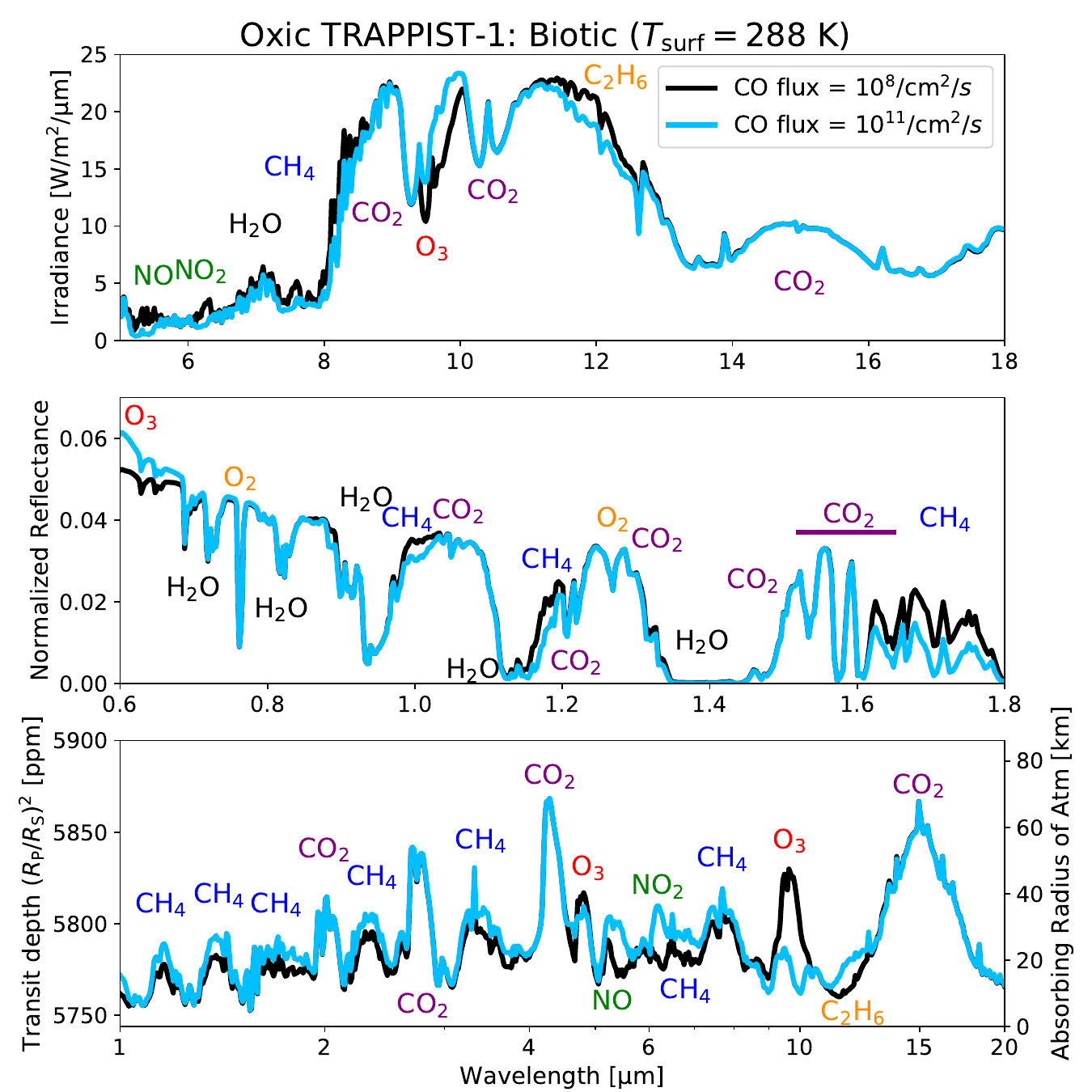}
         \caption{Oxic, biotic scenario ($T_\mathrm{surf} = \SI{288}{\kelvin}$): CO flux of $10^8$ (\textit{black}) and $10^{11} \, \si{molecules\per\centi\metre\squared\per\second}$ (\textit{blue}), corresponding to lightning flash rates of $\sim 0.9$ and $\sim 680$ times modern Earth's, respectively.}
         \label{Fig_Spectra_TR1_Oxic}
     \end{subfigure}
     \caption{Simulated spectra for the TRAPPIST-1 planet: Emitted (\textit{top}, MIR, $R = 400$), reflected (\textit{middle}, NIR, $R = 400$), and transmitted light (\textit{bottom}, NIR-MIR, $R = 200$). Note that MIR features in emission are not just dependent on the abundance of spectrally active gases, but also the temperature structure of the atmosphere.}
     \label{Fig_Spectra_TR1}
\end{figure*}

\begin{figure}
    \centering
    \includegraphics[width=\columnwidth]{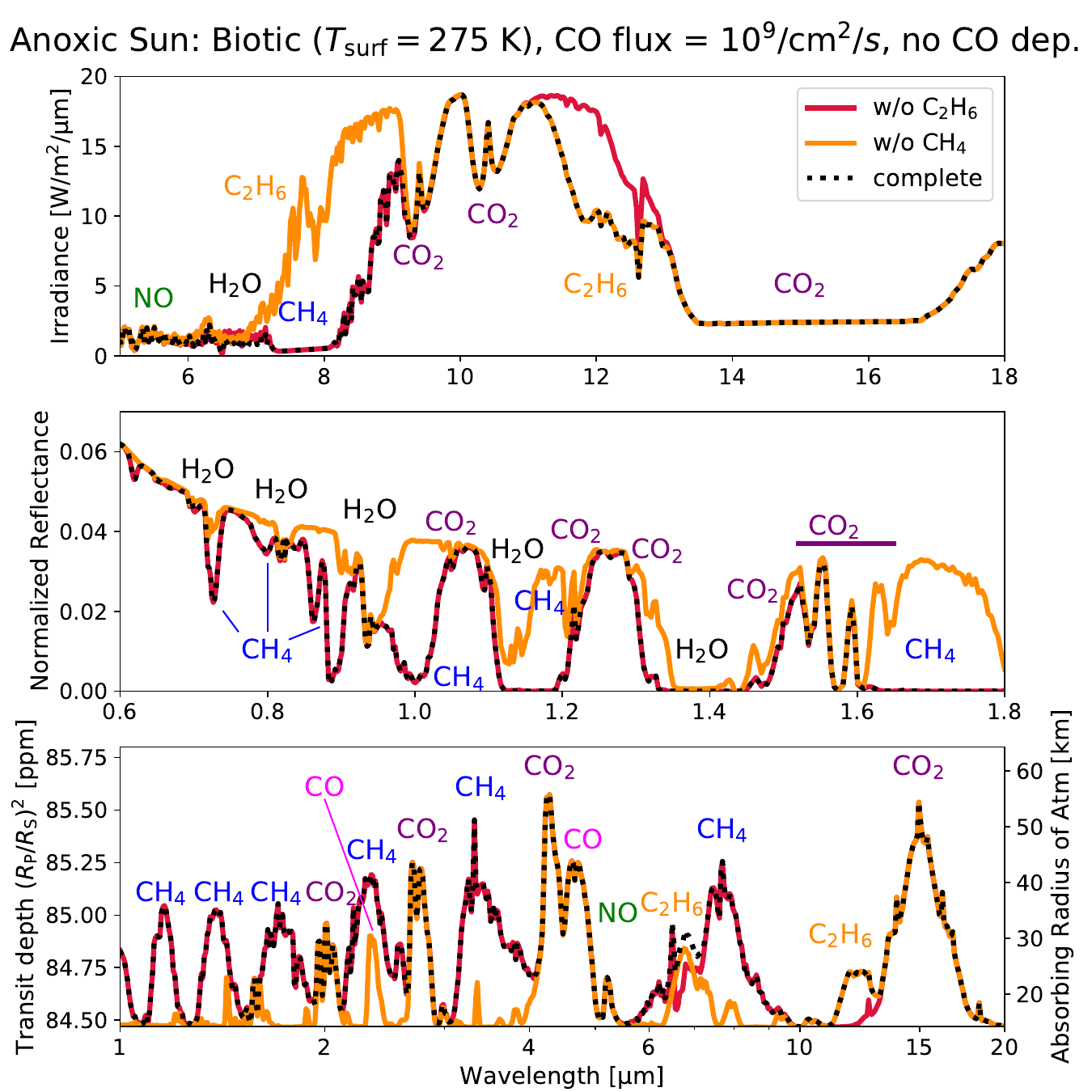}
    \caption{Simulated spectra for the planet orbiting the Sun: Anoxic, biotic scenario without CO deposition (CO flux $\sim 10^9 \, \si{molecules\per\centi\metre\squared\per\second}$). Spectra with individual species removed are shown (without \ce{C2H6}, \textit{red}, and without \ce{CH4}, \textit{orange}) in comparison to the full spectrum (\textit{black dotted}).}
    \label{Fig_Sepc_decomp}
\end{figure}

To investigate whether lightning produces observable signatures, we calculated emission, transmission, and reflectance spectra for a selection of simulations.
The most likely missions and observatories to directly image temperate terrestrial planets will either do so in the Vis/Near-IR range, such as NASA’s proposed Habitable Worlds Observatory or the Mid-IR, such as the proposed ESA Large Interferometer for Exoplanets (LIFE) mission. 
We have thus chosen the wavelength ranges and spectral resolving powers to accentuate features that would be plausibly detectable with these platforms.

Figure~\ref{Fig_Spectra_TR1} shows the calculated transmission spectrum for two scenarios of the TRAPPIST-1 planets: The anoxic, abiotic scenario ($T_\mathrm{surf} = \SI{275}{\kelvin}$; Fig.~\ref{Fig_Spectra_TR1_Anoxic_275_abiotic}) and the oxic, biotic scenario ($T_\mathrm{surf} = \SI{288}{\kelvin}$; Fig.~\ref{Fig_Spectra_TR1_Oxic}).
The different shapes of the \ce{CO2} feature around $\SI{15}{\micro\metre}$ in the emission spectra are due to the different atmospheric profiles we used: 
for the oxic atmosphere which is very similar to modern Earth's atmosphere, we used Earth's atmospheric profile with its stratospheric temperature inversion.
Since the emission is probing different heights in the atmosphere it registers different temperatures, with the centre of the line at $\SI{15}{\micro\metre}$ probing furthest up in the atmosphere, where the temperature is higher than deeper in the atmosphere.
For the anoxic atmosphere, we assumed an isothermal atmosphere above the convective troposphere, such that the absorption feature appears flat.
The spectra are shown for two different CO fluxes, representing scenarios at lightning flash rates below and above where the atmospheric composition changes drastically:
For the anoxic/abiotic scenario, these are CO fluxes of $10^9 \, \si{molecules\per\centi\metre\squared\per\second}$ (\textit{black}), $2 \times 10^{10} \, \si{molecules\per\centi\metre\squared\per\second}$ (\textit{magenta}), and $10^{12} \, \si{molecules\per\centi\metre\squared\per\second}$ (\textit{blue}), corresponding to lightning flash rates of $\sim 2$, $\sim 35$, and $\sim 1500$ times that of modern Earth, respectively.
For the oxic/biotic scenario, CO fluxes of $10^8 \, \si{molecules\per\centi\metre\squared\per\second}$ (\textit{black}) and $10^{11} \, \si{molecules\per\centi\metre\squared\per\second}$ (\textit{blue}), corresponding to lightning flash rates of $\sim 0.9$ and $\sim 680$ times modern Earth's, respectively, are shown.
Most notable is how the increased concentration of \ce{NO} and \ce{NO2} is reducing the ozone features at $\SI{9.6}{\micro\metre}$ (transmission and emission) and $<\SI{0.65}{\micro\metre}$ (reflectance) in the scenario with the higher CO flux. 
The individual mixing ratio profiles for the two cases compared in Fig.~\ref{Fig_Spectra_TR1} can be found in the appendix (Fig.~\ref{Fig_A_PMR_Spectra_TR1_comp}).

In the case of the anoxic planet, the removal of the ozone feature would prevent a possible false-positive detection of life.
In contrast to the oxic scenario, the CO concentration is much higher here (for the low and high lightning flash rates), though, allowing for strong CO features in the transmission spectrum at $\SI{2.35}{\micro\metre}$ and $\SI{4.65}{\micro\metre}$.
This allows us to identify a \ce{CO2}-rich atmosphere and for a correct interpretation of the ozone feature as potentially from an abiotic source.
For the intermediate lightning flash rate ($\sim 35$ times modern Earth's) we find both the CO and the ozone features to be removed or very weak. 
However, in the absence of strong methane features, the missing CO feature should not be interpreted as a sign of life.
\PB{However, if in this scenario, an increased CH4 feature from volcanism is observed it could be misinterpreted as a biosignature.}
In the corresponding biotic scenario (Fig.~\ref{Fig_A_Anoxic_biotic_TR1_wdep_vs_nodep}), where the CO concentration is similarly low, the biotic methane flux produces strong features that should allow the detection of the present biosphere.

In the oxic/biotic scenario, lightning is decreasing the ozone concentration in the atmosphere, reducing the spectral ozone feature used to identify an oxygen-rich atmosphere.
In that case, other biosignatures are necessary to identify that the planet is inhabited, for example, the combination of \ce{CO2} and \ce{CH4} in the absence of CO, as suggested by \citet{Krissansen-Totton2018}.
We find an increased concentration of \ce{CH4} for the higher CO flux, most notably in the reflected spectrum ($1.6 - \SI{1.8}{\micro\metre}$), likely because the increased CO flux is taking up most of the available sinks (e.g., OH).
We also find an \ce{NO2} feature at $\sim \SI{6}{\micro\metre}$ and a \ce{C2H6} feature at $\sim \SI{11}{\micro\metre}$ (both in transmission and emission spectra) to appear for the higher CO flux scenario, indicating the presence of lightning activity and helping to explain the absence of an ozone feature.
This \ce{C2H6} feature is enhanced because other gases are soaking up the radical sinks that would otherwise destroy \ce{C2H6}.
In the reflected light, we find molecular oxygen features at $\SI{0.76}{\micro\metre}$ (\ce{O2} A-band) and $\SI{1.27}{\micro\metre}$ which are not affected by the lightning activity, potentially allowing for the identification of an oxygen-rich atmosphere if that wavelength range is observed.
In any way, to enhance the CO concentration to a detectable level in the oxic scenario, significantly higher lightning flash rates are necessary.

Spectra for other scenarios and the corresponding atmospheric mixing ratio profiles are shown in Appendix~\ref{Sec_Appendix_Spectra}.
For the oxic, biotic Sun scenario (Fig.~\ref{Fig_A_oxic_biotic_Sun}), the behaviour is very similar to the TRAPPIST-1 planet: 
with higher CO and NO fluxes, the ozone features are reduced but we find enhanced methane features.
The spectra for the anoxic, abiotic Sun planet (Fig.~\ref{Fig_A_Anoxic_abiotic_Sun}), however, look very different compared to the TRAPPIST-1 planet: 
because of the lower XUV radiation of the Sun, there is no abiotic ozone build-up in the atmosphere. 
Also, the CO concentration is much lower and the spectra mainly show \ce{CO2} features. 
Only the NO feature at $\SI{5.3}{\micro\metre}$ is stronger when the CO flux is increased.

For the anoxic, biotic scenarios, we are comparing the spectra for simulations with and without CO deposition (Fig.~\ref{Fig_A_Anoxic_biotic_Sun_wdep_vs_nodep} \& \ref{Fig_A_Anoxic_biotic_TR1_wdep_vs_nodep}):
as expected, the main difference is that without CO deposition, we find a strong CO feature at $\SI{4.65}{\micro\metre}$.
For the anoxic, biotic planet around the Sun without CO deposition, we calculated the spectra with individual species removed to show their individual impacts on the planetary spectrum (Fig.~\ref{Fig_Sepc_decomp}).
In particular, we looked at the effect of removing methane (orange lines) and \ce{C2H6} (red).
The complete spectra are shown in comparison (black dotted lines).
As expected, we see many strong methane features in all three spectra, masking for example the $\SI{2.35}{\micro\metre}$ CO feature.
For \ce{C2H6}, we find a strong feature between 11 and $\SI{13}{\micro\metre}$ in the emitted (top) and transmitted spectra (bottom).
Removing \ce{NO2} and \ce{HNO3} from the calculations did not change the spectra, suggesting features of these molecules are small and masked by other, more abundant molecules such as \ce{CH4}.

Whether these spectral features are eventually detectable or not depends on many factors such as the wavelength range and resolving power of the instrument, the duration of the observation, or the variability of the stellar radiation.
For example the ozone feature at $\SI{9.6}{\micro\metre}$ might not be detectable with {JWST} \citep{krissansen-totton_detectability_2018}.
Future missions, dedicated to studying the atmospheres of terrestrial exoplanets will be better suited to detect atmospheric constituents of small planets, such as the {Habitable Worlds Observatory (HWO)} recommended by the US 2020 Astronomy \& Astrophysics Decadal Survey \citep{national_academies_of_sciences_engineering_pathways_2021} with capabilities informed by the {Habitable Exoplanet Observatory} \citep[HabEx,][]{gaudi_habitable_2020} and {Large UltraViolet Optical and Infrared} \citep[LUVOIR,][]{the_luvoir_team_luvoir_2019} mission concepts. 
This mission would be able to detect \ce{O3} and \ce{O2} features in the UV and visible and \ce{CH4} in the near-IR.
The proposed {Large Interferometer For Exoplanets} \citep[LIFE,][]{alei_large_2022,quanz_large_2022} would be able to detect \ce{O3} and \ce{CH4} features in the MIR.

\section{Discussion}
\label{Sec_Discussion}

When conducting our experiments and photochemical simulations, we encountered several unknowns and had to make assumptions that allowed us to simulate the atmospheric chemistry of terrestrial exoplanets. 
In this section, we want to discuss three of these uncertainties in more detail: the lightning flash rate on terrestrial exoplanets, the composition and stability of the atmosphere of Archean Earth and similar exoplanets, and the CO metabolism.

\subsection{Lightning Flash Rate}

In this work, we estimated the production rates of several gaseous and aqueous species by lightning in different atmospheric and planetary environments.
To do so, at several points in the paper, we used the global lightning flash rate on modern Earth of $44\pm \SI{5}{\per\second}$ \citep{Christian2003} with an energy of $\SI{6.7}{\giga\joule}$ per flash \citep{Price1997} to extrapolate from our laboratory experiments to an annual, global production.
Lightning rates and energies have been well studied on the modern Earth \citep[][and references therein]{Schumann2007,Hodosan2016}.
However, as discussed in the introduction, much uncertainty remains for the lightning flash rate on the early Earth, terrestrial planets, and exoplanets in general \citep{Wong2017,Hodosan2021,braam_lightning-induced_2022}.

We therefore treat the lightning flash rate as an independent variable in our photochemical simulations (while always using an energy of $\SI{6.7}{\giga\joule}$ per flash), simulating the atmospheric composition for a range of flash rates from less than 10\% to more than 1000 times modern Earth's flash rate ($44\pm \SI{5}{\per\second}$).
This allowed us to analyse more plausible scenarios of flash rates similar to modern Earth's but also to predict minimum flash rates for certain scenarios.
In some scenarios, small changes in the flash rate have very little effect on the composition of the atmosphere and thus the observability of certain features, such as the biotic scenarios for the anoxic planets around TRAPPIST-1 and the Sun.

In other scenarios, small changes in the flash rate, close to that of modern Earth, can have a significant impact on the planetary spectrum.
For example, our simulations of the anoxic, abiotic atmosphere of a planet orbiting TRAPPIST-1 show that lightning at modern Earth's flash rate is not able to remove the abiotic ozone feature caused by the dissociation of \ce{CO2} due to the star's strong XUV radiation.
However, at a flash rate ten times higher, the additional NO input and subsequent oxidation to \ce{NO2} enhances the recombination of CO and oxygen to \ce{CO2}, removing the abiotic ozone feature and preventing a false-positive biosignature detection.
This shows how sensitive the atmospheric composition and planetary spectrum can be to the lightning flash rate.

In addition, the lightning flash rate at which this threshold appears depends on the individual XUV spectrum of the planet's host star which is responsible for the \ce{CO2} dissociation and \ce{O2} production:
Larger stellar FUV/NUV flux ratios tend to drive higher abiotic \ce{O2} production rates \citep{harman_abiotic_2015} and TRAPPIST-1 (M8V) has a larger FUV/NUV ratio than the latest host star considered by \citet{harman_abiotic_2018}, an M4V dwarf.
Therefore, \citet{harman_abiotic_2018} find a flash rate equal to modern Earth's sufficient to remove the abiotic ozone feature in their simulations.

An additional factor that can influence the occurrence of lightning is cosmic rays as planets orbiting M dwarfs will experience more frequent and intense stellar flares that are associated with flares of charged particles.
In addition to their potential effect on the lightning flash rate, cosmic rays are also an important source of disequilibrium chemistry in the atmospheres of exoplanets.
Studies of Earth-like planets orbiting M dwarfs have shown that the enhanced cosmic ray flux leads to increased production of nitrogen oxides (NO, \ce{NO2}, \ce{HNO3}) as well as a decreased concentration of ozone, similar to our results for lightning \citep{grenfell_response_2012,tabataba-vakili_atmospheric_2016,Scheucher2020}.

\PB{Another uncertainty that is connected to the lightning flash rate is the effect of the flash properties.
The freeze-out temperature of a specific gas, and thus its concentration in the air parcel cycled through the flash, depends on the cooling timescale of the flash.
Therefore, the estimate of the freeze-out temperature depends on the properties of the flash considered in the experiment or calculations.
In addition, the chemical timescales and thus the freeze-out temperature depend on the specific location in the discharge where the trace gas is produced: 
Trace gases produced in the expanding shock wave of a lightning strike where chemical timescales are shorter will freeze out at a higher temperature \citep{chameides_implications_1979} than trace gases produced in the hot channel of the strike which is cooling more slowly \citep{Hill1980}.
It has since been suggested that the latter process is more important \citep{Uman2003a}, in particular in spark experiments \citep{Stark1996}, but a smaller contribution from the shock-wave could lead to an estimate of the freeze-out temperature somewhere between both extremes. 
The combination of these uncertainties explains the wide range of reported freeze-out temperatures for both NO \citep[$2300 - \SI{3500}{\kelvin}$;][]{gilmore_production_1975,Chameides1977,Kasting1981,Picone1981,borucki_lightning_1984} and CO \citep[$2000 - \SI{3500}{\kelvin}$;][]{Levine1979,chameides_implications_1979}, placing our estimates of $T_\mathrm{f} (\mathrm{NO}) \gtrsim \SI{3000}{\kelvin}$ and $T_\mathrm{f} (\mathrm{CO}) \simeq \SI{2430}{\kelvin}$ within the range of these estimates.}

\subsection{Atmospheric Stability}

In our photochemical simulations, we studied the kinetic chemistry and photochemistry of terrestrial, habitable-zone planets around two different types of stars: the Sun and the M dwarf TRAPPIST-1.
While we assumed a \ce{CO2} concentration of 4.6\% in the atmosphere (for both the anoxic and oxic atmospheres), corresponding to our experiments and necessary to keep the surface of the planets clement \citep{meadows_habitability_2018}, the \ce{CO2} partial pressure on the early Earth was likely higher \citep{lichtenegger_aeronomical_2010,johnstone_young_2021}.
Further, for planets around M dwarfs such as TRAPPIST-1, it will be much harder to hold on to their atmosphere because of the intense stellar wind and the enhanced and prolonged XUV activity of the host star \citep{lammer_loss_2011,airapetian_how_2017,johnstone_active_2021}.
Recent observations with JWST have shown that the innermost planet of the TRAPPIST-1 system, TRAPPIST-1~b, is likely a bare rock without any atmosphere \citep{greene_thermal_2023}.
Even if TRAPPIST-1~e has retained an atmosphere, the ongoing escape of hydrogen and nitrogen will continue to modify the atmosphere, potentially enhancing the oxygen fraction in the remaining atmosphere \citep{tian_thermal_2009,johnstone_extreme_2019}.
Connecting the upper atmosphere and the above-discussed escape processes to our photochemical model is beyond the scope of this study but should be investigated further in future studies. 

\PB{To obtain our results, we used the TRAPPIST-1 spectrum from \citet{peacock_predicting_2019}, though recent work has shown that photochemical results can differ when other versions of the TRAPPIST-1 spectrum are used \citep{cooke_degenerate_2023}.}
The here presented simulations are therefore not specific predictions for the atmospheric composition of TRAPPIST-1~e but rather an example to investigate the effect of lightning on different atmospheres around different types of host stars.
TRAPPIST-1 and the Sun are good end-members for several reasons, including the fact that TRAPPIST-1 is an immediate/near-term target, and terrestrial planets orbiting Sun-like stars are the targets for HWO.
In case TRAPPIST-1~e, or similar exoplanets in the habitable zones of M dwarfs, can indeed retain an atmosphere that is similar to our simulations, this study informs us about atmospheric signatures that one may expect to observe.

\subsection{Adapting CO Metabolisms}

In this study, we used biotic CO deposition and \ce{CH4} production rates that were independent of temperature, following previous work \citep{kharecha_coupled_2005,schwieterman_rethinking_2019}.
However, the metabolisms responsible for \ce{CH4} production and CO consumption might change depending on atmospheric composition, pressure, and temperature, and will likely vary substantially across a heterogeneous planetary surface. 
\citet{taubner_lipidomics_2023} show how the production of lipids and amino acids by methanogens can strongly depend on the temperature and nutrient supply.
In addition, the productivity of a biosphere might be limited by the availability of other nutrients like phosphorus or fixed nitrogen, rather than CO or stellar irradiation, as is the case for modern Earth \citep{moore_processes_2013,bristow_nutrients_2017}.
In particular, the availability of nickel can limit the CO consumption by acetogens \citep{dobbek_crystal_2001}.
Throughout Earth's history, the concentration of nickel in the oceans has varied, impacting the productivity of methanogens and acetogens \citep{konhauser_oceanic_2009,konhauser_archean_2015}.
These uncertainties suggest that in realistic atmospheres the biotic CO deposition rate might be lower than assumed in this study.
The biotic and abiotic deposition rates used here thus provide an upper and lower bound, respectively, on the efficiency of CO deposition.

\section{Conclusions}
\label{Sec_Conclusion}

We conducted spark discharge experiments to study the production of different gaseous and aqueous products, including potential (anti-)biosignatures, by lightning in atmospheres of \ce{N2}, \ce{CO2}, and \ce{H2}.
In contrast to previous studies that focused on individual or small numbers of products, we studied the production of a wide range of gaseous and aqueous compounds in a range of different atmospheres.
This allowed us to investigate trends in our experiments concerning the oxidation state of lightning products and the influence of water vapour.
In particular, we were interested in the effect of lightning on the production of potential (anti-)biosignatures in the context of current and upcoming observations of exoplanetary atmospheres.

Our results confirm that in a slightly reducing or oxidising atmosphere of a planet that has surface water and a habitable surface temperature, lightning will produce more oxidised than reduced nitrogen products.
We confirm predictions by \citet{harman_abiotic_2018} that in the absence of other forms of oxygen, water vapour is responsible for a baseline production of NO and other oxidised forms of nitrogen by lightning. 
In return, this allows us to predict that for the kind of atmospheres studied here, lightning is not able to produce a false-positive \ce{NH3} or \ce{CH4} biosignature.
It is also unlikely that lightning can produce a false-positive \ce{N2O} biosignature.

We then used the CO and NO production rates determined in our experiments to calculate the atmospheric composition over a range of different lightning flash rates with a photochemical model.
We applied this model to anoxic (\ce{CO2 - N2}) and oxic (\ce{O2 - CO2 - N2}) atmospheres on Earth-sized planets in the habitable zone of the Sun and TRAPPIST-1.
In particular, we conducted simulations with and without an assumed biosphere on the planet.
We also calculated simulated spectra to identify signatures for the different scenarios.

We find that lightning is not able to produce a false-positive CO anti-biosignature on an inhabited planet.
In an oxygen-rich atmosphere, however, lightning rates only a few times higher than modern Earth's can mask the \ce{O3} biosignature. 
Enhanced NO, \ce{NO2}, and \ce{C2H6} features might help to identify these oxygen-rich atmospheres with increased lightning activity.

Similarly, in an anoxic, abiotic atmosphere of a planet orbiting a late M dwarf, lightning at flash rates ten times or more than that of modern Earth can remove the abiotic ozone feature produced by \ce{CO2} photolysis, preventing a false-positive biosignature detection.
However, this also suggests that lightning might not be able to prevent all false-positive \ce{O2} scenarios for \ce{CO2}-rich terrestrial planets orbiting ultracool M dwarfs.
In summary, our work provides new constraints for the full characterisation of atmospheric and surface processes on exoplanets.

\begin{acknowledgements}
    We thank Terry Smith, Nathan Rochelle-Bates, and Abu S. Bayida for support in analysing our data, and Helmut Lammer and Manuel Scherf for comments on the manuscript.
    P.B. acknowledges a St Leonard’s Interdisciplinary Doctoral Scholarship from the University of St Andrews. 
    E.E.S. acknowledges funding from a Royal Society research grant (RGS\textbackslash R1\textbackslash 211184) and from a NERC Frontiers grant (NE/V010824/1). 
    Ch.H. is part of the CHAMELEON MC ITN EJD which received funding from the European Union’s Horizon 2020 research and innovation programme under the Marie Sklodowska-Curie grant agreement number 860470. 
    E.W.S. acknowledges support from the NASA Interdisciplinary Consortia for Astrobiology Research (ICAR) program via the Alternative Earths team with funding issued under grant No. 80NSSC21K0594 and the CHAMPs (Consortium on Habitability and Atmospheres of M-dwarf Planets) team with funding issued under grant No. 80NSSC21K0905. 
    E.W.S. additionally acknowledges support from the Virtual Planetary Laboratory, which is a member of the NASA Nexus for Exoplanet System Science and funded via NASA Astrobiology Program grant No. 80NSSC18K0829.
    In order to meet institutional and research funder open access requirements, any accepted manuscript arising shall be open access under a Creative Commons Attribution (CC BY) reuse licence with zero embargo.
    Machine-readable tables of the data presented in this work are available online at \url{https://doi.org/10.17630/8b72510f-62a8-43dc-94f1-af9b7766f817}.
\end{acknowledgements}

\begin{appendix}

\setcounter{table}{0}
\renewcommand{\thetable}{A\arabic{table}}

\setcounter{figure}{0}
\renewcommand{\thefigure}{A\arabic{figure}}

\section{Analytical Methods}
\label{Sec_Appendix_Methods}

\subsection{NO Gas Abundance Measurements}
\label{Sec_methods_gas}

The gas in the flask was analysed with a quadrupole mass spectrometer (Hiden Analytical ExQ Quantitative Gas Analyser) in `multiple ions detection' mode to monitor the abundance of several mass/charge ratios (m/z) (12, 14, 15, 16, 17, 18, 20, 28, 30, 32, 40, 46, 48) and thus detect \ce{N2}, \ce{O2}, \ce{CO2}, \ce{NO}, \ce{H2O}, and \ce{O3}, as described in \citet{barth_isotopic_2023}.
Due to mass interferences, we were not able to measure CO at m/z 30 (same as \ce{NO}), \ce{N2O} at m/z 30 (same as NO) and m/z 44 (same as \ce{CO2}), and \ce{NO2} at m/z 30 and 46 (the \ce{CO2} isotopologue \ce{^{16}O^{12}C^{18}O} produces an m/z 46 peak).

\subsection{CH\textsubscript{4} and N\textsubscript{2}O Gas Abundance Measurements}
\label{Sec_methods_ch4n2o}

Concentrations of methane (\ce{CH4}) and nitrous oxide (\ce{N2O}) were measured with a Thermo Fisher gas chromatograph (Trace Ultra) equipped with an electron capture detector (ECD) for \ce{N2O} and a flame-ionisation detector (FID) for \ce{CH4}. 
The packed column was held at $60^\circ$C, the ECD at a base temperature of $180^\circ$C and a central temperature of $330^\circ$C, and the FID at $300^\circ$C throughout the run. 
$\SI{10}{\milli\litre}$ of sample gas were extracted from the experiment with a gas-tight, lockable syringe through the septum, and this volume was injected undiluted into the GC to fill and flush out the $\SI{2}{\milli\litre}$ sample loop. 
Helium was used as a carrier gas at a pressure of $\SI{450}{\kilo\pascal}$. 
Auxiliary gases for the detectors were \ce{N2} ($\SI{10}{\milli\litre\per\minute}$), \ce{H2} ($\SI{35}{\milli\litre\per\minute}$) and zero air ($\SI{350}{\milli\litre\per\minute}$).
Analysis of the measurements is limited by the availability of only one standard gas mixture (\ce{CH4}: $\SI{10}{ppm}$, \ce{N2O}: $\SI{2.5}{ppm}$).
The lowest methane concentration that still produced a recognisable peak was about 0.2~ppm.
However, this limit could not be tested with a low standard.
Repeated measurements of the standard produced a standard deviation of 0.6~ppm, a relative error of 6\%, which we assume for the error of each measurement.

\subsection{CO Gas Abundance Measurements}

A limited number of samples was analysed for their concentration of CO at the School of Natural and Environmental Sciences, Newcastle University, with an SRI 8610C gas chromatograph with a Hg reduction gas detector (RGD).
The following run parameters were used: 
$\SI{25}{psi}$ \ce{N2} carrier gas at $\SI{20}{\milli\litre\per\minute}$; column oven at $80^\circ$C; $\SI{5}{\minute}$ run time with the RGD at $280^\circ$C and in low sensitivity setting.
All samples were diluted by a factor of 100 by adding $\SI{120}{\micro\litre}$ of the sample to an exetainer vial filled with $\SI{12}{\milli\litre}$ of \ce{N2}. 
$\SI{3}{\milli\litre}$ of \ce{N2} were then injected into the diluted samples in the exetainer, the gas was mixed using a $\SI{5}{\milli\litre}$ gas-tight syringe, and then $\SI{3}{\milli\litre}$ of the gas were injected into a $\SI{0.5}{\milli\litre}$ sample loop on the GC. 
The sample loop then injected $\SI{0.5}{\milli\litre}$ of the gas into the column (molecular sieve 5A packed 60/80 mesh, 6 ft length).
The centre of the CO peak leaves the column after approximately 3.0 to 3.1 minutes, depending on the concentration. 
No evidence of any interfering peaks in standards or samples was found.
$\SI{1}{ppm}$ and $\SI{10}{ppm}$ standards were prepared by diluting a certified $\SI{100}{ppm}$ CO in \ce{N2} standard (Calgaz Ltd). 
A four-point calibration (0, 1, 10, $\SI{100}{ppm}$) was performed. 
Multiple $\SI{100}{ppm}$ standard runs were conducted to determine the precision of the measurement to 3.5\%.
The error of both dilution steps is estimated to be 8.3\%, giving a combined error of the CO concentration measurements of 12.3\%. 

\subsection{Aqueous Nitrate and Nitrite Analyses}

To determine the concentrations of nitrate (\ce{NO3^-}) and nitrite (\ce{NO2^-}) in our solutions, we used a Thermo Scientific Dionex ICS-6000 ion chromatograph equipped with a Dionex AS-AP autosampler, a $\SI{25}{\milli\metre}$ Dionex IonPac AS17-C separation column ($\SI{2}{\milli\metre}$ bore), a $\SI{25}{\milli\metre}$ Dionex IonPac AG17-G guard column ($\SI{2}{\milli\metre}$ bore), and a Dionex ADRS-600 $\SI{2}{\milli\metre}$ suppressor. 
The flow rate was held constant at $\SI{0.5}{\milli\litre\per\minute}$ while the concentration of the KOH eluent solution was ramped up from $\SI{1}{\milli\molar}$ to $\SI{40}{\milli\molar}$ over 20~minutes.

\subsection{Ammonium and Ammonia}

We follow a colourimetric method \citep{Solorzano1969, Cleaves2008} to measure the ammonium concentration in our samples.
Three stocks of reagents were prepared in larger quantities and stored for several months:
(1) sodium citrate buffer ($\SI{7.6}{\gram}$ trisodium citrate (\ce{Na3C6H5O7}) and $\SI{0.4}{\gram}$ sodium hydroxide (\ce{NaOH}) in $\SI{500}{\milli\litre}$ of water), (2) phenol alcohol ($\SI{1}{\milli\litre}$ liquefied phenol (\ce{C6H5OH}) in $\SI{90}{\milli\litre}$ of 100\% ethanol (\ce{C2H5OH}), brought up to $\SI{100}{\milli\litre}$ with water), and (3) aqueous sodium nitroprusside ($\SI{0.15}{\gram}$ of sodium nitroprusside (\ce{C5FeN6Na2O}) in $\SI{200}{\milli\litre}$ of water).
Each day, an oxidising solution of $\SI{10}{\milli\litre}$ of the sodium citrate buffer with $\SI{0.1}{\milli\litre}$ of aqueous sodium hypochlorite (\ce{ClNaO}, with 10-15\% available chlorine) was prepared fresh (amount adjusted to number of samples).
For our standards, we used a $\SI{1}{\milli\molar}$ stock of \ce{NH4Cl}, diluted to 1, 2, 5, 10, 20, 50, and $\SI{100}{\micro\molar}$.
For analyses, $\SI{1}{\milli\litre}$ of sample or standard were transferred into a $\SI{15}{\milli\litre}$ Falcon centrifuge tube, followed by $\SI{0.5}{\milli\litre}$ phenol alcohol, $\SI{0.5}{\milli\litre}$ aqueous sodium nitroprusside, and $\SI{1}{\milli\litre}$ of the oxidising solution.
The mixture is incubated for 60-80 minutes at room temperature, allowing it to develop a blue colour.
Absorption was measured at $\SI{640}{\nano\metre}$ with a Thermo Fisher Evolution 220 UV-Vis spectrophotometer.
According to \citet{Solorzano1969}, there is no interference between ammonium and other nitrogen compounds or seawater, which contains nitrate.
From the ammonium concentration in the water, we can estimate the ammonia concentration in the gas phase.
The ratio between the concentration of ammonium in the liquid phase and the pressure of ammonia gas is given by Henry's law constant for \ce{NH3} which is $H = (73 \pm 5) \, \si{\molar\per\bar}$ for temperatures between 20 and $23^\circ$C \citep{edwards_vapor-liquid_1978,burkholder_chemical_2019}.

\subsection{Urea}
For the quantification of urea (\ce{CO(NH2)2}) in solution, we followed the colourimetric method described by \citet{Cleaves2008}.
The following stocks of reagents were prepared:
(1) Acidic ferric chloride (AFC, $\SI{0.208}{\gram}$ \ce{FeCl3}$\bullet$6\ce{H2O} + $\SI{20}{\milli\litre}$ concentrated \ce{H2SO4} + $\SI{2.5}{\milli\litre}$ 85\% \ce{H3PO4} in $\SI{250}{\milli\litre}$ of DI-water),
(2) diacetyl monoxime (DAM, $\SI{2.5}{\gram}$ in $\SI{100}{\milli\litre}$ DI-water),
and (3) thiosemicarbazide (TSC, $\SI{0.25}{\gram}$ in $\SI{100}{\milli\litre}$ DI-water).
Each day, $\SI{12}{\milli\litre}$ of DAM stock were combined with $\SI{5}{\milli\litre}$ of TSC stock and $\SI{33}{\milli\litre}$ DI-water.
Then $\SI{8}{\milli\litre}$ of this combined DAM-TSC stock were added to $\SI{40}{\milli\litre}$ AFC to prepare the color reagent.
$\SI{2}{\milli\litre}$ of color reagent are added to $\SI{200}{\micro\litre}$ of sample in a $\SI{15}{\milli\litre}$ centrifuge tubes.
The closed tubes are then heated in boiling water for 15~minutes and then analysed at $\SI{520}{\nano\metre}$ with the UV-Vis spectrophotometer.
We found the detection limit of this method to be $\SI{1}{\micro\molar}$.
Analyses were calibrated with a $\SI{1}{\milli\molar}$ stock solution of urea, diluted to 1, 2, 5, 10, 20, 50, and $\SI{100}{\micro\molar}$. 

\subsection{Cyanide and HCN}

For the quantification of aqueous cyanide (\ce{CN^-}) we followed the method described by \citet{Cacace2007}, which requires three stock solutions of reagents:
(1) Borax buffer: $\SI{4.767}{\gram}$ of sodium tetraborate decahydrate are dissolved in $\SI{500}{\milli\litre}$ DI-water.
The solution is then brought to a pH of 10.8 by adding approximately $\SI{2}{\milli\litre}$ of $\SI{10}{\molar}$ NaOH.
This buffer can be stored in the fridge for several months.
(2) Combined reagents: $\SI{0.2996}{\gram}$ copper(II) nitrate trihydrate are dissolved in $\SI{20}{\milli\litre}$ DI-water.
Then, $\SI{0.1861}{\gram}$ ethylenediaminetetraacetic acid (EDTA) is added and dissolved (stir).
Finally, $\SI{0.1777}{\gram}$ of phenolphthalin are added and the solution is topped up to $\SI{50}{\milli\litre}$ with DI-water.
(3) Stabilising solution: $\SI{0.324}{\gram}$ triethanolamine hydrochloride and $\SI{1.25}{\gram}$ sodium sulfite are dissolved in $\SI{100}{\milli\litre}$ DI-water.
Both the combined reagents and the stabilising solution should be made fresh every few weeks as the absorption efficiency will deteriorate.
For the analysis, $\SI{0.2}{\milli\litre}$ of sample are added to $\SI{9}{\milli\litre}$ of borax buffer, followed by $\SI{0.2}{\milli\litre}$ of the combined reagents.
After three minutes, $\SI{0.2}{\milli\litre}$ of stabilising solution are added.
The sample was then analysed at $\SI{553}{\nano\metre}$ with the UV-Vis spectrophotometer.
Measurements were calibrated with a $\SI{1000}{ppm}$ stock solution of KCN, diluted to 5, 8, 10, and $\SI{20}{ppm}$. 
The detection limit was $\SI{5}{ppm}$. 
Similar to ammonia, from the cyanide concentration in the water, we can estimate the HCN concentration in the gas phase using the respective Henry's law constant.
For HCN, this constant at 20 to 23$^\circ$C is $H = (13 \pm 4) \, \si{\molar\per\bar}$ \citep{ma_temperature_2010}.

\setcounter{table}{0}
\renewcommand{\thetable}{B\arabic{table}}

\setcounter{figure}{0}
\renewcommand{\thefigure}{B\arabic{figure}}

\section{Additional Results from Photochemical Simulations}
\label{Sec_Appendix_Spectra}

\begin{figure*}
     \centering
     \begin{subfigure}[b]{\columnwidth}
         \centering
         \includegraphics[width=\textwidth]{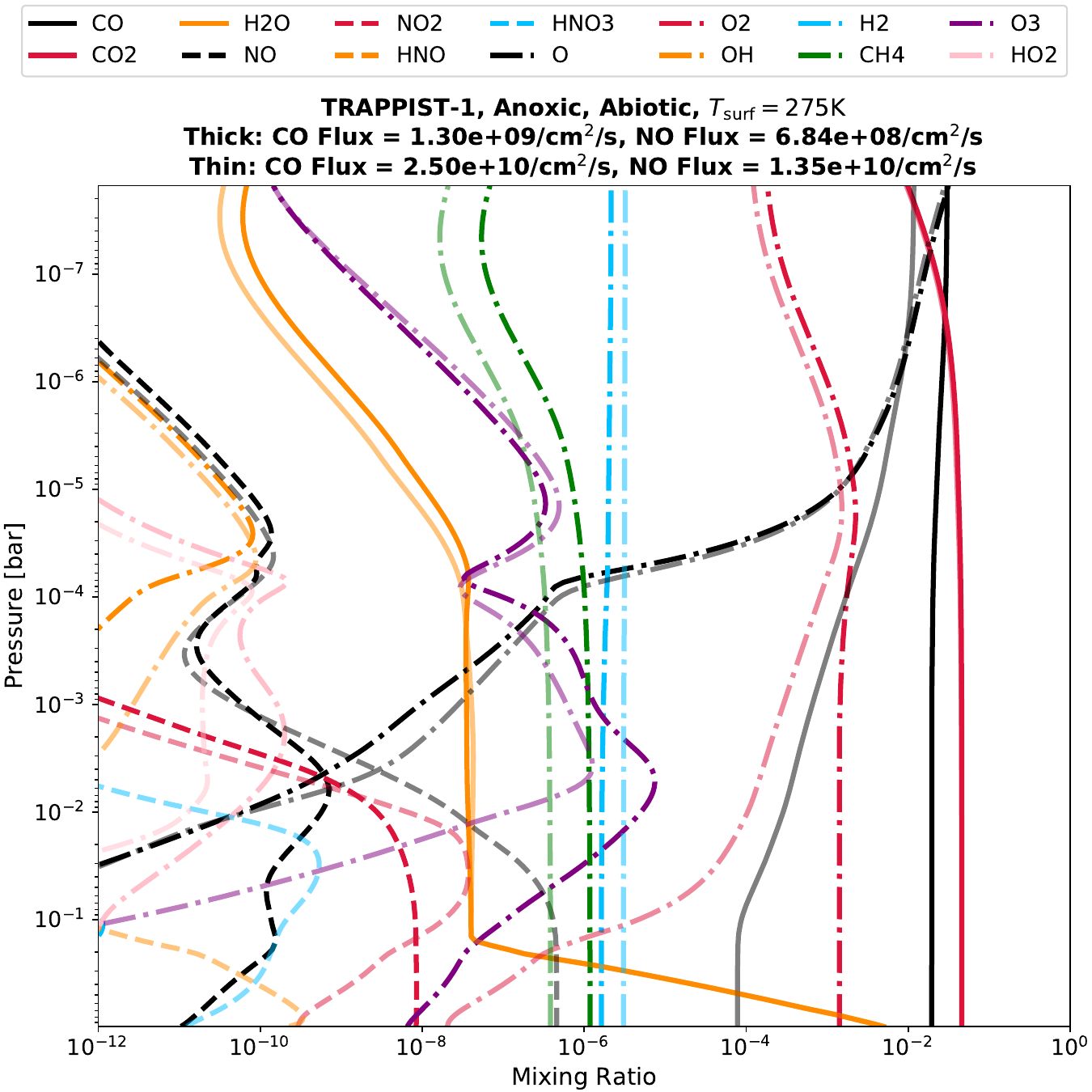}
         \caption{TRAPPIST-1, Anoxic, abiotic scenario ($T_\mathrm{surf} = \SI{275}{\kelvin}$): CO flux of $10^9$ (\textit{thick}) and $10^{11} \, \si{molecules\per\centi\metre\squared\per\second}$ (\textit{thin}), corresponding to lightning flash rates of $\sim 2$ and $\sim 160$ times modern Earth's, respectively.}
         \label{Fig_A_TR1_Anoxic_275_abiotic_PMR_spec}
     \end{subfigure}
     \hfill
     \begin{subfigure}[b]{\columnwidth}
         \centering
         \includegraphics[width=\textwidth]{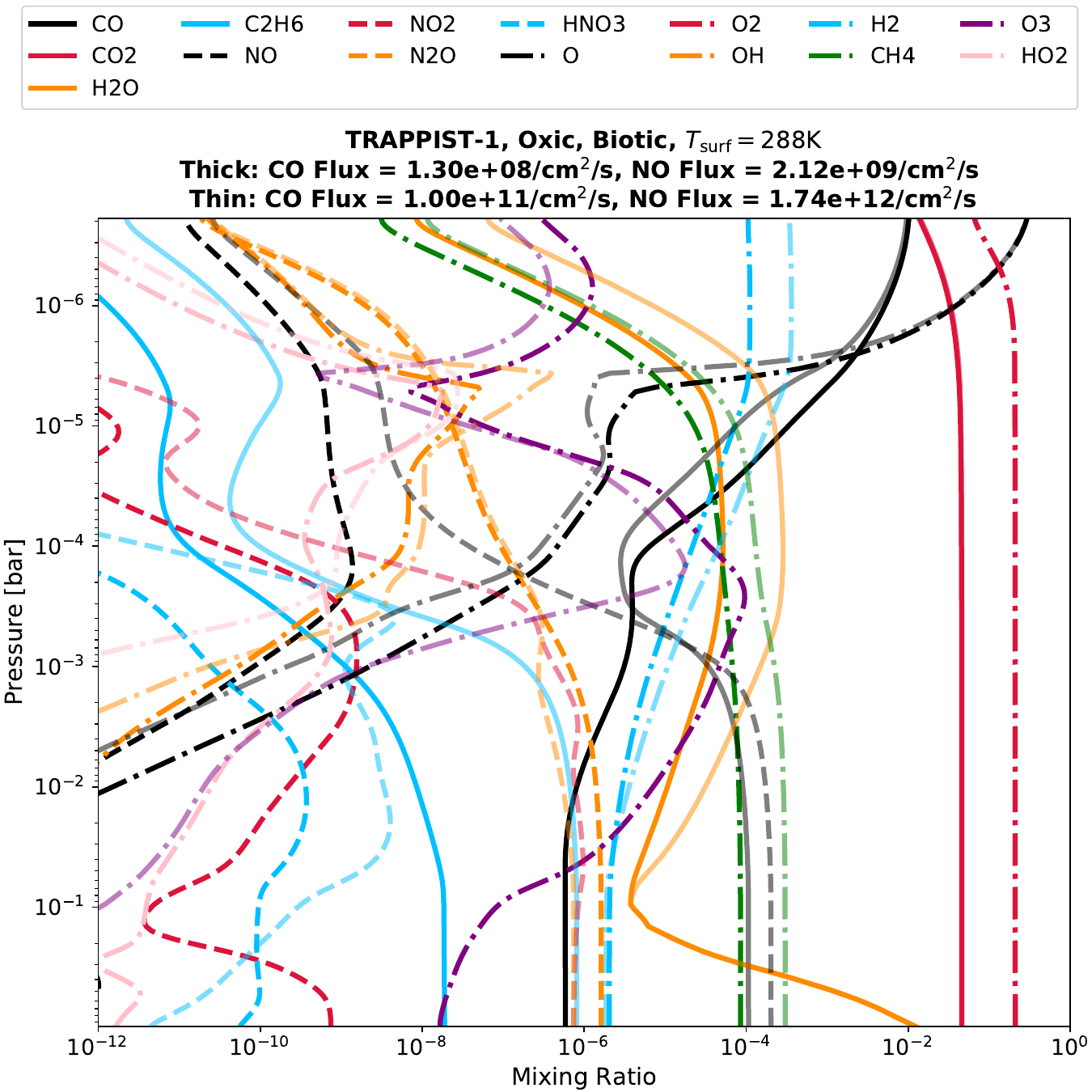}
         \caption{TRAPPIST-1, Oxic, biotic scenario ($T_\mathrm{surf} = \SI{288}{\kelvin}$): CO flux of $10^8$ (\textit{thick}) and $10^{11} \, \si{molecules\per\centi\metre\squared\per\second}$ (\textit{thin}), corresponding to lightning flash rates of $\sim 0.9$ and $\sim 680$ times modern Earth's, respectively.}
         \label{Fig_A_TR1_Oxic_noO_PMR_spec}
     \end{subfigure}
     \caption{Atmospheric mixing ratios of most abundant species for scenarios shown in spectra in Fig.~\ref{Fig_Spectra_TR1}.}
     \label{Fig_A_PMR_Spectra_TR1_comp}
\end{figure*}

\begin{figure*}
     \centering
     \begin{subfigure}[b]{\columnwidth}
         \centering
         \includegraphics[width=\textwidth]{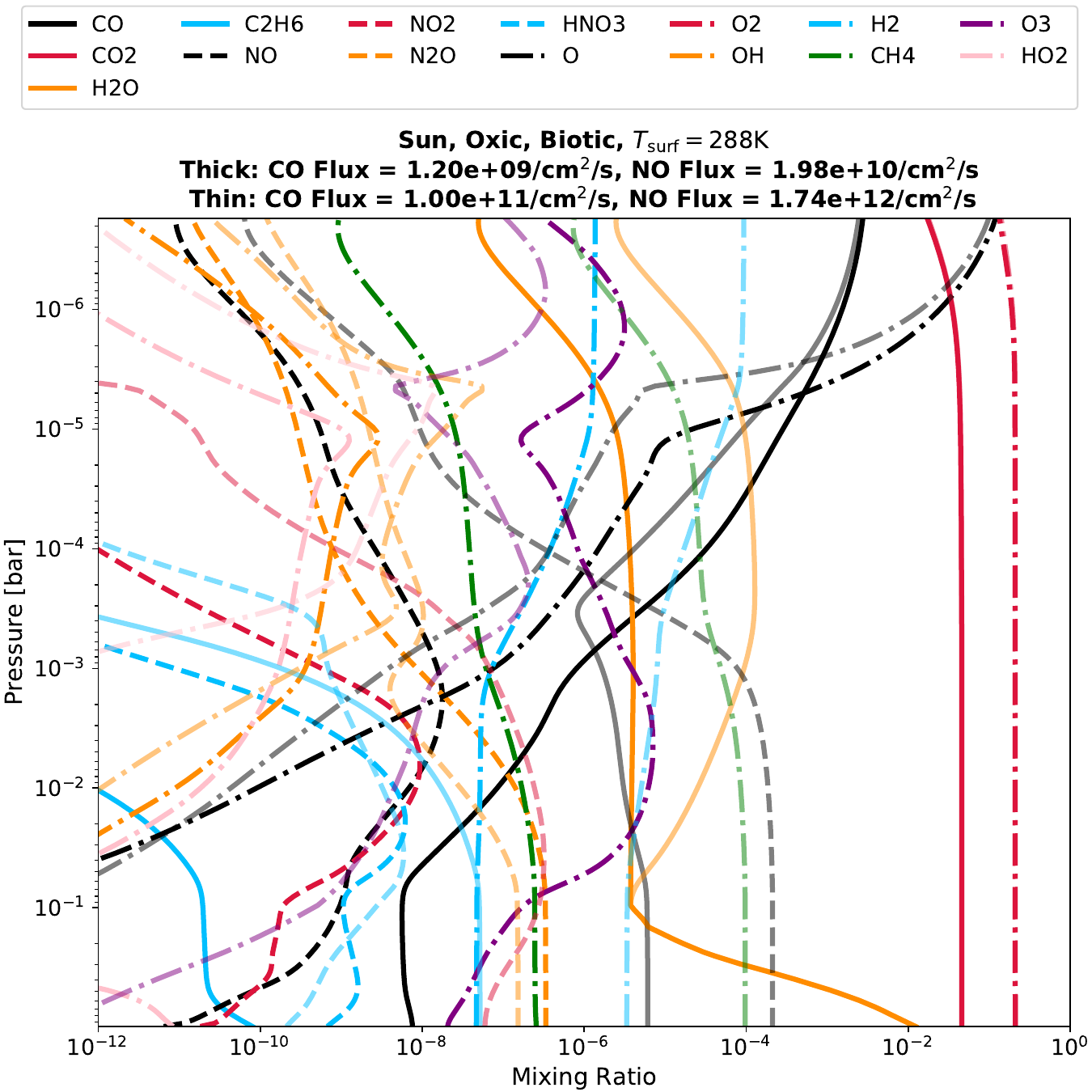}
         \caption{Atmospheric mixing ratios of most abundant species.}
         \label{Fig_A_Sun_Oxic_noO_PMR_spec}
     \end{subfigure}
     \hfill
     \begin{subfigure}[b]{\columnwidth}
         \centering
         \includegraphics[width=\textwidth]{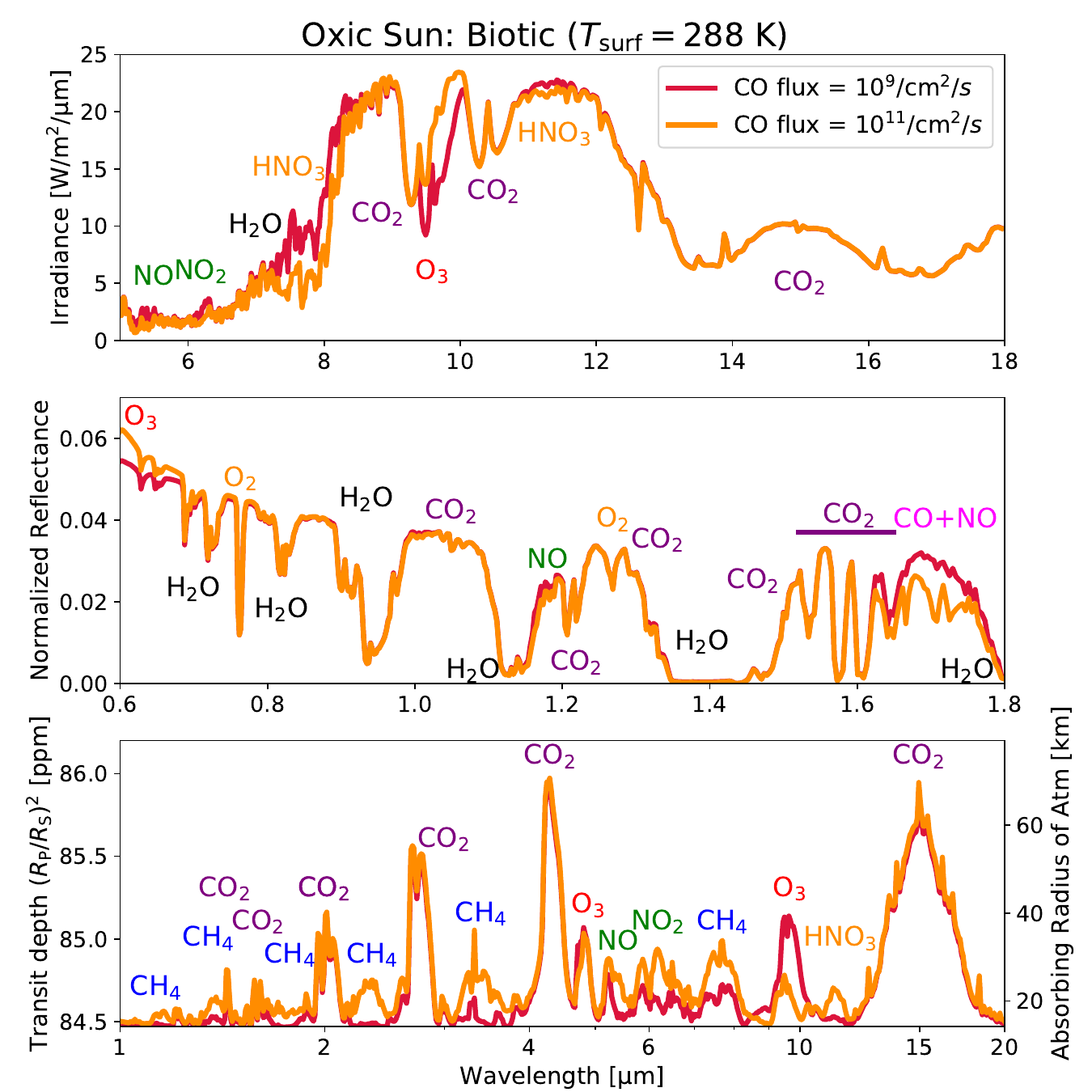}
         \caption{Simulated spectra: Emitted (\textit{top}, MIR, $R = 400$), reflected (\textit{middle}, NIR, $R = 400$), and transmitted light (\textit{bottom}, NIR-MIR, $R = 200$).}
         \label{Fig_A_Spectra_Sun_Oxic}
     \end{subfigure}
     \caption{Oxic, biotic Sun scenario ($T_\mathrm{surf} = \SI{288}{\kelvin}$): CO flux of $10^9$ and $10^{11} \, \si{molecules\per\centi\metre\squared\per\second}$, corresponding to lightning flash rates of $\sim 9$ and $\sim 680$ times modern Earth's, respectively.}
     \label{Fig_A_oxic_biotic_Sun}
\end{figure*}

\begin{figure*}
     \centering
     \begin{subfigure}[b]{\columnwidth}
         \centering
         \includegraphics[width=\textwidth]{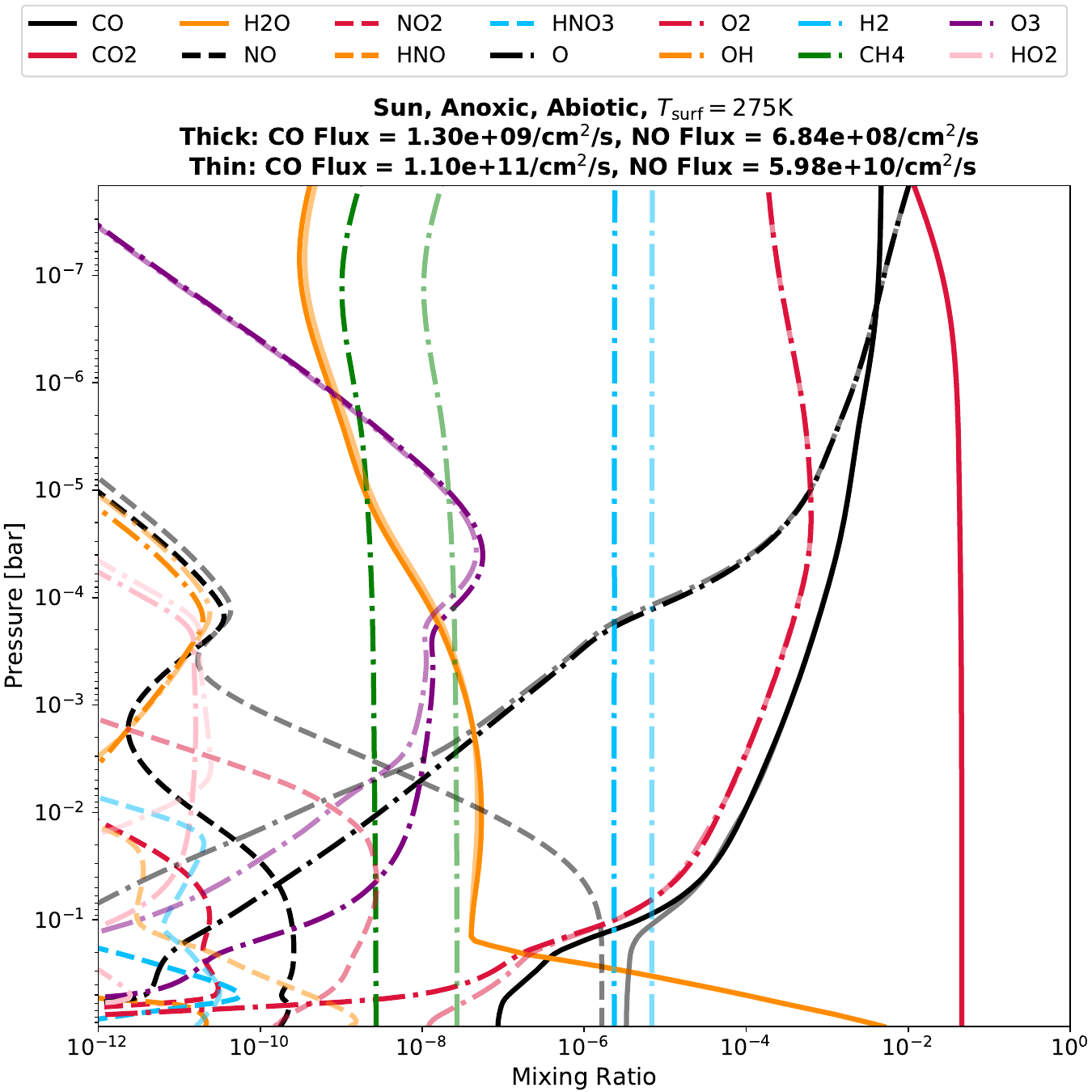}
         \caption{Atmospheric mixing ratios of most abundant species.}
         \label{Fig_A_Sun_Anoxic_275_abiotic_PMR_spec}
     \end{subfigure}
     \hfill
     \begin{subfigure}[b]{\columnwidth}
        \centering
         \includegraphics[width=\textwidth]{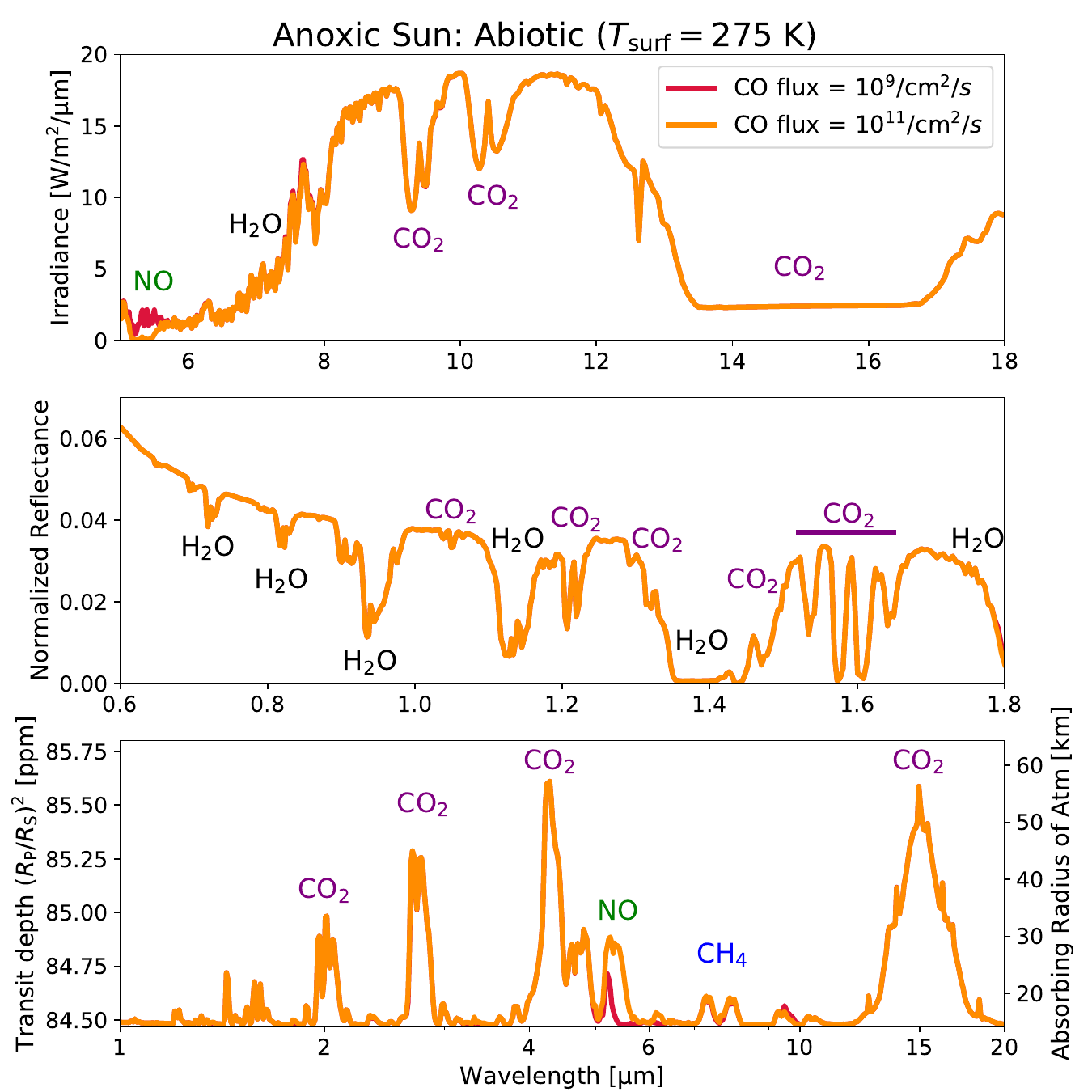}
         \caption{Simulated spectra: Emitted (\textit{top}, MIR, $R = 400$), reflected (\textit{middle}, NIR, $R = 400$), and transmitted light (\textit{bottom}, NIR-MIR, $R = 200$).}
         \label{Fig_A_Spectra_Sun_Anoxic_275_abiotic}
     \end{subfigure}
     \caption{Anoxic, abiotic Sun scenario ($T_\mathrm{surf} = \SI{275}{\kelvin}$): CO flux of $10^9$ and $10^{11} \, \si{molecules\per\centi\metre\squared\per\second}$, corresponding to lightning flash rates of $\sim 2$ and $\sim 160$ times modern Earth's, respectively.}
     \label{Fig_A_Anoxic_abiotic_Sun}
\end{figure*}

\begin{figure*}
     \centering
     \begin{subfigure}[b]{\columnwidth}
         \centering
         \includegraphics[width=\textwidth]{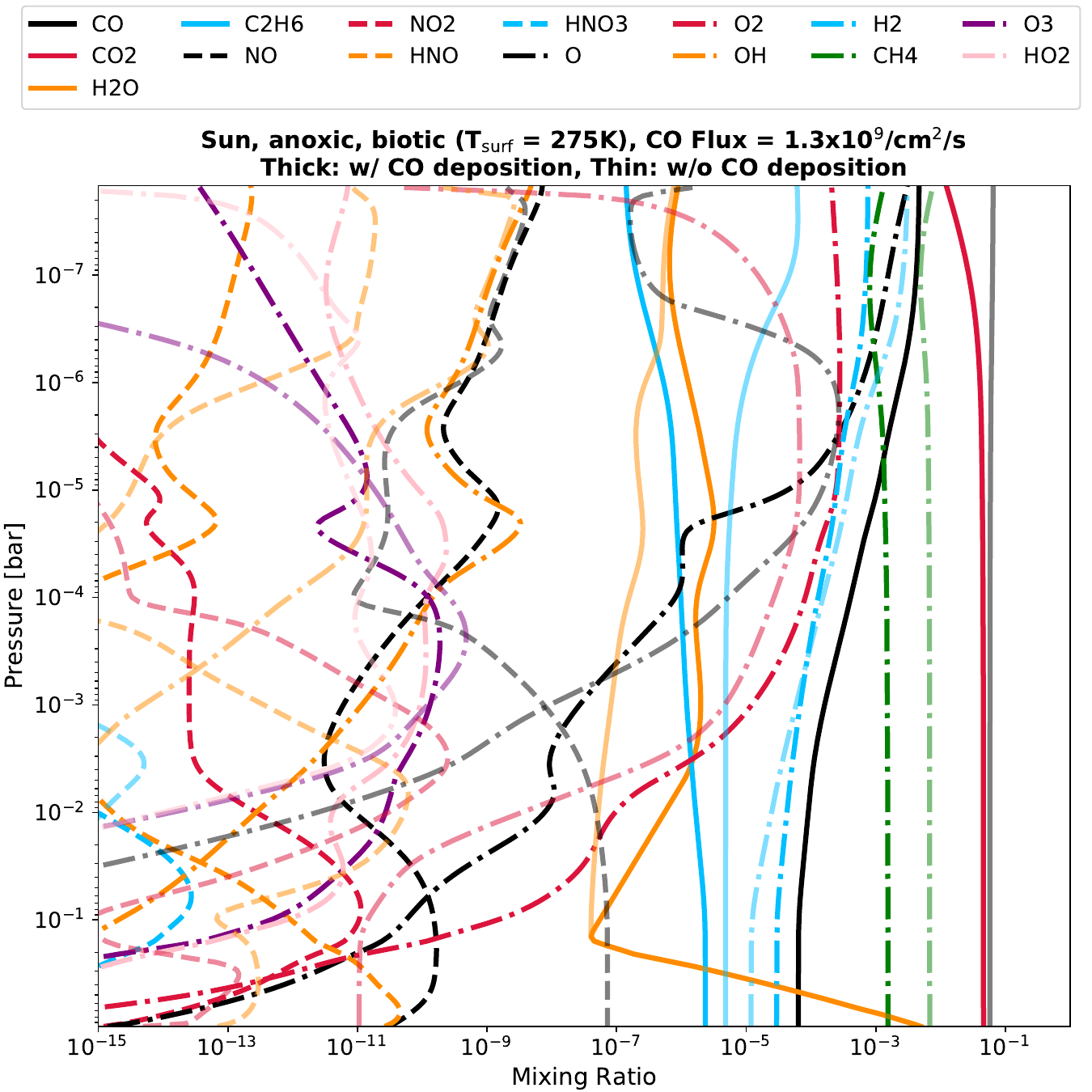}
         \caption{Atmospheric mixing ratios of most abundant species for the scenario with (\textit{thick}) and without (\textit{thin}) CO deposition.}
         \label{Fig_A_Sun_Anoxic_275_biotic_wdep_vs_ndep_PMR}
     \end{subfigure}
     \hfill
     \begin{subfigure}[b]{\columnwidth}
         \centering
         \includegraphics[width=\textwidth]{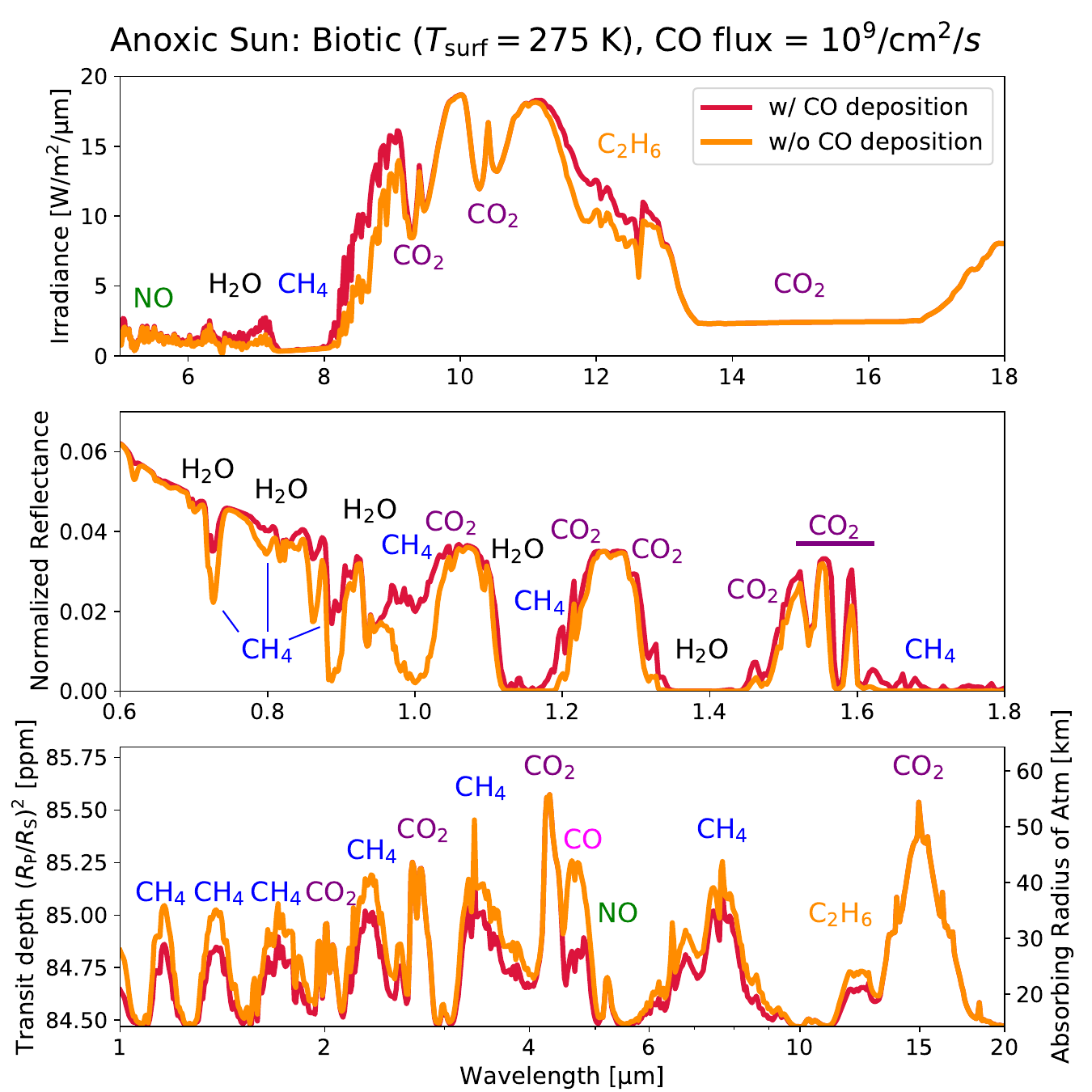}
         \caption{Simulated spectra: Emitted (\textit{top}, MIR, $R = 400$), reflected (\textit{middle}, NIR, $R = 400$), and transmitted light (\textit{bottom}, NIR-MIR, $R = 200$) for scenario with (\textit{red}) and without (\textit{orange}) CO deposition.}
         \label{Fig_A_comb_anoxic_sun_biotic_Spectra}
     \end{subfigure}
     \caption{Anoxic, biotic Sun scenario ($T_\mathrm{surf} = \SI{275}{\kelvin}$) with CO flux of $10^9 \, \si{molecules\per\centi\metre\squared\per\second}$, corresponding to lightning flash rate of $\sim 2$ times modern Earth's. Comparison between scenarios with and without CO deposition.
     }
     \label{Fig_A_Anoxic_biotic_Sun_wdep_vs_nodep}
\end{figure*}

\begin{figure*}
     \centering
     \begin{subfigure}[b]{\columnwidth}
         \centering
         \includegraphics[width=\textwidth]{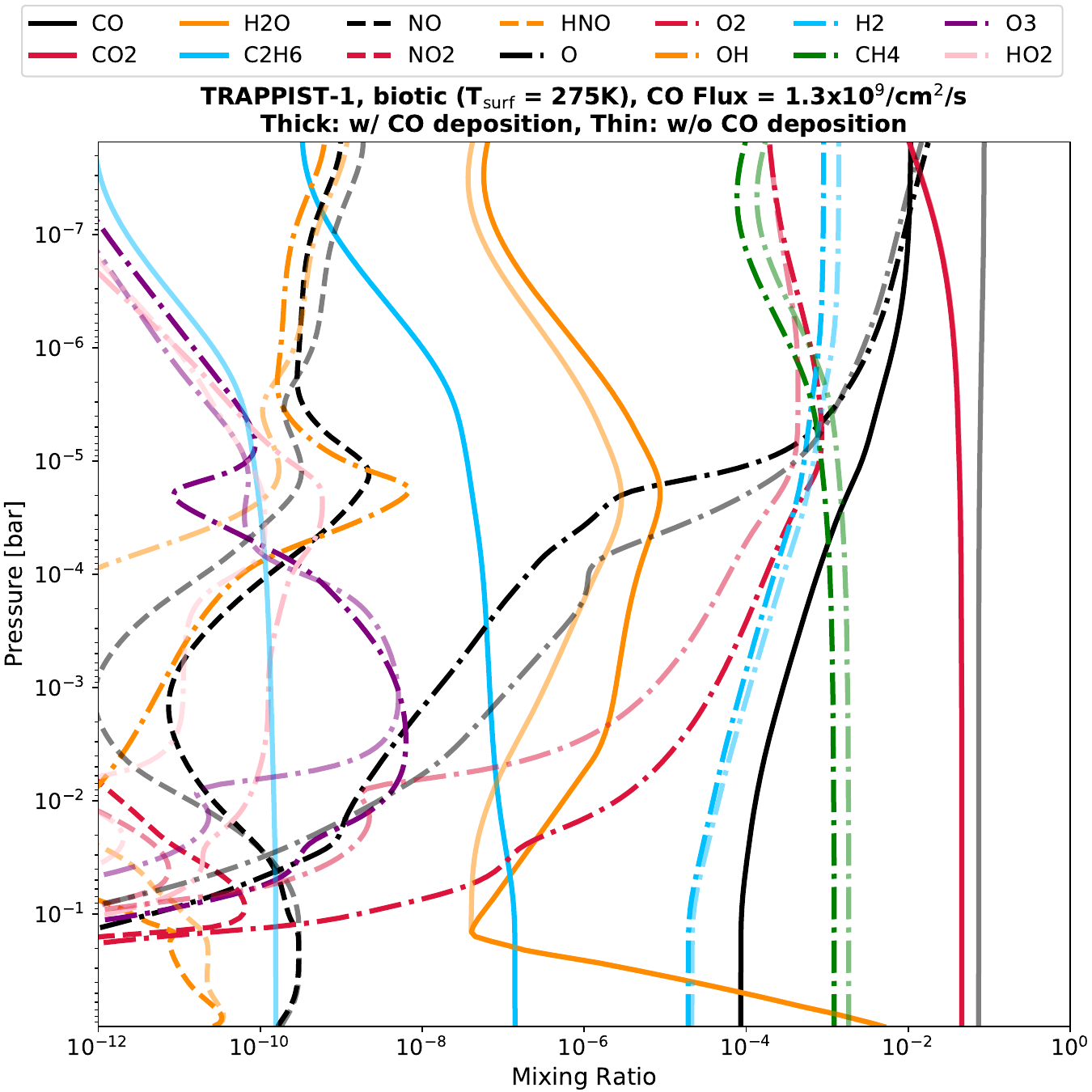}
         \caption{Atmospheric mixing ratios of most abundant species for the scenario with (\textit{thick}) and without (\textit{thin}) CO deposition.}
         \label{Fig_A_TR1_Anoxic_275_biotic_wdep_vs_ndep_PMR}
     \end{subfigure}
     \hfill
     \begin{subfigure}[b]{\columnwidth}
         \centering
         \includegraphics[width=\textwidth]{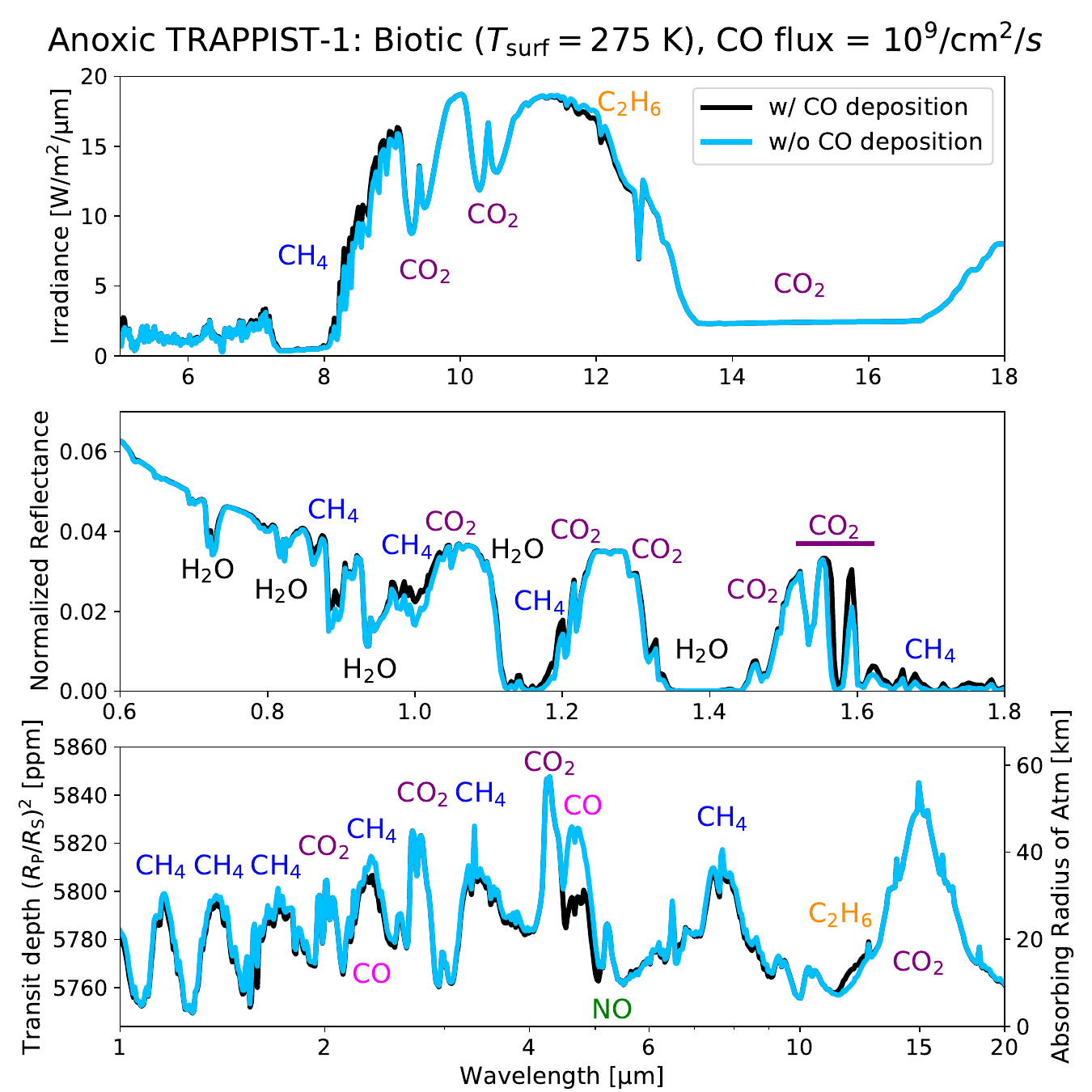}
         \caption{Simulated spectra: Emitted (\textit{top}, MIR, $R = 400$), reflected (\textit{middle}, NIR, $R = 400$), and transmitted light (\textit{bottom}, NIR-MIR, $R = 200$) for scenario with (\textit{black}) and without (\textit{blue}) CO deposition.}
         \label{Fig_A_comb_anoxic_tr1_biotic_Spectra}
     \end{subfigure}
     \caption{Anoxic, biotic TRAPPIST-1 scenario ($T_\mathrm{surf} = \SI{275}{\kelvin}$) with CO flux of $10^9 \, \si{molecules\per\centi\metre\squared\per\second}$, corresponding to lightning flash rate of $\sim 2$ times modern Earth's. Comparison between scenarios with and without CO deposition.
     }
     \label{Fig_A_Anoxic_biotic_TR1_wdep_vs_nodep}
\end{figure*}

\begin{figure*}
     \centering
     \begin{subfigure}[b]{\columnwidth}
         \centering
         \includegraphics[width=\textwidth]{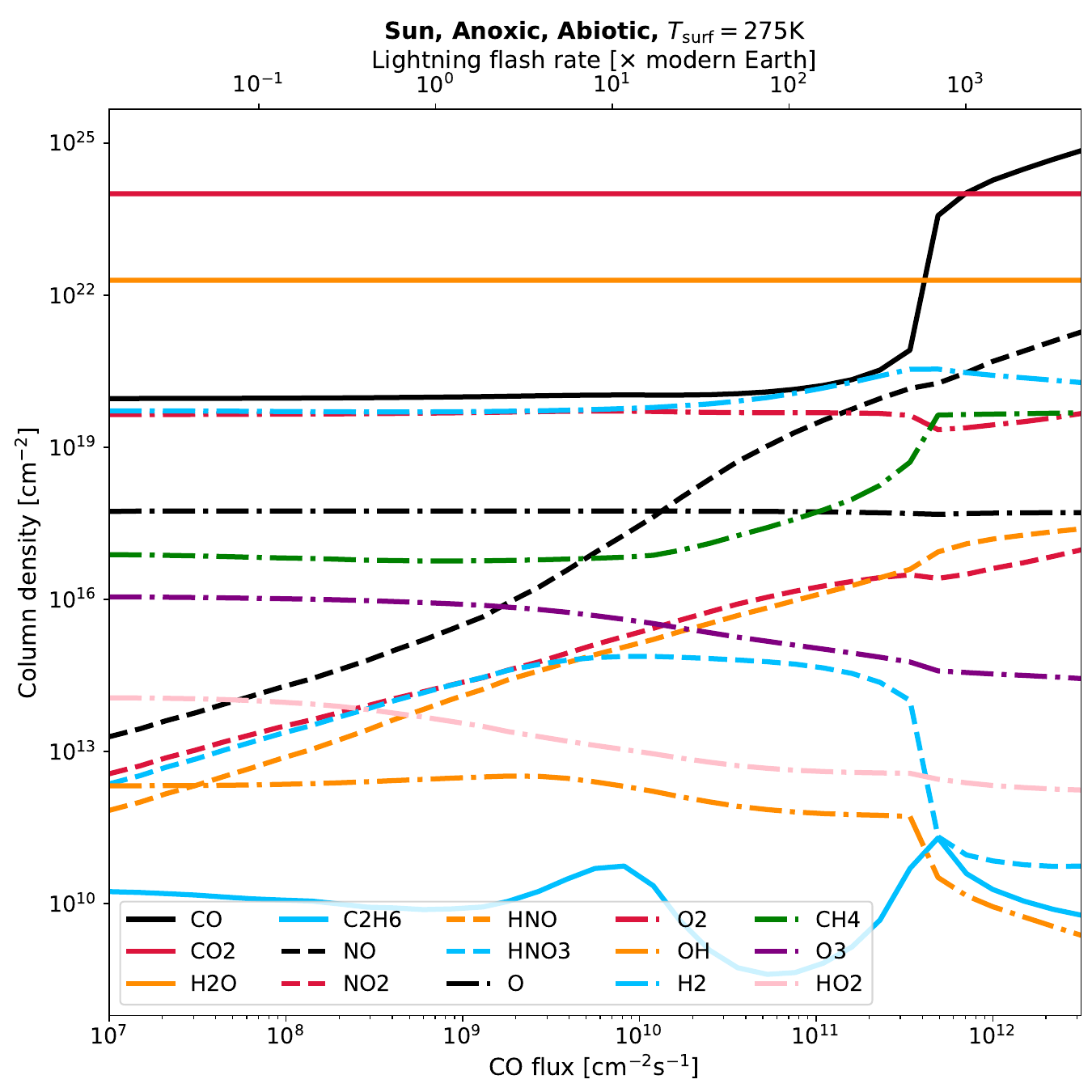}
         \label{Sun_Anoxic_275_abiotic_ncol}
     \end{subfigure}
     \hfill
     \begin{subfigure}[b]{\columnwidth}
         \centering
         \includegraphics[width=\textwidth]{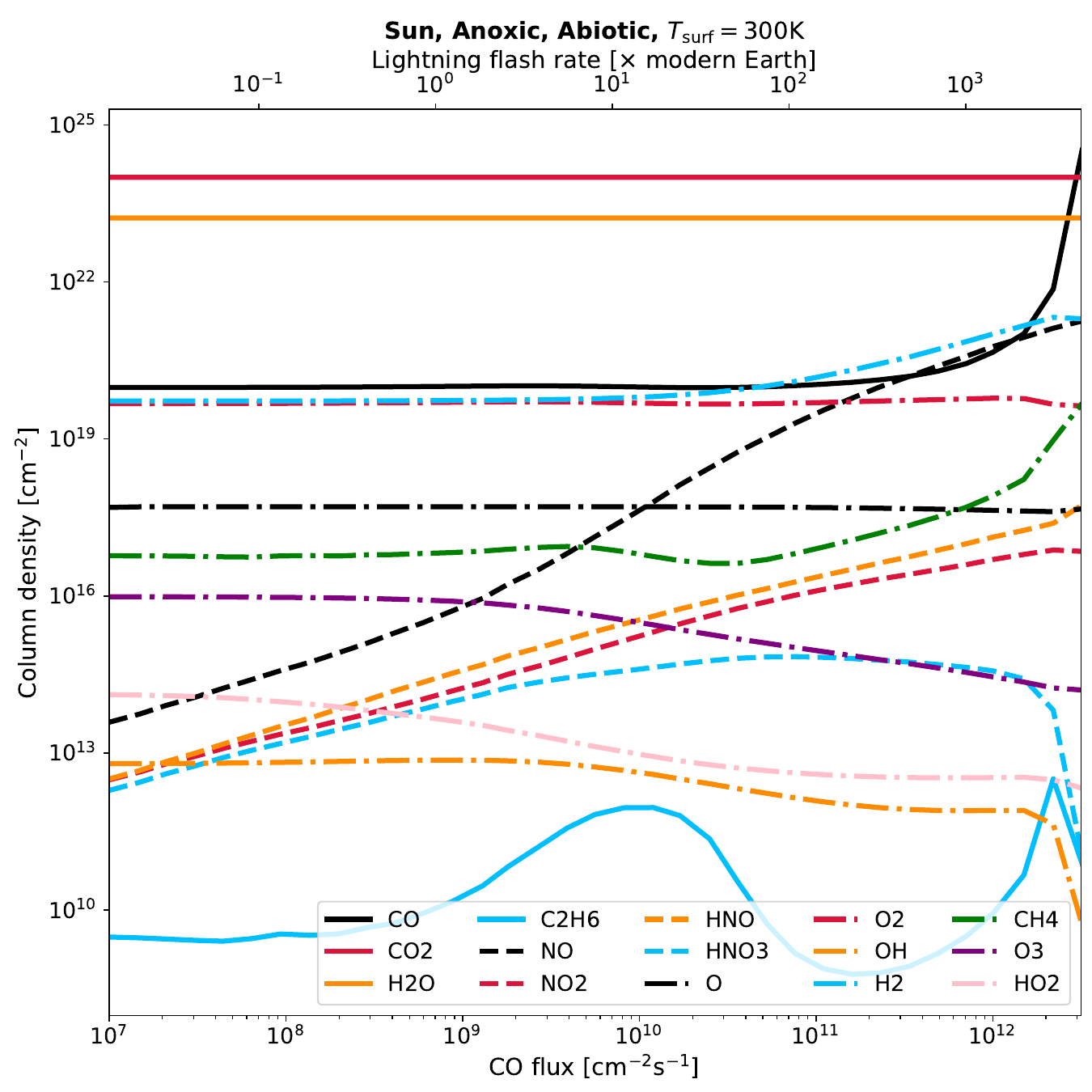}
         \label{Sun_Anoxic_300_abiotic_ncol}
     \end{subfigure}
     
     \begin{subfigure}[b]{\columnwidth}
         \centering
         \includegraphics[width=\textwidth]{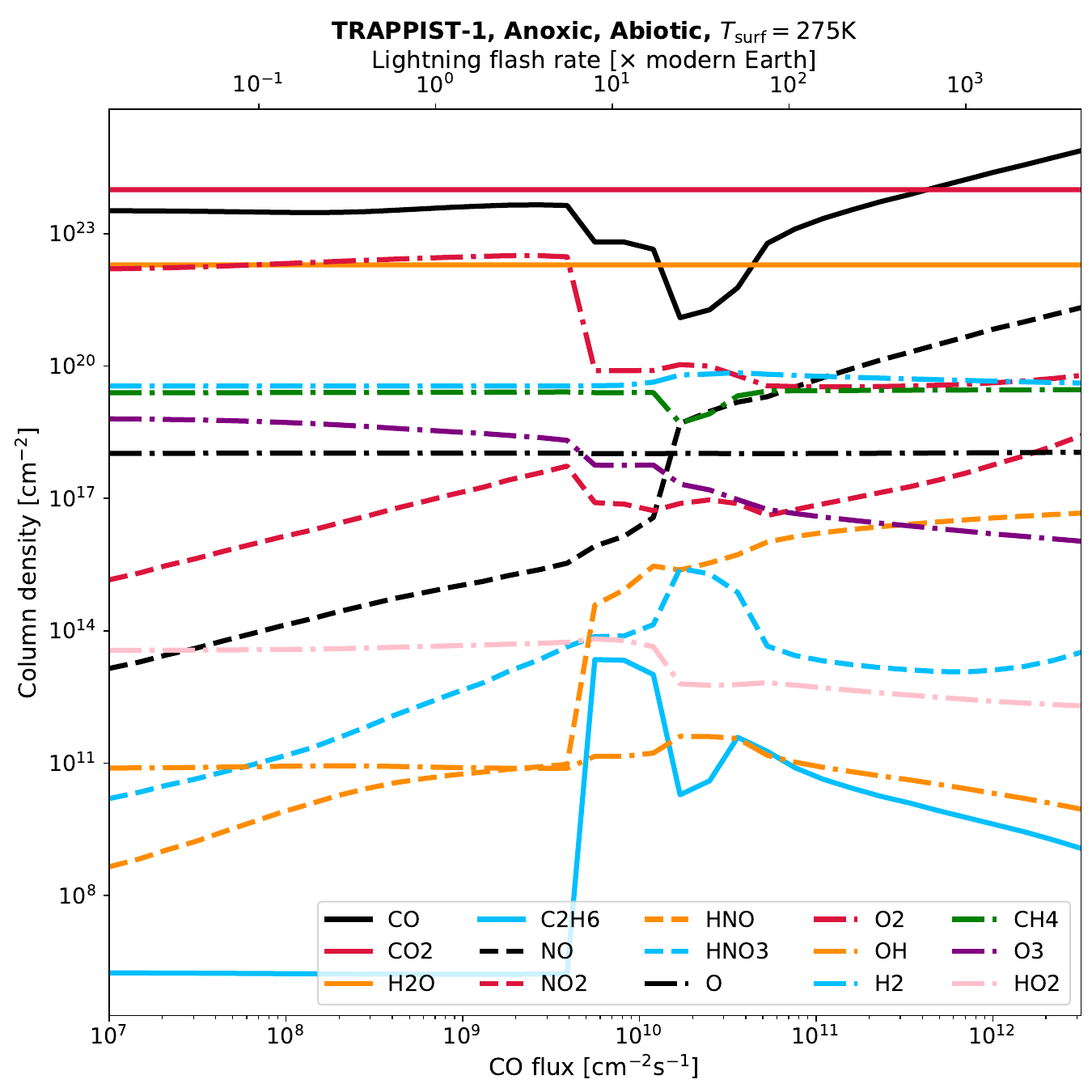}
         \label{Fig_A_TR1_Anoxic_275_abiotic_ncol}
     \end{subfigure}
     \hfill
     \begin{subfigure}[b]{\columnwidth}
         \centering
         \includegraphics[width=\textwidth]{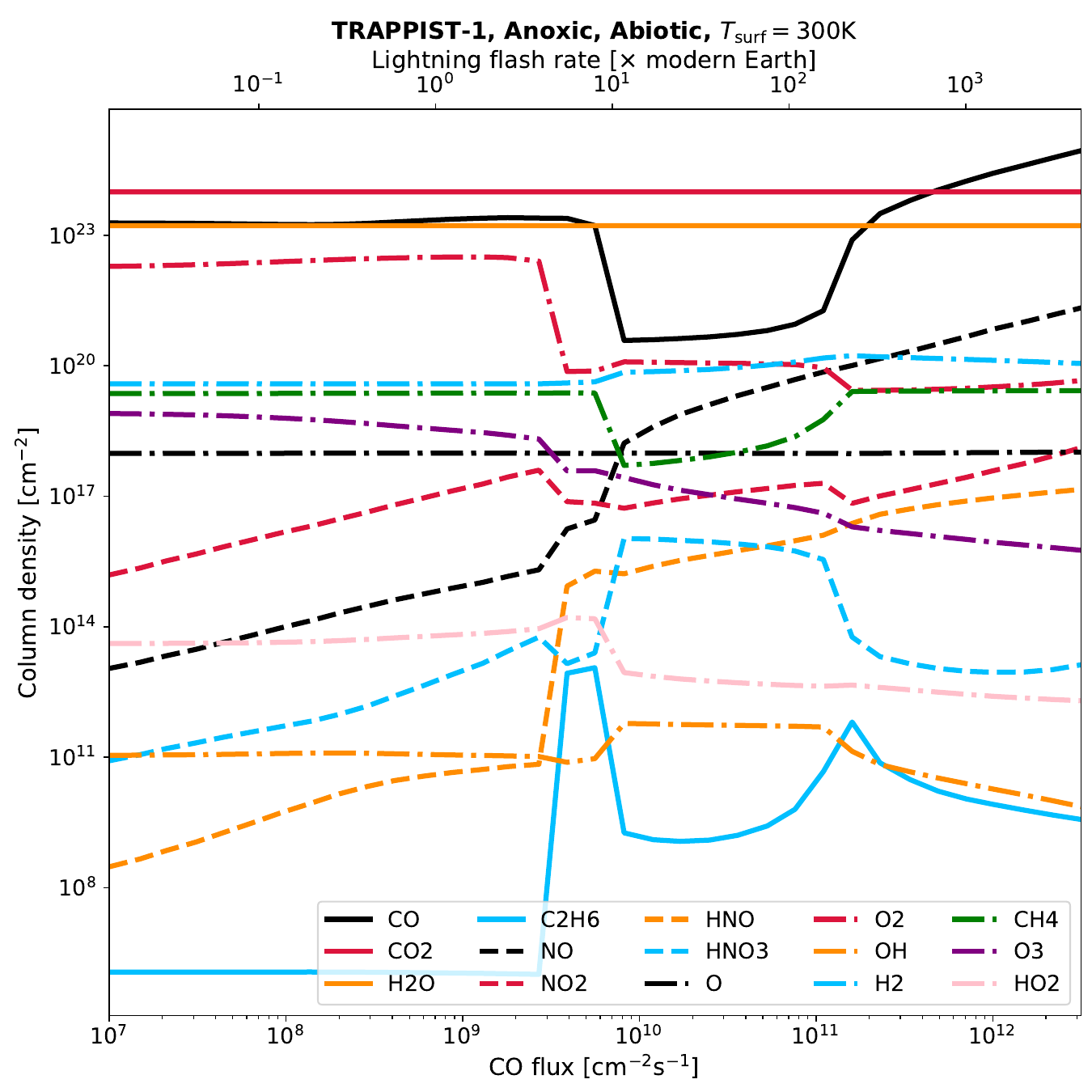}
         \label{Fig_A_TR1_Anoxic_300_abiotic_ncol}
     \end{subfigure}
        \caption{Column densities for a range of CO and NO fluxes. Anoxic abiotic cases.}
        \label{Fig_A_Anoxic_abiotic_ncol}
\end{figure*}

\begin{figure*}
     \centering
     \begin{subfigure}[b]{\columnwidth}
         \centering
         \includegraphics[width=\textwidth]{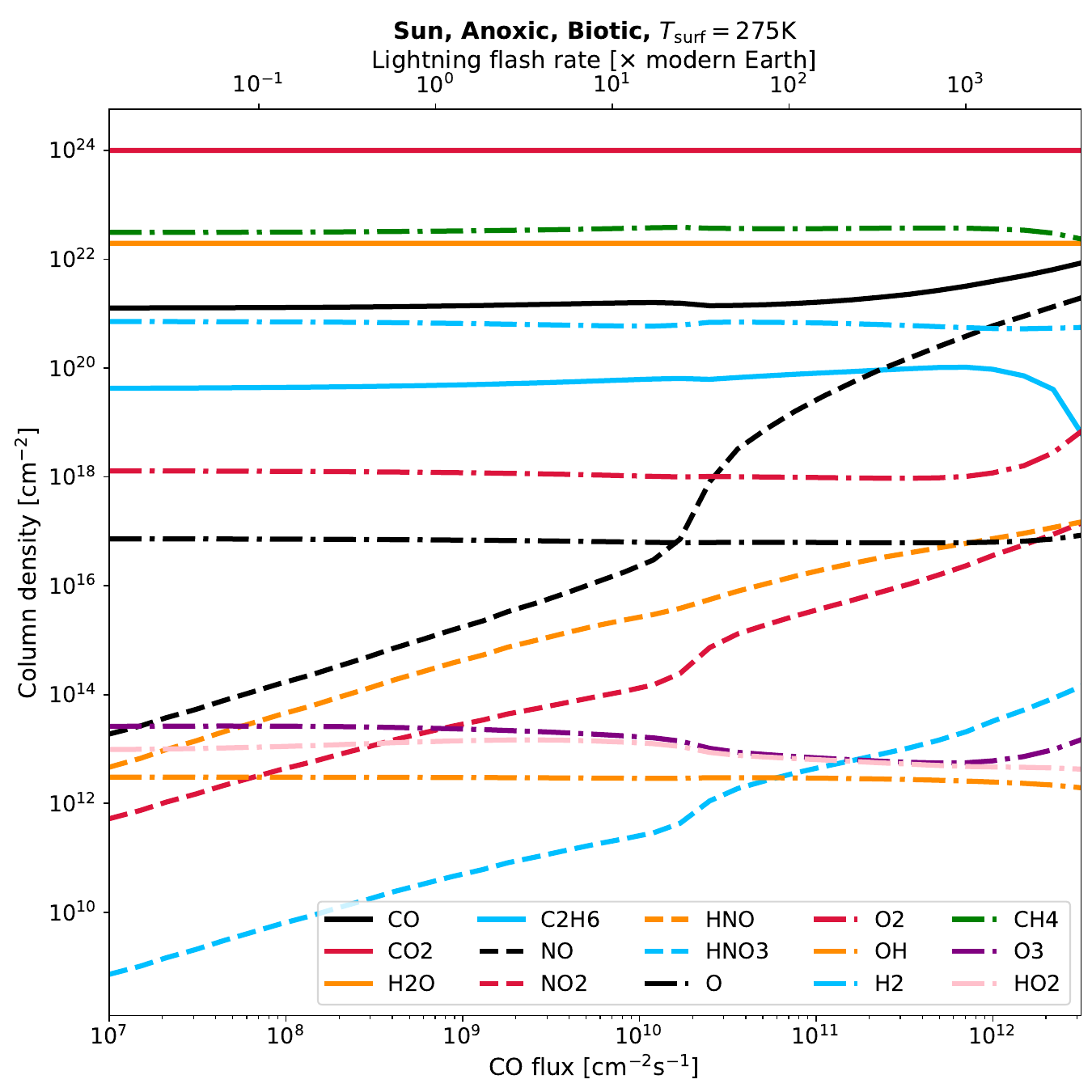}
         \label{Sun_Anoxic_275_biotic_ncol}
     \end{subfigure}
     \hfill
     \begin{subfigure}[b]{\columnwidth}
         \centering
         \includegraphics[width=\textwidth]{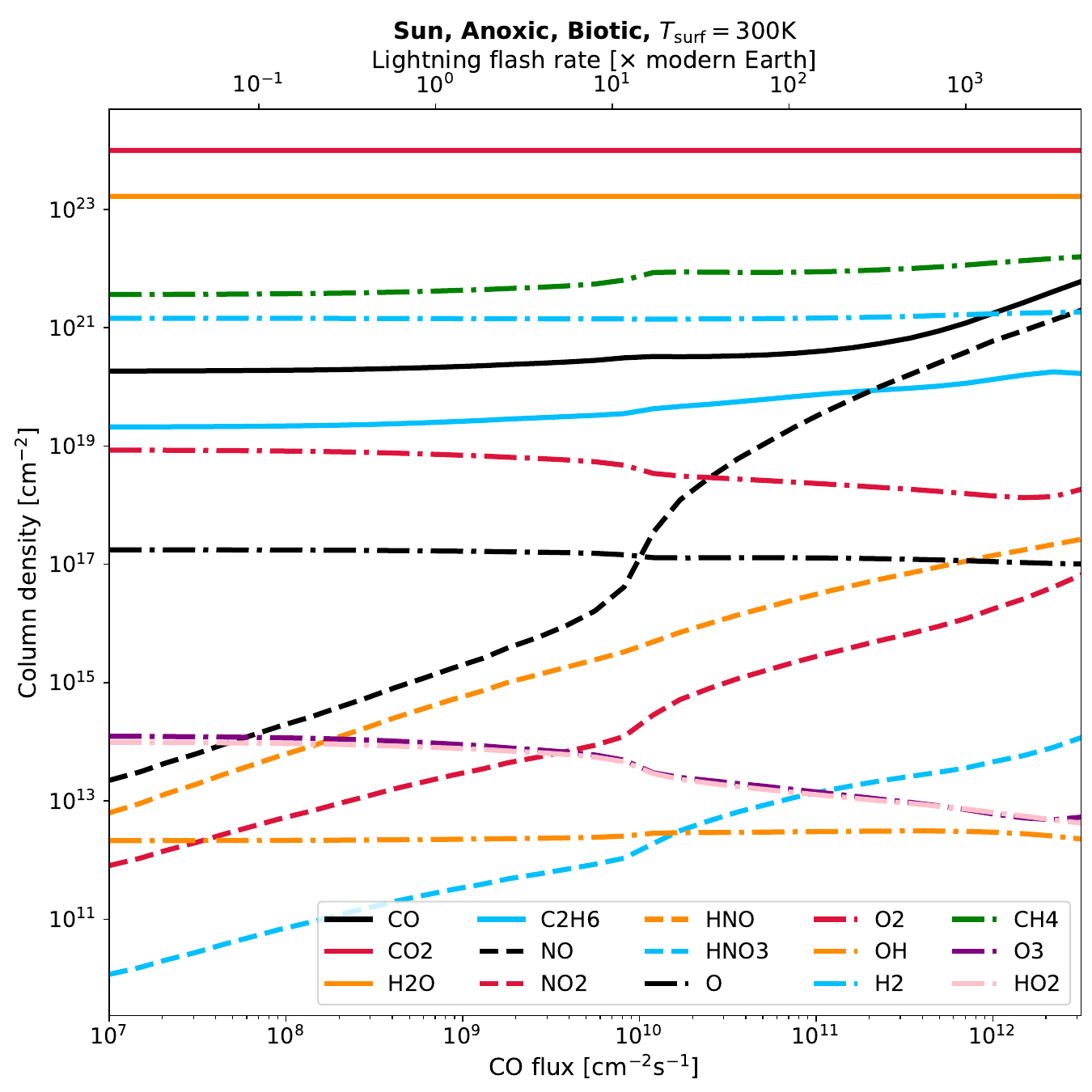}
         \label{Sun_Anoxic_300_biotic_ncol}
     \end{subfigure}
     
     \begin{subfigure}[b]{\columnwidth}
         \centering
         \includegraphics[width=\textwidth]{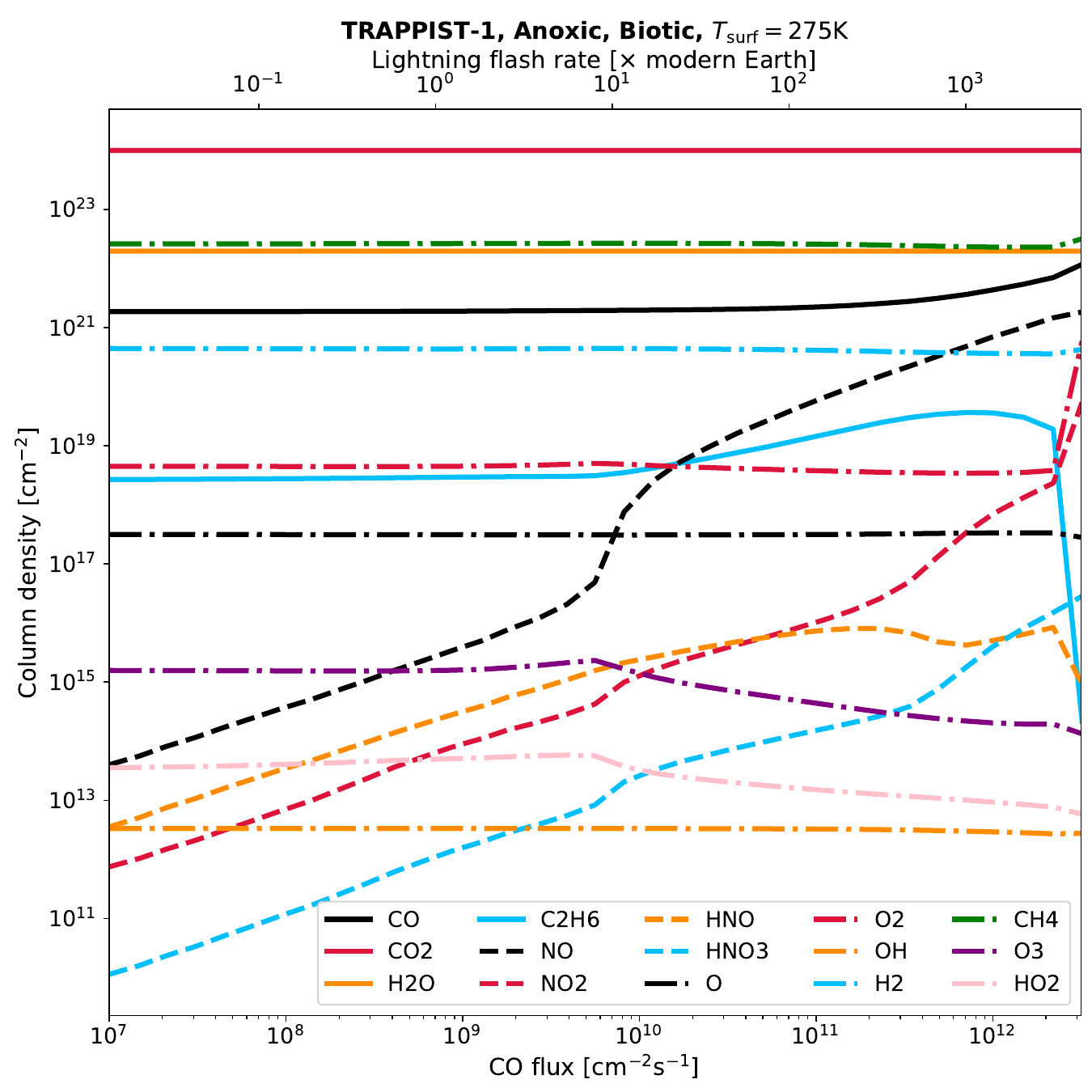}
         \label{TR1_Anoxic_275_biotic_ncol}
     \end{subfigure}
     \hfill
     \begin{subfigure}[b]{\columnwidth}
         \centering
         \includegraphics[width=\textwidth]{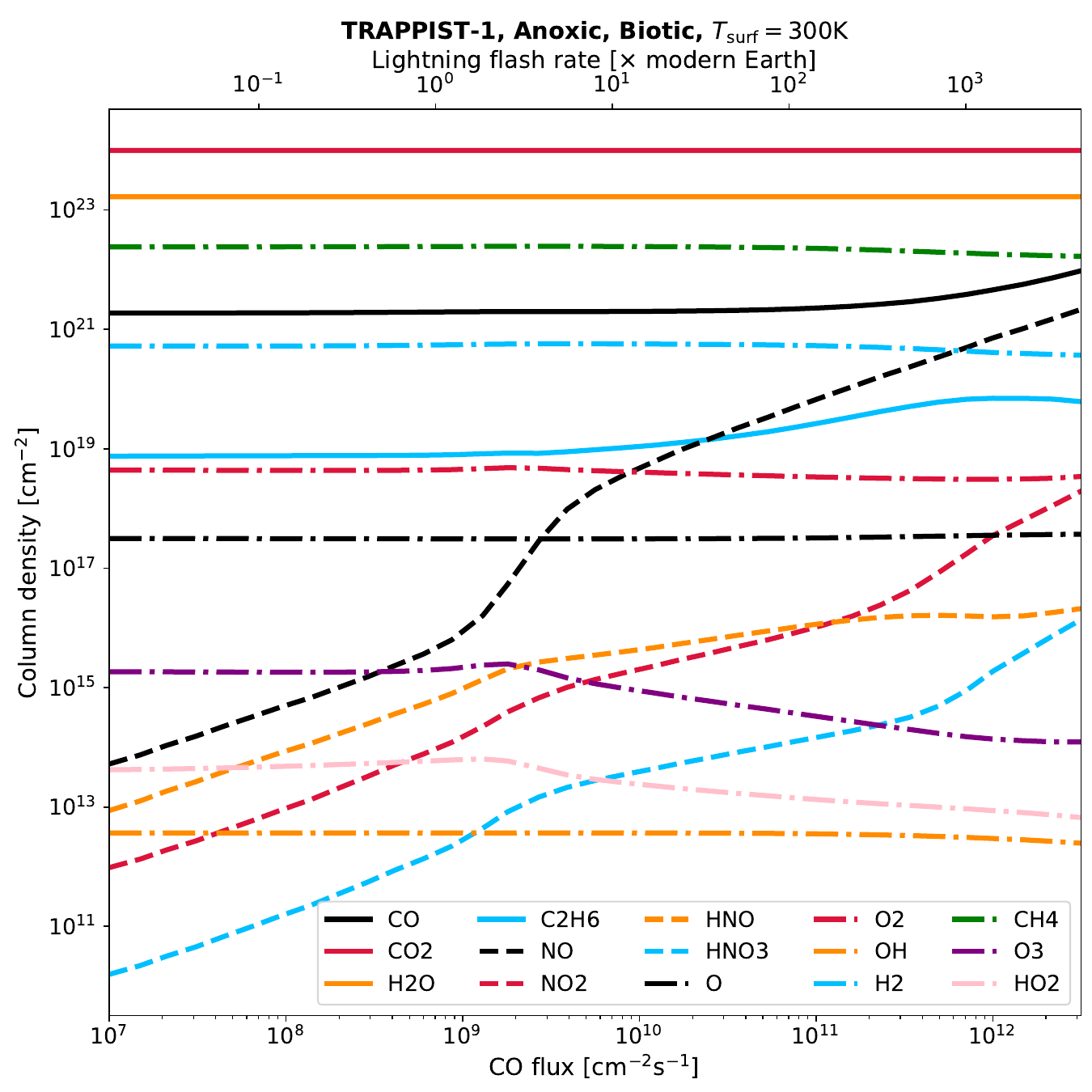}
         \label{TR1_Anoxic_300_biotic_ncol}
     \end{subfigure}
        \caption{Column densities for a range of CO and NO fluxes. Anoxic biotic cases.}
        \label{Fig_anoxic_biotic_ncol}
\end{figure*}

\begin{figure*}
     \centering
     \begin{subfigure}[b]{\columnwidth}
         \centering
         \includegraphics[width=\textwidth]{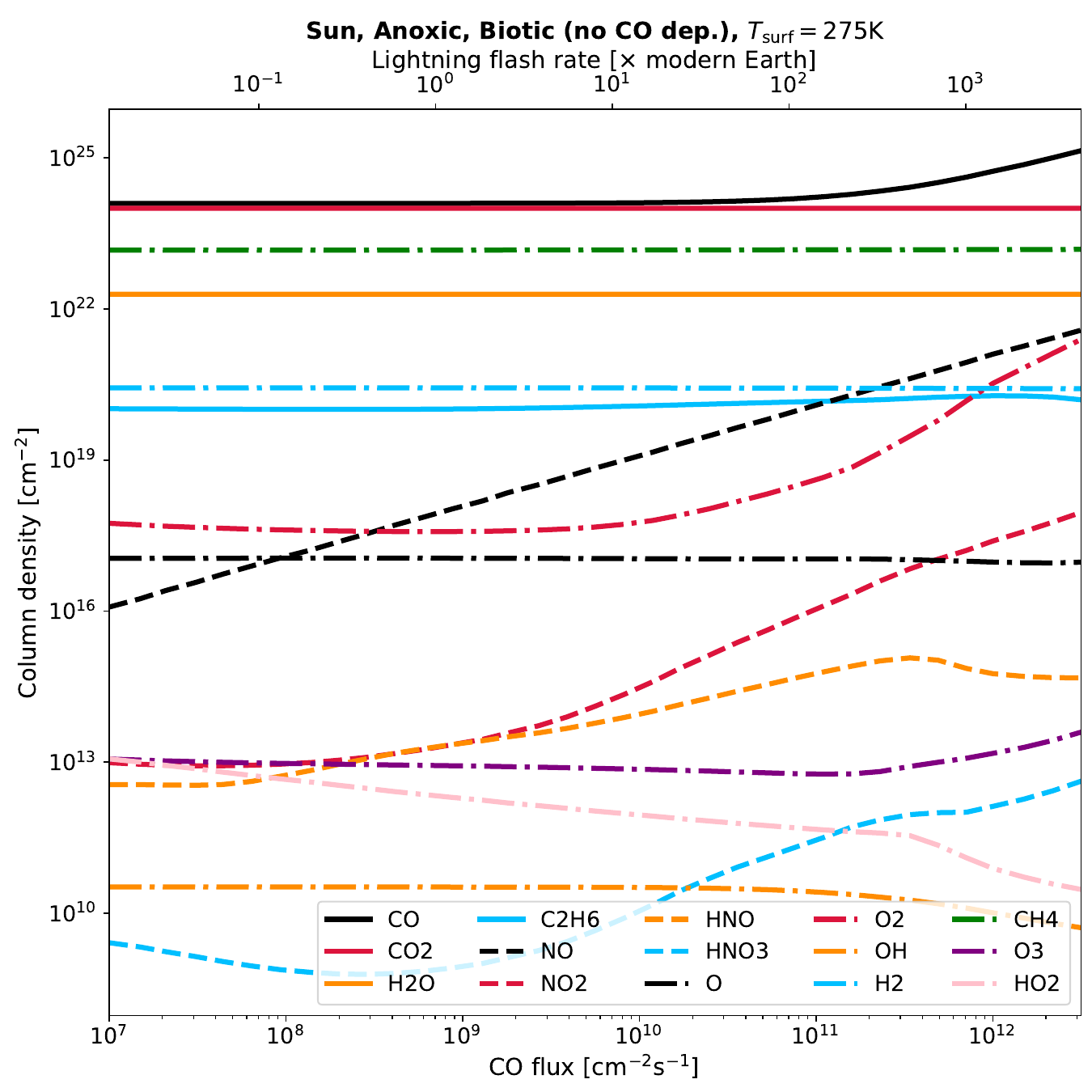}
         \label{Sun_Anoxic_275_biotic_nodep_ncol}
     \end{subfigure}
     \hfill
     \begin{subfigure}[b]{\columnwidth}
         \centering
         \includegraphics[width=\textwidth]{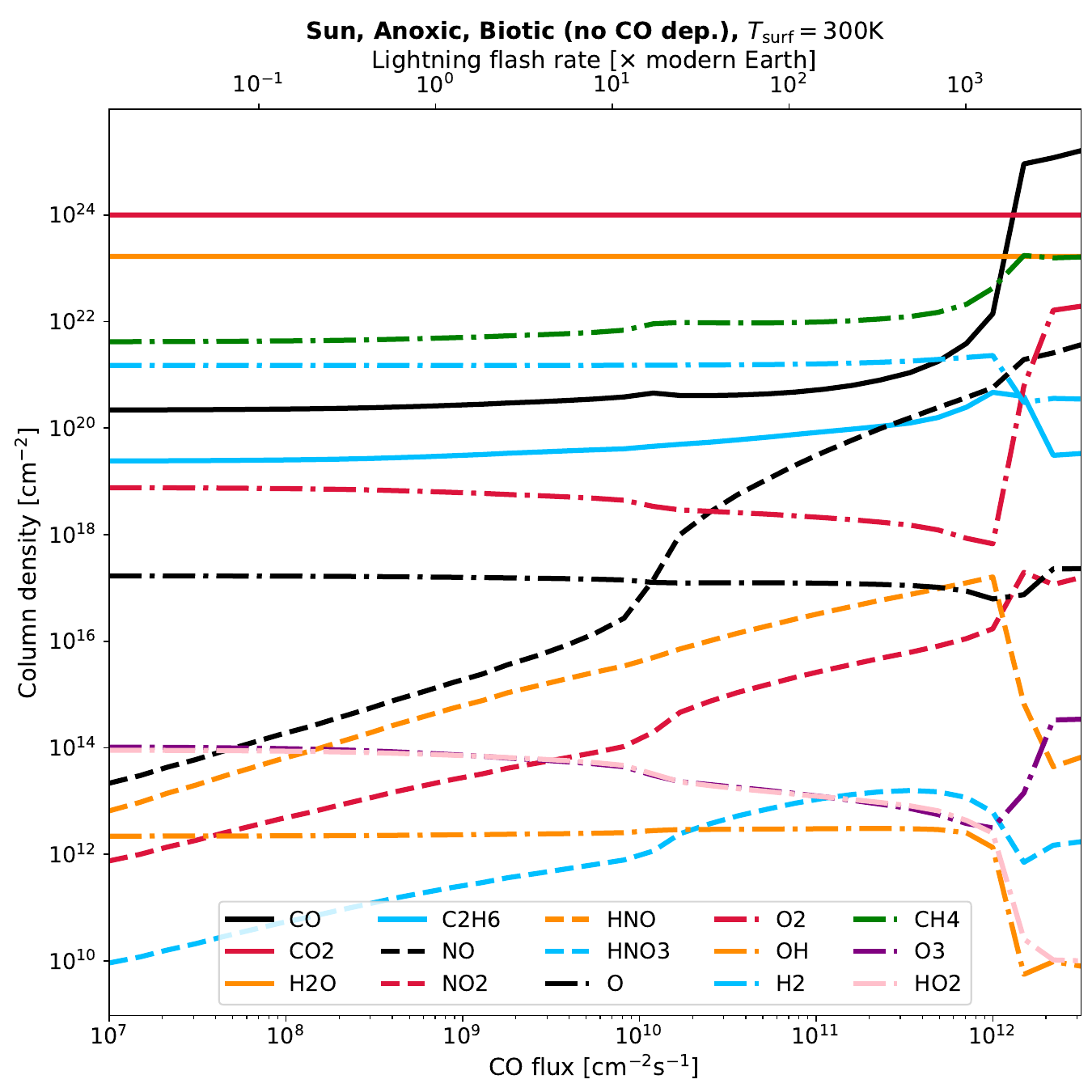}
         \label{Sun_Anoxic_300_biotic_nodep_ncol}
     \end{subfigure}
     
     \begin{subfigure}[b]{\columnwidth}
         \centering
         \includegraphics[width=\textwidth]{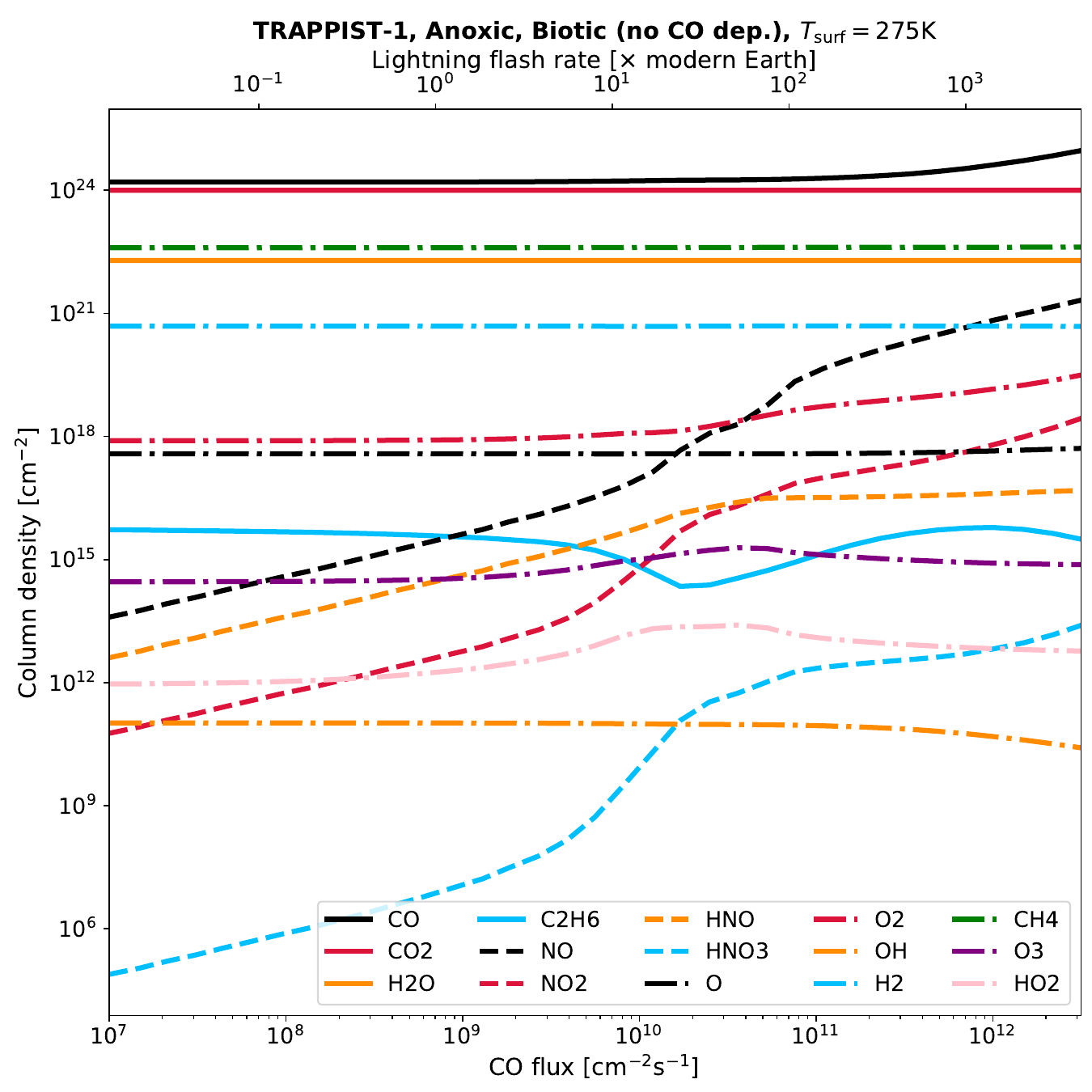}
           \label{TR1_Anoxic_275_biotic_nodep_ncol}
     \end{subfigure}
     \hfill
     \begin{subfigure}[b]{\columnwidth}
         \centering
         \includegraphics[width=\textwidth]{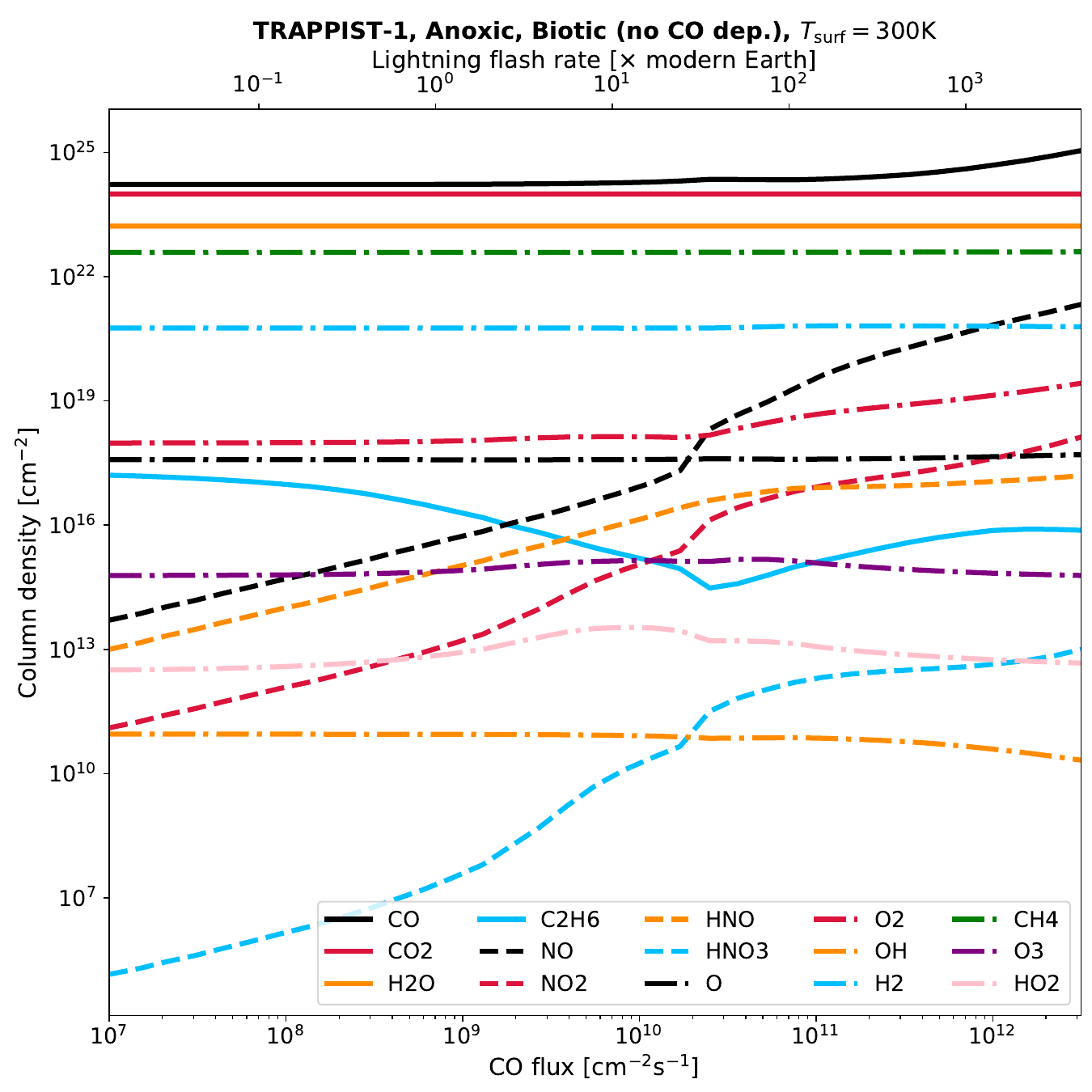}
         \label{TR1_Anoxic_300_biotic_nodep_ncol}
     \end{subfigure}
        \caption{Column densities for a range of CO and NO fluxes. Anoxic biotic (no CO deposition) cases.}
        \label{Fig_anoxic_biotic_nodep_ncol}
\end{figure*}

\begin{figure*}
     \centering
     \begin{subfigure}[b]{\columnwidth}
         \centering
         \includegraphics[width=\textwidth]{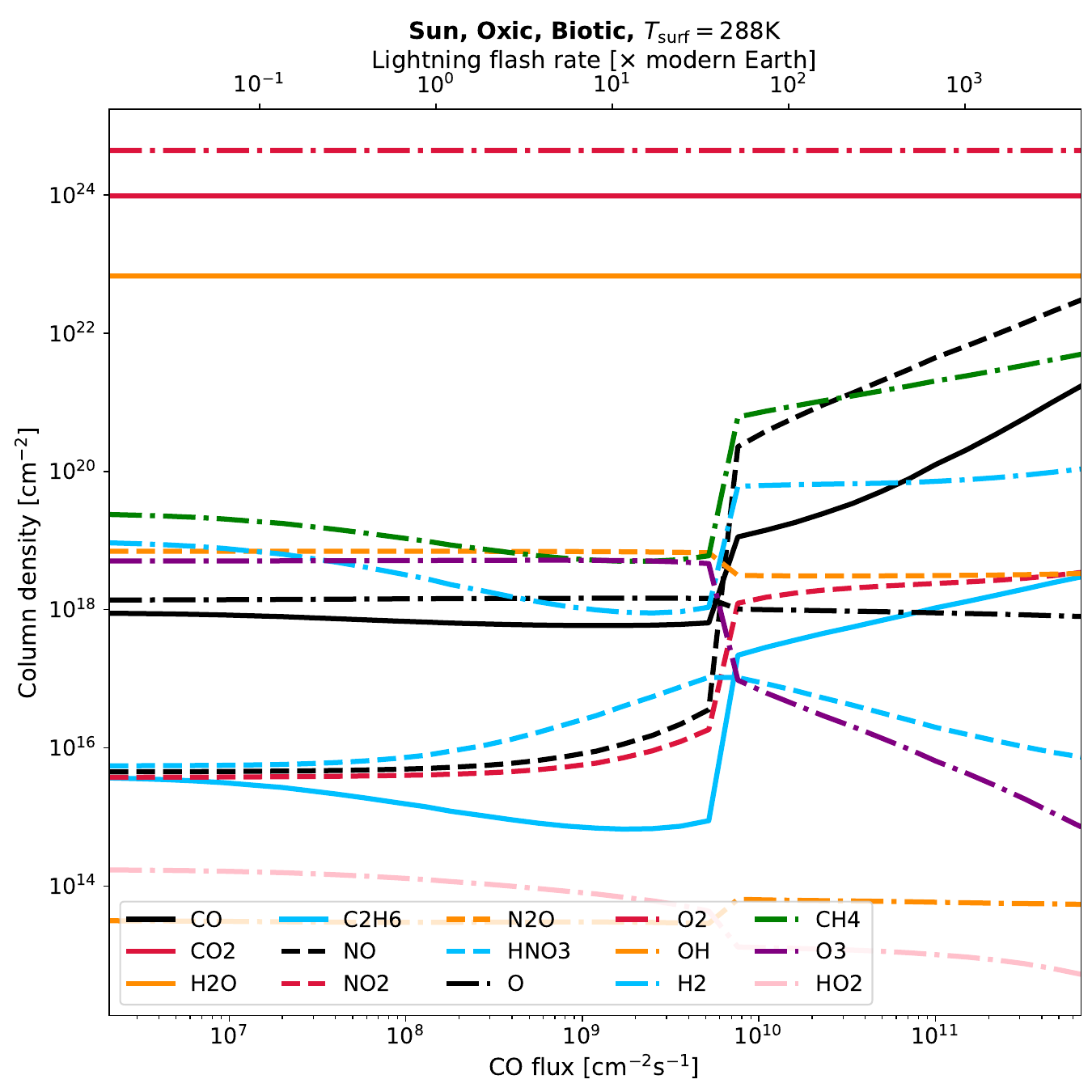}
         \label{Sun_Oxic_noO_ncol}
     \end{subfigure}
     \hfill
     \begin{subfigure}[b]{\columnwidth}
         \centering
         \includegraphics[width=\textwidth]{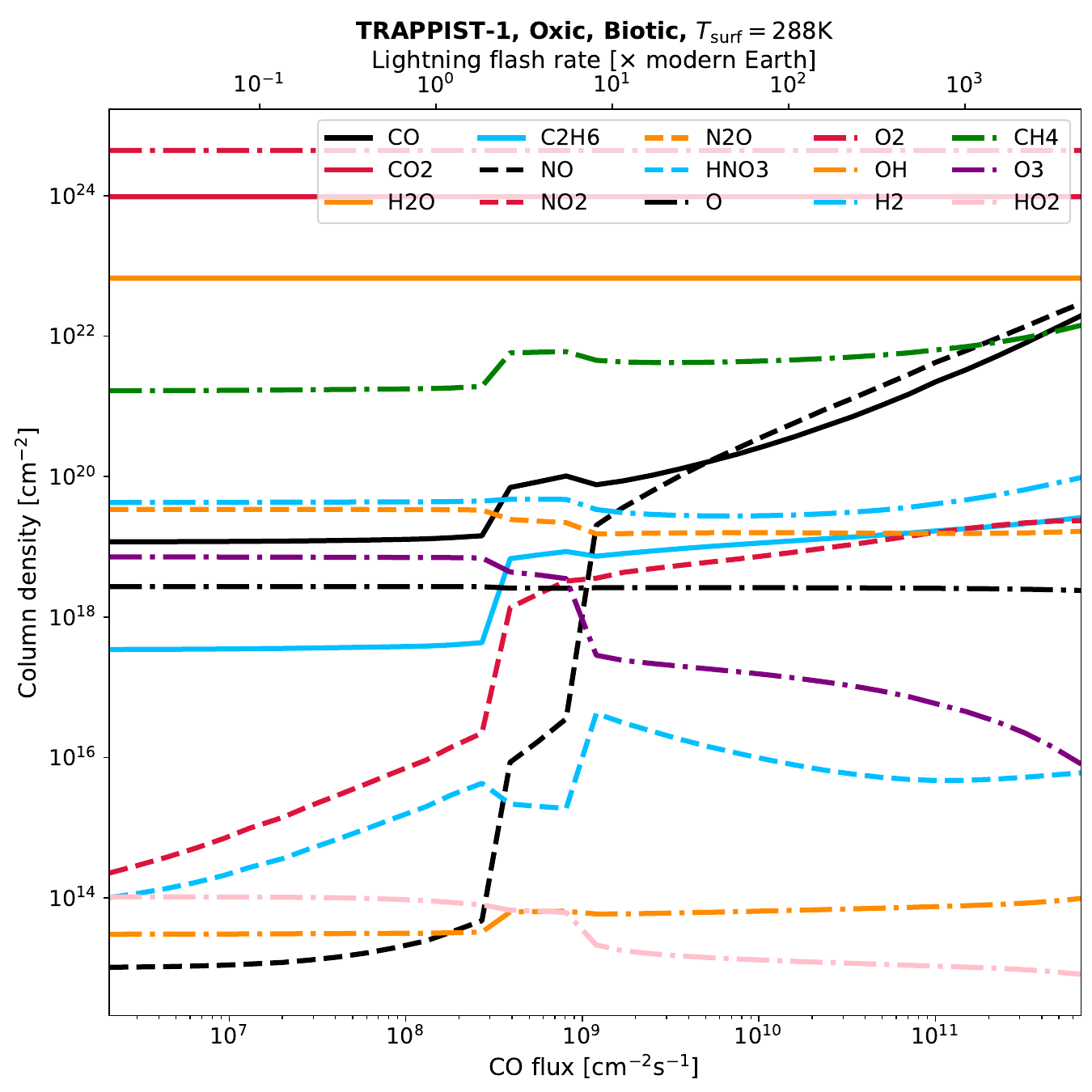}
         \label{TR1_Oxic_noO_ncol}
     \end{subfigure}
     \caption{Column densities for a range of CO and NO fluxes. Oxic biotic cases.}
     \label{Fig_A_Oxic_biotic_ncol}
\end{figure*}

This section includes a table with the detailed parameters of the photochemical model used (Table~\ref{Tab_Photochem_Parameters}).
It further contains the atmospheric mixing ratio profiles for the spectra shown in Fig.~\ref{Fig_Spectra_TR1} (Fig.~\ref{Fig_A_PMR_Spectra_TR1_comp}) and further spectra for different scenarios with the corresponding atmospheric mixing ratio profiles (Fig.~\ref{Fig_A_oxic_biotic_Sun} - \ref{Fig_A_Anoxic_biotic_TR1_wdep_vs_nodep}), as well as the column densities of the major atmospheric constituents for the whole range of CO fluxes and all scenarios presented in this paper (Fig.~\ref{Fig_A_Anoxic_abiotic_ncol} - \ref{Fig_A_Oxic_biotic_ncol}).

\begin{table*}[ht]
    \begin{center}
    \caption{Input parameters for our photochemical model}
    \begin{tabular}{w{c}{2cm}w{c}{3cm}w{c}{5cm}w{c}{5cm}}
    	\toprule
    	\multirow{2}{*}{Species} & \multirow{2}{*}{Mixing ratio} & Flux & Deposition velocity \\ 
        & & [$\si{molecules\per\centi\metre\squared\per\second}$] &[$\si{\centi\metre\per\second}$] \\
    	\noalign{\smallskip}
	    \hline \hline
	    \noalign{\smallskip}
    	\ce{CO2}                          & 4.6\%              & -                                          & -                                     \\
    	\midrule
    	\multirow{2}{*}{\ce{O2}}          & Anoxic: 0\%        & \multirow{2}{*}{-}                         & \multirow{2}{*}{-}                    \\
    	                                  & Oxic: 21\%         &                                            &                                       \\
    	\midrule
    	\ce{H2O}${}^{a}$                  & rel. hum.          & -                                          & -                                     \\
    	\midrule
    	\multirow{2}{*}{\ce{CH4}}         & \multirow{2}{*}{-} & Abiotic: $10^8$                            & \multirow{2}{*}{-}                    \\
    	                                  &                    & Biotic: $10^{11}$                          &                                       \\
    	\midrule
    	\multirow{2}{*}{\ce{CO}${}^{b}$}  & \multirow{2}{*}{-} & Anoxic: $10^7 - 3.2\times10^{12}$          & Abiotic: $10^{-8}$                    \\
    	                                  &                    & Oxic: $2.1\times10^6 - 6.7\times10^{11}$   & Biotic: $1.2 \times 10^{-4}$          \\
    	\midrule
    	\multirow{2}{*}{\ce{NO}${}^{b,c}$}& \multirow{2}{*}{-} & Anoxic: $5.4\times10^6 - 1.7\times10^{12}$ & \multirow{2}{*}{$3 \times 10^{-4}$}   \\
    	                                  &                    & Oxic: $3.5\times10^7 - 1.1\times10^{13}$   &                                       \\
    	\midrule 
    	\ce{NO2}                          & -                  & -                                          & $3 \times 10^{-3}$                    \\
    	\midrule
    	\ce{HNO}                          & -                  & -                                          & 1                                     \\
    	\midrule
    	\ce{HNO3}${}^{c}$                 & -                  & -                                          & 0.2                                   \\
    	\midrule
    	\multirow{2}{*}{\ce{H2}${}^{b}$}  & \multirow{2}{*}{-} & Anoxic: $10^{10}$ (anoxic)                 & \multirow{2}{*}{$2.4 \times 10^{-4}$} \\
    	                                  &                    & Oxic: 0                                    &                                       \\
    	\midrule
    	\multirow{2}{*}{\ce{H2S}${}^{b}$} & \multirow{2}{*}{-} & Anoxic: $3.5 \times 10^8$                  & \multirow{2}{*}{0.02}                 \\
    	                                  &                    & Oxic: $2 \times 10^8$                      &                                       \\
    	\midrule
    	\multirow{2}{*}{\ce{SO2}${}^{b}$} & \multirow{2}{*}{-} & Anoxic: $3.5 \times 10^9$                  & \multirow{2}{*}{1}                    \\
    	                                  &                    & Oxic: $9 \times 10^9$                      &                                       \\
    	\bottomrule
    \end{tabular}
    \\
    ${}^{a}$ Depending on surface temperature, ${}^{b}$ Distributed over lower $\SI{10}{\kilo\metre}$ of atmosphere, ${}^{c}$ Only dry deposition 
    \label{Tab_Photochem_Parameters}
    \end{center}
\end{table*}


\end{appendix}

\end{document}